\documentclass[12pt]{article}
\usepackage[ascii]{inputenc}
\usepackage{amsmath,amssymb,amsfonts,amsthm}
\usepackage[margin=2.4cm]{geometry}
\usepackage{graphicx}
\usepackage{bbm}
\usepackage[caption=false]{subfig}
\usepackage[pdftex,bookmarks=false,colorlinks=true,linkcolor=blue,
citecolor=blue,filecolor=black,urlcolor=blue]{hyperref}

\providecommand{\ud}{\mathrm{d}}
\numberwithin{equation}{section}

\begin{document}

\title{Shocks, rarefaction waves, and current fluctuations \\ for anharmonic chains}

\author{Christian B. Mendl\footnote{Geballe Laboratory for Advanced Materials, Stanford University, 476 Lomita Mall, California 94305, and Stanford Institute for Materials and Energy Sciences, SLAC National Accelerator Laboratory, Menlo Park, California 94025, USA. Email: \href{mailto:mendl@stanford.edu}{mendl@stanford.edu}}
$\,$ and Herbert Spohn\footnote{Zentrum Mathematik and Physik Department,
Technische Universit\"at M\"unchen,
Boltzmannstra{\ss}e 3, 85747 Garching bei M\"unchen, Germany.
Email: \href{mailto:spohn@tum.de}{spohn@tum.de}}}

\date{July 14, 2016}

\maketitle

\begin{abstract}
The nonequilibrium dynamics of anharmonic chains is studied by imposing an initial domain-wall state, in which the two half lattices are prepared in equilibrium with distinct parameters. We analyse the Riemann problem for the corresponding Euler equations and, in specific cases, compare with molecular dynamics. Additionally, the fluctuations of time-integrated currents are investigated. In analogy with the KPZ equation, their typical fluctuations should be of size $t^{1/3}$ and have a Tracy-Widom GUE distributed amplitude. The proper extension to anharmonic chains is explained and tested through molecular dynamics. Our results are calibrated against the stochastic LeRoux lattice gas.
\end{abstract}

\newpage

\tableofcontents

\newpage

\section{Introduction}

The cold atom community has revived the study of the approach to thermal equilibrium for large isolated quantum systems. We refer to \cite{RDO08,EF16} and references therein on previous work. Most accessible, both numerically and experimentally, are lattice systems in one dimension. One dimension is peculiar, since there are models with an extensive number of locally conserved fields. Examples of such quantum integrable systems are the XXZ spin chain and the continuum Lieb-Liniger $\delta$-Bose gas. Obviously the pathway to equilibrium will depend crucially on whether the system is integrable or not \cite{BCH11,SM13,LLMM15}. But in addition there is also the dependence on initial conditions which is potentially overwhelming. One could prepare the system already in thermal equilibrium and study the response to small initial perturbations \cite{KIM13}. These are the much investigated time response and correlation functions in equilibrium. The initial state could be translation invariant, to some extent thereby suppressing the mostly slow spatial variations \cite{BCH11}. Recently initial domain-wall states have become very popular \cite{VKM15,CDY16}. Such a state is obtained by joining two distinct thermal states at a single point (and at two points in case of periodic boundary conditions). Domain-wall states will be the main focus of our contribution.

Browsing the introductions to the papers mentioned above, one might have the impression that the approach to equilibrium for classical systems in one dimension is a well-covered topic. We study here Fermi-Pasta-Ulam type anharmonic chains with domain-wall initial conditions and are not aware of any previous systematic study. The structure of equilibrium time-correlations for such chains has been elucidated only recently \cite{Spohn2014, MeSp13}. In particular one now understands the link to anomalous transport which is most directly observed when coupling the chain to thermal reservoirs at distinct temperatures, see \cite{L16}. As in the quantum world, there are integrable chains, in our context the most famous one being the Toda chain. But the KAM theorem signals in addition the possibility that, as a function of the energy, the structure may change from integrable to chaotic. This energy threshold is fascinating from the perspective of nonlinear dynamics and has attracted considerable attention \cite{PCCFC05}. We hope that a better understanding of classical models also serves as an incentive to look for related phenomena in quantum systems.

The parameters of the initial domain-wall state will be chosen such that in the accessible part of phase space the chain dynamics is sufficiently chaotic. Then one would expect that the conserved fields as computed from the chain dynamics are approximated by the respective solution of the macroscopic Euler equations, for times limited by diffusive effects. How well such expectations work out will have to be studied. The Euler equations are based on the notion of local thermodynamic equilibrium. The microscopic local conservation laws are deduced from the chain dynamics and are then averaged in the stipulated local equilibrium state so to arrive at a closed set of equations for the conserved fields. In particular, to reach non-trivial predictions, the thermal average of the microscopic currents is not allowed to vanish. This is ensured if the interaction potential depends only on positional differences, as $V(q_{j+1} - q_j)$, implying momentum conservation. Upon adding an on-site potential, $V_\mathrm{os}(q_j)$, momentum conservation would be broken, all Euler currents would vanish, and the evolution of the initial step profile is determined by diffusive effects only.

The Euler equations are a system of $n$ hyperbolic conservation laws, $n=3$ for our case of anharmonic chains. They are of the generic form
\begin{equation}\label{1.1}
\partial_t u_\alpha + \partial_x \mathsf{j}_\alpha(\vec{u}) = 0,
\end{equation}
$\alpha = 1,\dots,n$, $\vec{u} = (u_1,\dots,u_n)$, with given current functions $\vec{\mathsf{j}}$. In the mathematical literature the domain-wall initial data are known as Riemann problem for Eq.~\eqref{1.1}, which means
\begin{equation}\label{1.2}
\vec{u}(x,0) = \vec{u}_\ell \text{ for } x < 0, \qquad \vec{u}(x,0) = \vec{u}_\mathrm{r} \text{ for } x > 0.
\end{equation}
For a wide class of current functions, there is a unique entropy solution to \eqref{1.1} with initial conditions \eqref{1.2}, see the exposition~\cite{Bressan2013}, Sections 4--8. This solution scales ballistically as
\begin{equation}\label{1.3}
\vec{u}(x,t) = \vec{u}_\mathrm{dw}(x/t),
\end{equation}
where $\vec{u}_\mathrm{dw}$ is bounded and continuous except for isolated jumps, possibly. There is a well developed theory of how to compute $\vec{u}_\mathrm{dw}$, at least in principle \cite{Bressan2013}, Sections 1--3. In our case the current functions are determined through the microscopic particle model, and hence of a very particular form. Thus our task is twofold. Firstly we have to investigate the solution to the Riemann problem. Secondly such predictions should be compared with numerical simulations of the dynamics.

Let us return for a moment to the distinction between integrable and non-integrable systems, both starting from a domain-wall initial state. As supported by a variety of studies on quantum integrable models \cite{BCH11,SM13}, one expects that \eqref{1.3} still holds in the integrable case. Thus at first sight there seems to be little difference. Of course, the macroscopic profiles are computed by using completely different methods for the two cases. But the real distinguishing feature is the appearance of shocks. An ideal gas with step-initial conditions shows ballistic spreading but no shocks. The entropy solution for the Euler equations \eqref{1.1} is a mathematical shorthand for the limit of small dissipation, which is meaningful only if the underlying dynamics is sufficiently chaotic. Merely invoking the conservation laws, the Euler equations admit stable and unstable shocks. Such unphysical solutions to \eqref{1.1} are removed by requiring a positive entropy production at the shock, as will be illustrated in the examples below. In a local region away from the shock, the local state is (to very good approximation) in thermal equilibrium.

Our study adds current fluctuations as an item, which can no longer be based on the Euler equations \eqref{1.1}. Most simple-mindedly, one would consider the fluctuations of the time-integrated current across the origin. For anharmonic chains the current is a three-vector. In most cases one would find Gaussian fluctuations of size $\sqrt{t}$, thus not so interesting from a theoretical perspective. A more global picture emerges by considering the current integrated along the ray $\{ x = \mathsf{v} t \}$ for some prescribed velocity $\mathsf{v}$. The ray is chosen to lie in the interior of a rarefaction wave. In addition, one has to consider a computable but particular linear combination of the three currents. As will be discussed, then the integrated current fluctuations are of size $t^{1/3}$, smaller as for all other linear combinations, and the statistics is given by the Tracy-Widom distribution known from random matrix theory.

Our paper consists of three, at first sight somewhat unrelated parts. We start with a stochastic model with two conserved fields, as always with domain-wall initial conditions. For our particular system the validity of the Euler equations has been established mathematically \cite{FritzToth2004}. Thus the model is used to explain the method by which one obtains the solution of the Riemann problem and as a numerical benchmark. We proceed to anharmonic chains, first with a general interaction potential. Analytically and numerically we then consider two specific choices for the potential, which generate sufficiently chaotic dynamics (as known from previous studies). In the third part we discuss the fluctuations of time-integrated currents.

\section{Riemann problem for the LeRoux lattice gas}

A prototypical stochastic lattice gas is the totally asymmetric simple exclusion process (TASEP). Particles are located on $\mathbb{Z}$, at most one particle per site. Independently, after an exponentially distributed waiting time, a particle hops one step to the right, provided the target site is empty. Clearly, the particle number is the only conserved field.

To move towards several conservation laws, we look for a minimal extension of the TASEP to a model with two conserved fields. In the literature a standard generalization is known as LeRoux stochastic lattice gas. (This name goes back to Fritz and T\'{o}th \cite{FritzToth2004}, who prove the hydrodynamic limit globally in time. Apparently, LeRoux first wrote down this particular system of conservation laws \cite{SerreVol1, SerreVol2}.) The LeRoux lattice gas has two types of particles with label $\pm 1$. Subject to the exclusion rule, the 1 particles jump to the right and the $-1$ particles to the left, both according to the TASEP rule. Furthermore, a neighboring pair $1,-1$ is exchanged to $-1,1$ with rate $2$, which leads to the simplification that the stationary measures are Bernoulli. A generalization of LeRoux is the Arndt-Heinzel-Rittenberg (AHR) model \cite{AHR1998,FerrariSasamotoSpohn2013}.

More formally, we introduce occupation variables $\eta_j = -1, 0, 1$, $j \in \mathbb{Z}$. The only allowed exchanges are
\begin{center}
\begin{tabular}{cc}
$1,\hspace{9pt}0\ \to \ \hspace{7pt}\,0,1$ & at\,\,rate 1,\\
$0,-1\            \to \ -1,0$ & at\,\,rate 1,\\
$1,-1\            \to \ -1,1$ & at\,\,rate 2.
\end{tabular}
\end{center}
Note that in our convention the labels of the components are interchanged in comparison to \cite{FritzToth2004}. Clearly, the only conserved fields are the two particle numbers. The invariant Bernoulli measures are parametrized by the average densities $\rho_1$ and $\rho_{-1}$. The hydrodynamic equations simplify when written in terms of the average number of holes and the average velocity, i.e.,
\begin{equation}
\label{eq:avr_hole_vel_def}
\rho = 1 - \rho_1 - \rho_{-1} ,\qquad v = \rho_1 - \rho_{-1}
\end{equation}
with
\begin{equation}
\lvert v \rvert \leq 1, \quad 0 \leq \rho \leq 1 - \lvert v \rvert.
\end{equation}
We refer to $(\rho,v)$ as states, more appropriately, but also more lengthy, as steady state parameters, resp. as thermodynamic states in case of anharmonic chains. The single-site probabilities of the steady states are 
\begin{equation}
\label{eq:probLeRoux}
\mathbbm{P}_{\rho,v}(\eta_j = 0) = \rho, \qquad \mathbbm{P}_{\rho,v}(\eta_j = \pm 1) = \tfrac{1}{2}(1 - \rho \pm v).
\end{equation}
Averages will be denoted by $\langle \cdot \rangle_{\rho,v}$, the subscripts being omitted when obvious from the context. Since the steady states are of product form, their average current is easily computed. Thus, on a large space-time scale the conserved fields are governed by the entropy solution of
\begin{equation}
\label{eq:LeRouxEuler}
\partial_t \vec{u} + \partial_x \vec{\mathsf{j}}(\vec{u}) = 0,\quad \vec{u} = (\rho, v)
\end{equation}
with the current vector
\begin{equation}
\label{eq:currentLeRoux}
\vec{\mathsf{j}}(\vec{u}) = -(\rho\,v, \rho + v^2 - 1).
\end{equation}

To discuss the solution to the Riemann problem, we follow fairly closely the conventions of Ref.~\cite{Bressan2013}. One rewrites \eqref{eq:LeRouxEuler} in semilinear form as
\begin{equation}
\label{eq:LeRouxEuler1}
\partial_t \vec{u} + A(\vec{u})\partial_x \vec{u} = 0,
\end{equation}
where 
\begin{equation}
\label{eq:LeRoux_A}
A = \frac{\partial \vec{\mathsf{j}}(\vec{u})}{\partial \vec{u}} =
-\begin{pmatrix}
v & \rho\\
1 & 2 v
\end{pmatrix}.
\end{equation}
The eigenvalues of $A$ are
\begin{equation}
\label{eq:LeRoux_c}
c_{\sigma} = -\tfrac{3}{2}v + \tfrac{1}{2}\sigma \sqrt{4 \rho + v^2}, \qquad \sigma = \pm 1,
\end{equation}
and the corresponding right and left eigenvectors, $A \psi_{\sigma} = c_\sigma \psi_{\sigma}$, $A^\mathrm{T} \tilde{\psi}_{\sigma} = c_\sigma \tilde{\psi}_{\sigma}$, are given by
\begin{equation}
\label{eq:psiLeRoux}
\psi_\sigma =
Z_{\sigma}^{-1} \begin{pmatrix}
2 \sigma \rho \\ \sigma v - \sqrt{4 \rho + v^2}
\end{pmatrix},\qquad
\tilde{\psi}_\sigma =
\tilde{Z}_{\sigma}^{-1} \begin{pmatrix}
2 \sigma \\ \sigma v - \sqrt{4 \rho +v^2}
\end{pmatrix}.
\end{equation}
Here $Z_{\sigma}$ and $\tilde{Z}_{\sigma}$ are positive normalization constants. For the Riemann problem their explicit form is not needed. Setting $D = (\partial_\rho,\partial_v)$, one obtains for the change of $c_{\sigma}$ along the vector fields $\psi_{\sigma}$,
\begin{equation}
\label{eq:Leroux_gradient_c_rarefaction}
\psi_\sigma \cdot D c_{\sigma} = 2\, Z_{\sigma}^{-1} \big(\sqrt{4 \rho +v^2} - \sigma v\big) \ge 0,
\end{equation}
and strictly positive for $\rho > 0$.

\subsection{Rarefaction waves}

\begin{figure}[!ht]
\centering
\includegraphics[width=0.5\textwidth]{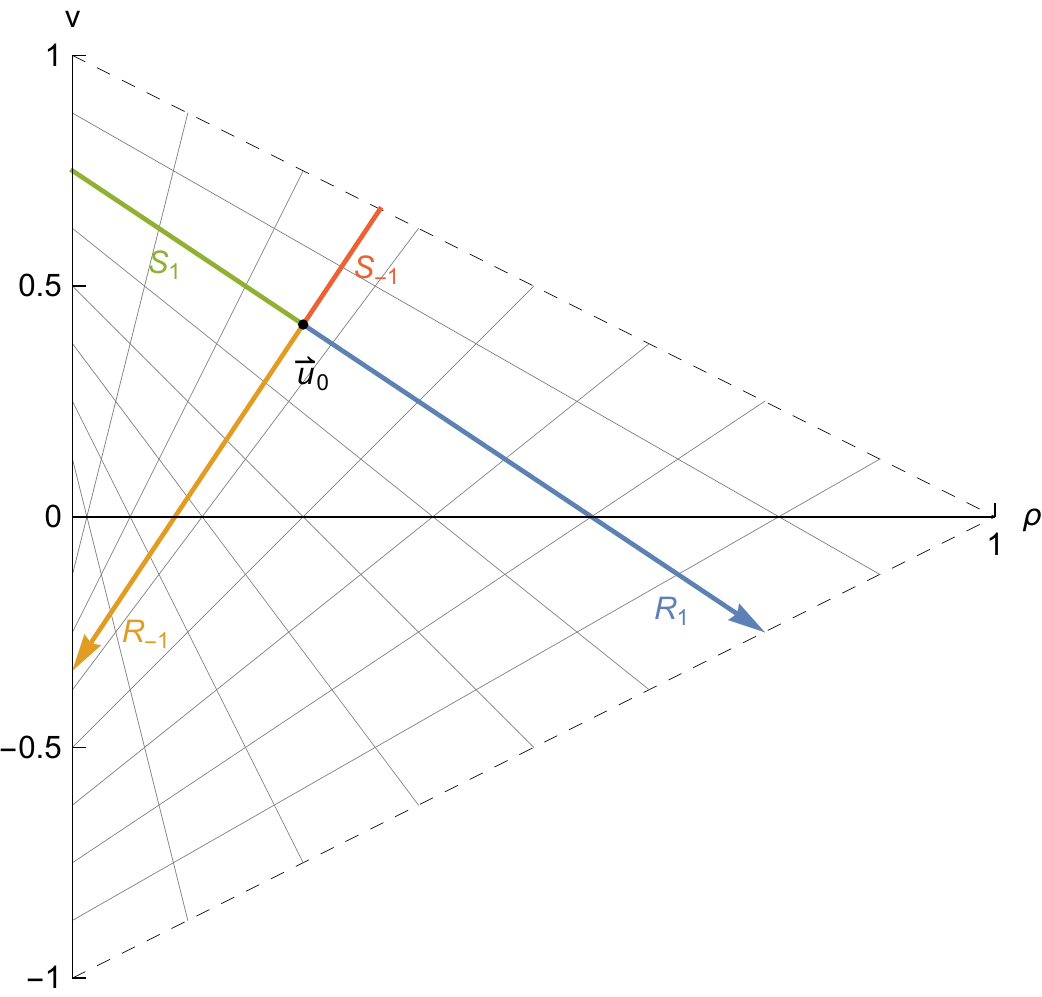}
\hspace{10pt}
\caption{Rarefaction and shock curves through a point $\vec{u}_0$ in state space for the LeRoux system according to Eqs.~\eqref{eq:LeRouxR_v_solution} and \eqref{eq:LeRoux_shock_solution_b}. Traversal in arrow direction corresponds to a rarefaction wave and increasing $c_{\sigma}$, and traversal in opposite arrow direction to a shock curve.}
\label{fig:LeRoux_rarefaction_shock}
\end{figure}

The rarefaction curves, $R_{\sigma}$, are obtained by solving the Cauchy problem in $\vec{u}$-space, 
\begin{equation}
\label{eq:LeRouxCauchy}
\partial_{\tau} \vec{u} = \psi_{\sigma}(\vec{u})
\end{equation}
for $\sigma = \pm 1$, with $\psi_{\sigma}$ the right eigenvectors of $A$, see \eqref{eq:psiLeRoux}. The normalization has been absorbed into the $\tau$-parameter. The integral curves are then determined by

\begin{equation}
\partial_{\tau} \rho = 2 \sigma \rho, \quad \partial_{\tau} v = \sigma v - \sqrt{4 \rho +v^2}.
\end{equation}
It follows that
\begin{equation}
\label{eq:LeRouxR_v_rho}
\frac{dv}{d\rho} = - \frac{1}{2\sigma\rho} \big( \sqrt{4 \rho +v^2} - \sigma v \big),
\end{equation}
which is negative for $R_1$ and positive for $R_{-1}$. The solution of \eqref{eq:LeRouxR_v_rho} is
\begin{equation}
\label{eq:LeRouxR_v_solution}
v_{\sigma} = \sigma\left( b_{\sigma} - b_{\sigma}^{-1}\rho \right), \qquad 0 < b_{\sigma} \leq 1,
\end{equation}
as visualized in Fig.~\ref{fig:LeRoux_rarefaction_shock}. Maximally, $R_1$ starts at $\vec{u}_{1,\ell} = (0,b_1)$ and ends at $\vec{u}_{1,\mathrm{r}} = (b_1, b_1-1)$, whereas $R_{-1}$ starts at $\vec{u}_{-1,\ell} = (b_{-1}, 1 - b_{-1})$ and ends at $\vec{u}_{-1,\mathrm{r}} = (0,-b_{-1})$. The local eigenvalue is
\begin{equation}
\label{eq:LeRouxR_c}
c_{\sigma} = -\tfrac{3}{2}v + \tfrac{1}{2}\sigma \sqrt{4 \rho + v^2} = \sigma \left( b_{\sigma}^{-1} 2 \rho - b_{\sigma} \right).
\end{equation}
To convert the solution from $\vec{u}$-space to position space, we set $c_{\sigma} = x/t$. The solution is self-similar and we may assume $t = 1$. Then
\begin{equation}
\label{eq:LeRouxR_solution}
\rho_{\sigma}(x) = \tfrac{1}{2} b_{\sigma} (b_{\sigma} + \sigma x),\quad v_{\sigma}(x) = \tfrac{1}{2}(b_{\sigma} - \sigma x).
\end{equation}
The boundary speeds of $R_1$ are
\begin{equation}
c_{1,\ell} = c_1(\vec{u}_{1,\ell}) = -b_1,\quad c_{1,\mathrm{r}} = c_1(\vec{u}_{1,\mathrm{r}}) = 2 - b_1,
\end{equation}
and of $R_{-1}$
\begin{equation}
c_{-1,\ell} = c_{-1}(\vec{u}_{-1,\ell}) = b_{-1} - 2,\quad c_{-1,\mathrm{r}} = c_{-1}(\vec{u}_{-1,\mathrm{r}}) = b_{-1}.
\end{equation}
Eq.~\eqref{eq:LeRouxR_solution} as a function of $x/t$ describes solutions of the Euler equation \eqref{eq:LeRouxEuler}. The two solutions with $\sigma = \pm 1$ are mirror images of each other.

\subsection{Shock curves}

Shock curves are determined by the Rankine-Hugoniot jump condition
\begin{equation}
\lambda (\vec{u} - \vec{u}_0) = \vec{\mathsf{j}}(\vec{u}) - \vec{\mathsf{j}}(\vec{u}_0).
\end{equation}
Hence in our case
\begin{align}
-\lambda (\rho - \rho_0) &= \rho v - \rho_0 v_0, \\
-\lambda (v - v_0) &= \rho - \rho_0 + v^2 - v_0^2.
\end{align}
According to the first equation the shock speed is
\begin{equation}
\label{eq:LeRoux_shock_speed_RH}
\lambda = -\frac{\rho v - \rho_0 v_0}{\rho - \rho_0} .
\end{equation}
We eliminate $\lambda$ with the result
\begin{equation}
\label{eq:LeRoux_shock_condition}
(\rho - \rho_0)^2 = (v - v_0)(v \rho_0 - v_0 \rho).
\end{equation}
If $\rho_0 > 0$, the two solutions for $v$ are
\begin{equation}
\label{eq:LeRoux_shock_solution}
v_{\sigma} = v_0 + (1 - \hat{\rho}) (c_{\sigma,0} + v_0), \quad \hat{\rho} = \rho / \rho_0
\end{equation}
with $\sigma = \pm 1$ and $c_{\sigma,0} = c_{\sigma}(\vec{u}_0)$ the sound speed \eqref{eq:LeRoux_c} on the left side of the shock. In particular, one recovers $v_{\sigma} = v_0$ for $\hat{\rho} = 1$. In fact, the rarefaction curves \eqref{eq:LeRouxR_v_solution} coincide with the solution to the Rankine-Hugoniot equations \eqref{eq:LeRoux_shock_solution}. This can be seen by defining $b_{\sigma}$ implicitly via
\begin{equation}
v_{\sigma,0} = \sigma \left( b_{\sigma} - b_{\sigma}^{-1} \rho_0 \right),
\end{equation}
then
\begin{equation}
\label{eq:LeRoux_shock_solution_b}
v_{\sigma} = \sigma \left( b_{\sigma} - b_{\sigma}^{-1} \rho \right).
\end{equation}
Inserting into \eqref{eq:LeRoux_shock_speed_RH} one arrives at the shock speed
\begin{equation}
\label{eq:LeRoux_shock_speed}
\lambda_{\sigma} = \tfrac{1}{2} (c_{\sigma,0} + c_{\sigma}) = \sigma \left( b_{\sigma}^{-1} (\rho + \rho_0) - b_{\sigma} \right).
\end{equation}

The coincidence of rarefaction and shock curves is the defining property of the Temple class \cite{Temple1982}. The LeRoux system is a further member of this class. In our context the interest results from a maximally simple underlying particle dynamics.

The \emph{Lax admissibility condition} states that characteristics must move ``towards'' the shock:
\begin{equation}
c_{\sigma,0} \ge \lambda \ge c_{\sigma}.
\end{equation}
Since $\lambda$ is the mean value of the sound speeds, the condition simplifies to $c_{\sigma} \le c_{\sigma,0}$. Together with $c_{\sigma} = \sigma(b^{-1} 2 \rho - b)$, this is equivalent to $\rho \le \rho_0$ for $\sigma = 1$ and $\rho \ge \rho_0$ for $\sigma = -1$. The stable, physically admissible part of the Rankine-Hugoniot curve are the shock curves $S_1$ and $S_{-1}$ as displayed in Fig~\ref{fig:LeRoux_rarefaction_shock}.

\subsection{General solution}

\begin{figure}[!ht]
\centering
\subfloat[path in state space]{\label{fig:LeRoux_rarefaction_shock_path}\includegraphics[width=0.4\textwidth]{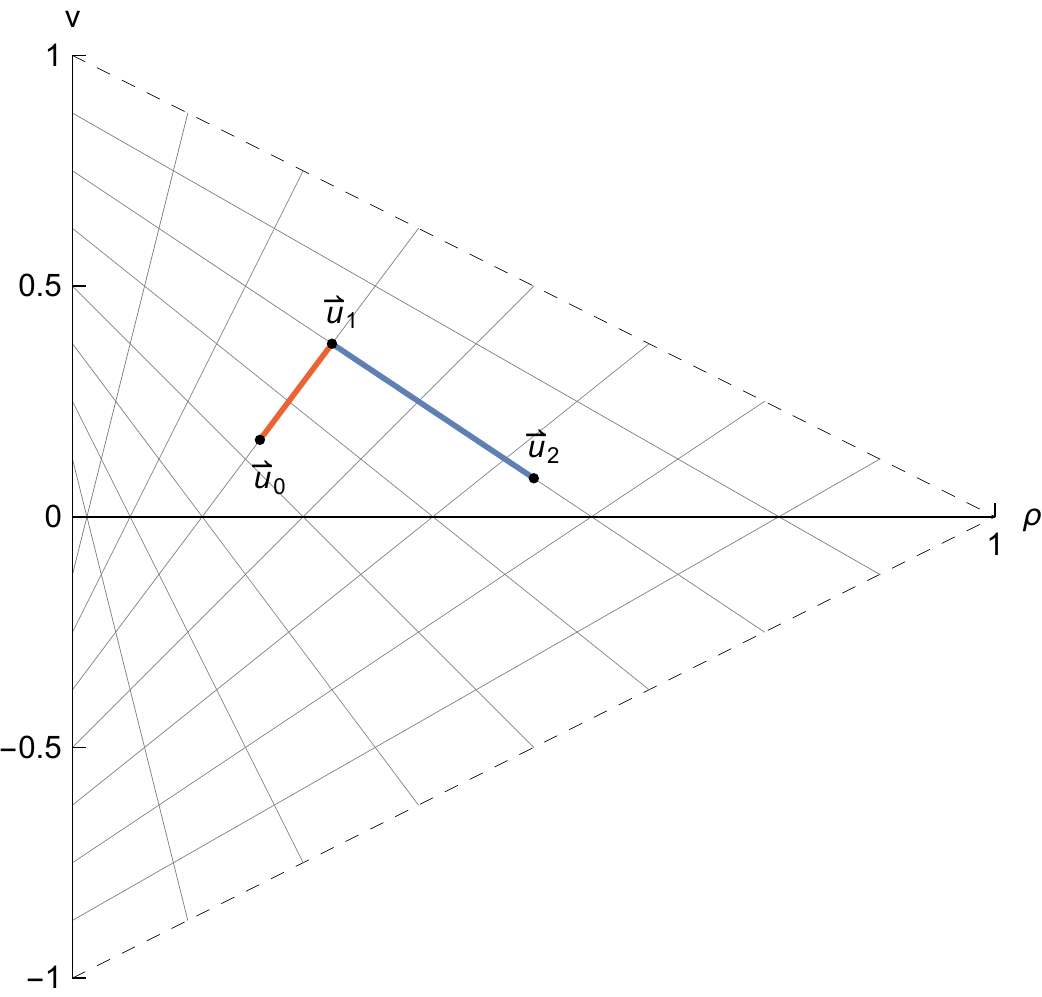}}
\hspace{15pt}
\subfloat[corresponding shock and rarefaction wave]{\label{fig:LeRoux_rarefaction_shock_xt}\includegraphics[width=0.45\textwidth]{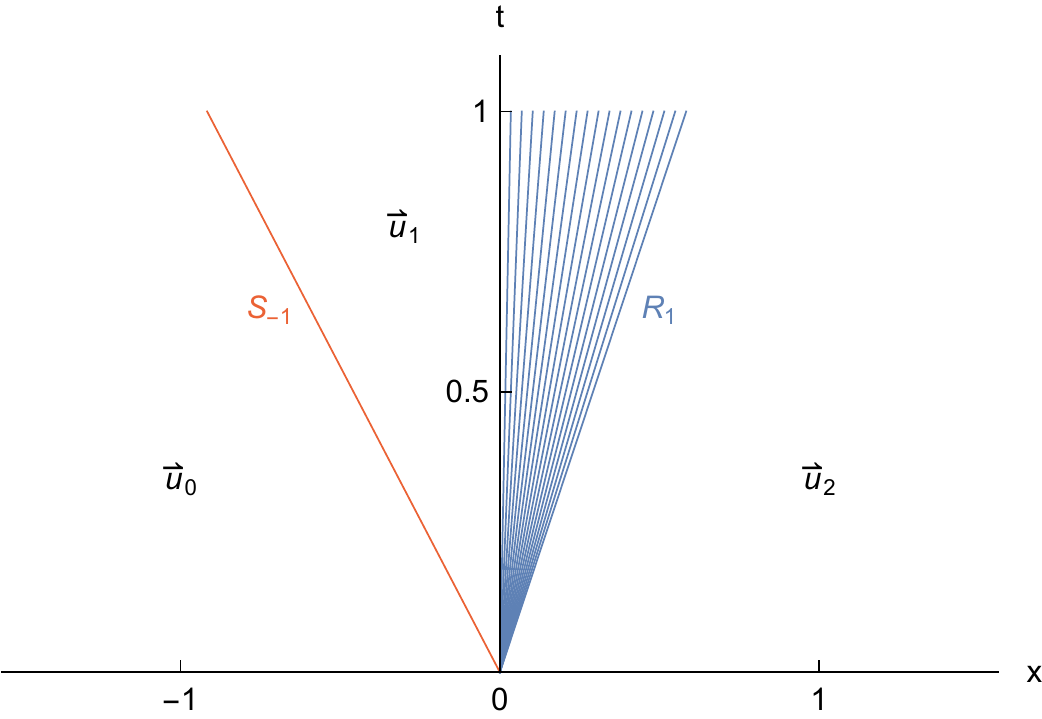}}
\caption{(a) Path in state space from initial state $\vec{u}_0$ via $\vec{u}_1$ to the final state $\vec{u}_2$, traversing along the red shock curve $S_{-1}$ first and then along the blue rarefaction curve $R_1$. (b) Corresponding shock and rarefaction waves in a $x$-$t$ diagram.}
\label{fig:LeRoux_path}
\end{figure}

The construction of the general solution is illustrated in Fig.~\ref{fig:LeRoux_path}. Starting from the asymptotic left value $\vec{u}_{\ell} =\vec{u}_0$, one first follows either the rarefaction curve $R_{-1}$ or the stable part of the shock curve $S_{-1}$ (shown as linear orange-red line in Fig.~\ref{fig:LeRoux_rarefaction_shock}) up to a unique intermediate state $\vec{u}_1$. Then $\vec{u}_1$ is connected by either a rarefaction curve $R_1$ or shock curve $S_1$ (blue-green in Fig.~\ref{fig:LeRoux_rarefaction_shock}) to the asymptotic right value $\vec{u}_2 = \vec{u}_{\mathrm{r}}$. Rarefaction curves correspond to traversal in arrow direction in Fig.~\ref{fig:LeRoux_rarefaction_shock}, equivalently increasing eigenvalue $c_{\sigma}$. This procedure splits the parameter domain into four distinct pieces according to rarefaction-rarefaction, shock-rarefaction, rarefaction-shock, and shock-shock. The two domain boundaries correspond to either a single rarefaction or a single shock, with no intermediate value $\vec{u}_1$. In Fig.~\ref{fig:LeRoux_rarefaction_shock_xt} we show the space-time plot corresponding to the case shock-rarefaction of Fig.~\ref{fig:LeRoux_rarefaction_shock_path}.

\subsection{Monte Carlo simulations}

We perform Monte Carlo simulations of the LeRoux model for $L = 4096$ sites with periodic boundary conditions. To obtain an initial domain wall state, we sample the single-site probability distribution \eqref{eq:probLeRoux} on the left half $j = -\frac{L}{2},\dots,-1$ using parameters $\vec{u}_{\ell} = (\rho_{\ell}, v_{\ell})$, and on the right half $j = 0,\dots, \frac{L}{2}-1$ using parameters $\vec{u}_{\mathrm{r}} = (\rho_{\mathrm{r}}, v_{\mathrm{r}})$. The dynamics is simulated by random exchanges at exponentially distributed waiting times up to $t_{\max} = 1024$. This procedure is realized $10^6$ times to compute average profiles, as shown below.

\begin{figure}[!htp]
\centering
\subfloat[density, $t = 0$]{\includegraphics[width=0.225\textwidth]{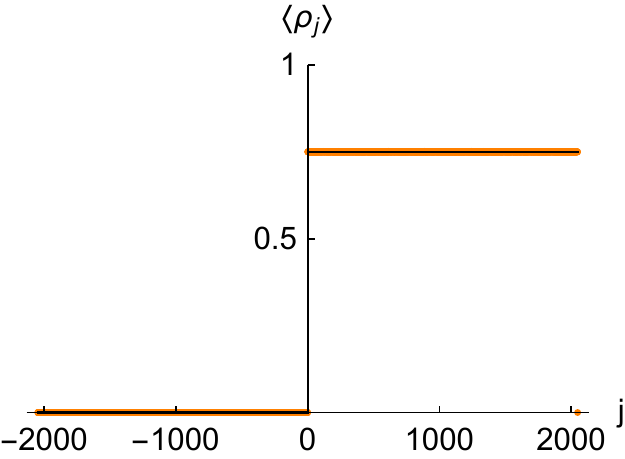}}
\hspace{0.02\textwidth}
\subfloat[density, $t = 256$]{\includegraphics[width=0.225\textwidth]{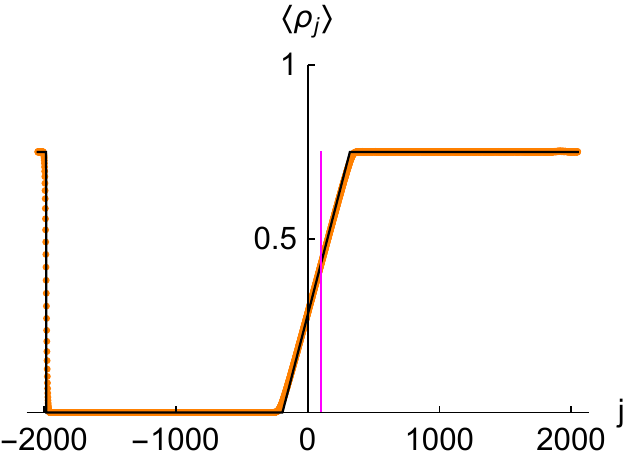}}
\hspace{0.02\textwidth}
\subfloat[density, $t = 512$]{\includegraphics[width=0.225\textwidth]{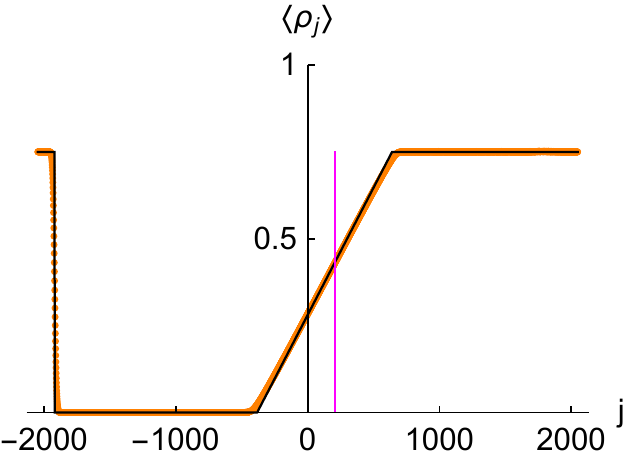}}
\hspace{0.02\textwidth}
\subfloat[density, $t = 1024$]{\includegraphics[width=0.225\textwidth]{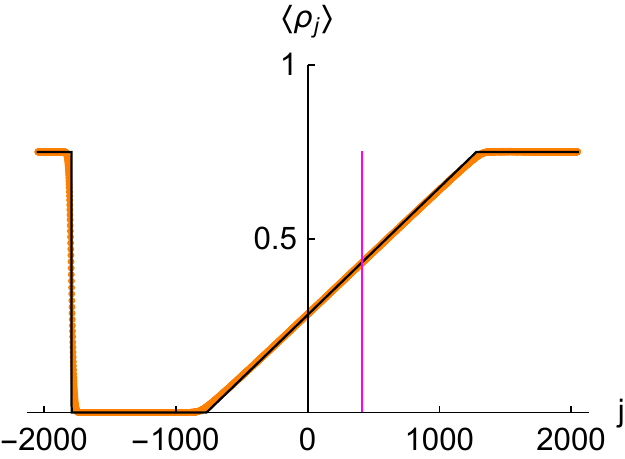}}\\
\subfloat[velocity, $t = 0$]{\includegraphics[width=0.225\textwidth]{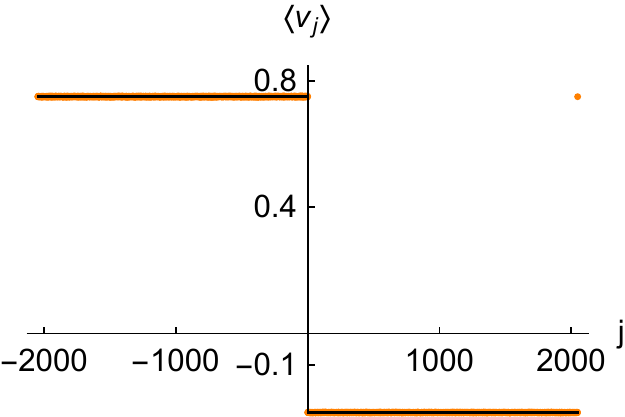}}
\hspace{0.02\textwidth}
\subfloat[velocity, $t = 256$]{\includegraphics[width=0.225\textwidth]{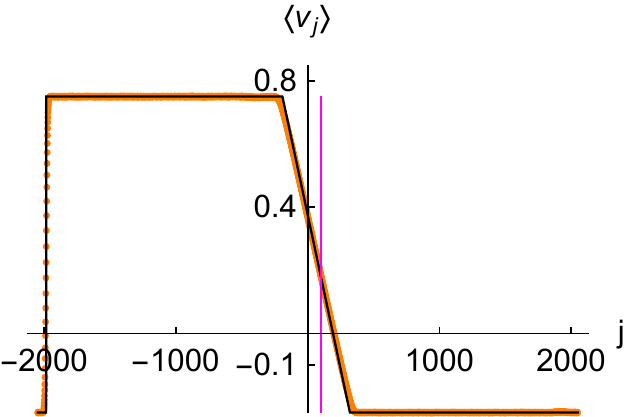}}
\hspace{0.02\textwidth}
\subfloat[velocity, $t = 512$]{\includegraphics[width=0.225\textwidth]{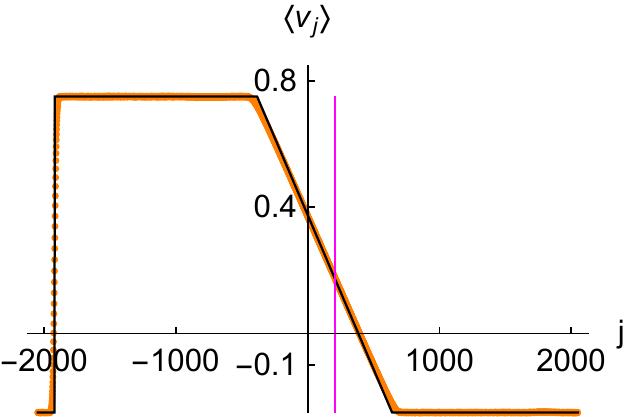}}
\hspace{0.02\textwidth}
\subfloat[velocity, $t = 1024$]{\includegraphics[width=0.225\textwidth]{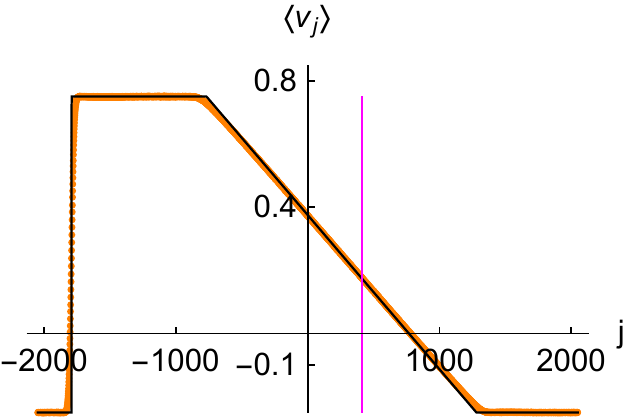}}
\caption{Density and velocity profiles at various times for the LeRoux model with domain-wall initial conditions $\vec{u}_{\ell} = (0,b_1)$ and $\vec{u}_{\mathrm{r}} = (b_1, b_1-1)$, $b_1 = \frac{3}{4}$ and $L = 4096$ sites. The orange dots are molecular dynamics results and the black line shows the theoretically predicted profile. The sharp jump on the left is a shock resulting from the periodic boundary conditions, and the sloped linear segment is a rarefaction wave.}
\label{fig:LeRoux_profiles}
\end{figure}

\begin{figure}[!htp]
\centering
\subfloat[density, $t = 0$]{\includegraphics[width=0.225\textwidth]{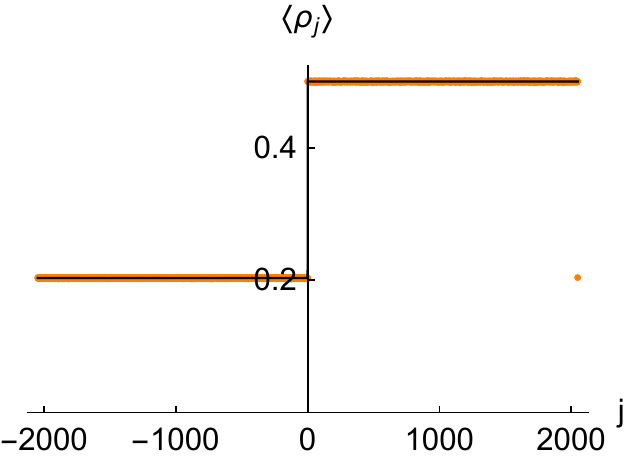}}
\hspace{0.02\textwidth}
\subfloat[density, $t = 256$]{\includegraphics[width=0.225\textwidth]{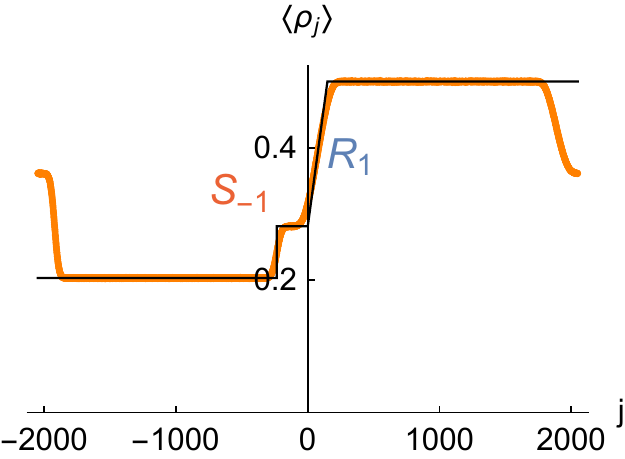}}
\hspace{0.02\textwidth}
\subfloat[density, $t = 512$]{\includegraphics[width=0.225\textwidth]{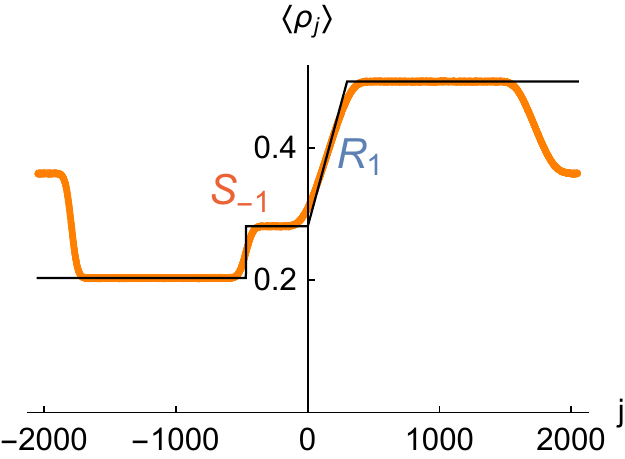}}
\hspace{0.02\textwidth}
\subfloat[density, $t = 1024$]{\includegraphics[width=0.225\textwidth]{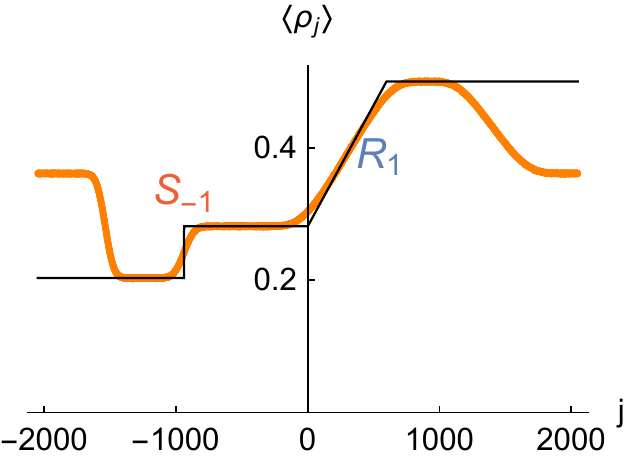}}\\
\subfloat[velocity, $t = 0$]{\includegraphics[width=0.225\textwidth]{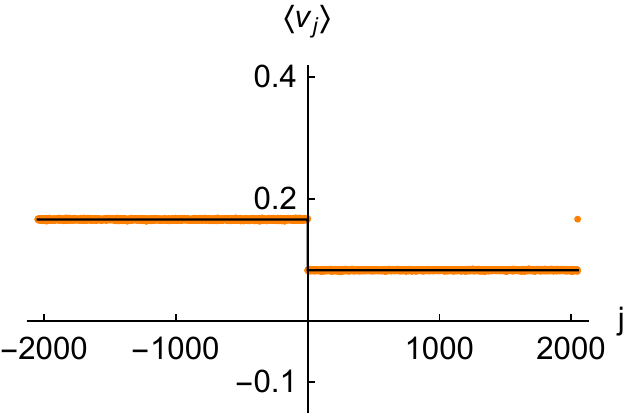}}
\hspace{0.02\textwidth}
\subfloat[velocity, $t = 256$]{\includegraphics[width=0.225\textwidth]{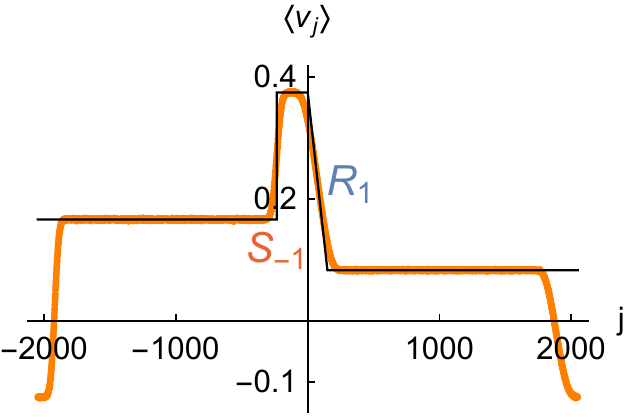}}
\hspace{0.02\textwidth}
\subfloat[velocity, $t = 512$]{\includegraphics[width=0.225\textwidth]{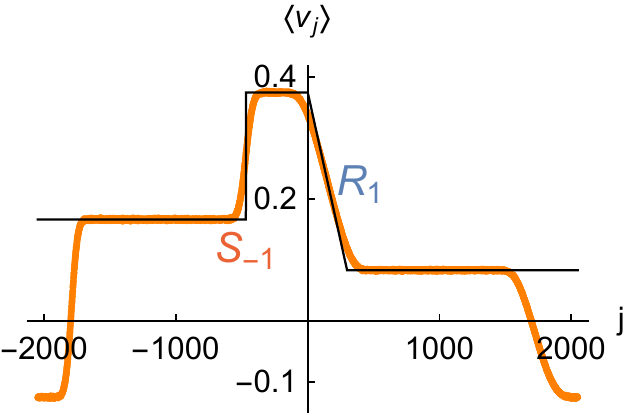}}
\hspace{0.02\textwidth}
\subfloat[velocity, $t = 1024$]{\includegraphics[width=0.225\textwidth]{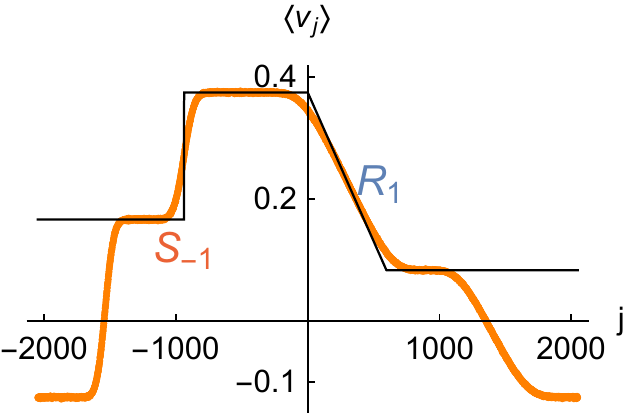}}
\caption{Density and velocity profiles at various times for the LeRoux model with domain-wall initial conditions corresponding to Fig.~\ref{fig:LeRoux_path}. The shock wave $S_{-1}$ and rarefaction wave $R_1$ are indicated in red and blue, respectively. The theoretical prediction (black lines) refers only to the Riemann problem centered at the origin. The outer features of the MC profiles arise from periodic boundary conditions.}
\label{fig:LeRoux_profiles_general}
\end{figure}

First, we illustrate the special case of a single rarefaction wave $R_1$ connecting $\vec{u}_{\ell}$ to $\vec{u}_{\mathrm{r}}$, which are chosen maximally as $\vec{u}_{\ell} = (0,b_1)$ and $\vec{u}_{\mathrm{r}} = (b_1, b_1-1)$ with $b_1 = \frac{3}{4}$. This is the particular choice in \cite{TWMendlSpohn2016}, and corresponds to the extremal points of the green-blue line segment in Fig.~\ref{fig:LeRoux_rarefaction_shock}. Besides the $\vec{u}_{\ell}\vert\vec{u}_{\mathrm{r}}$ Riemann problem centered at $j = 0$ and generating rarefaction $R_1$, the periodic boundary condition translates to an additional $\vec{u}_{\mathrm{r}}\vert\vec{u}_{\ell}$ Riemann problem centered at $j = \frac{L}{2}$. Since the solutions of the rarefaction and shock curves coincide except for orientation, one concludes that the solution of the $\vec{u}_{\mathrm{r}}\vert\vec{u}_{\ell}$ Riemann problem is a single shock curve $S_1$. In other words, one traverses the green-blue line segment in Fig.~\ref{fig:LeRoux_rarefaction_shock} in opposite direction. The numerical Monte Carlo profiles, shown as orange dots in Fig.~\ref{fig:LeRoux_profiles}, agree very well with the theoretical prediction (solid black lines). In particular, note the sharp jump at the shock. This shock curve is located at $j = -\frac{L}{2} + \lambda_1 t$ and moves to the right, with shock speed $\lambda_1 = 1 - b_1$.

Next, we perform molecular dynamics simulations corresponding to the general case in Fig.~\ref{fig:LeRoux_path}. Specifically, the left $\vec{u}_{\ell} = \vec{u}_0 = (\frac{13}{64}, \frac{1}{6})$, the intermediate $\vec{u}_1 = (\frac{9}{32}, \frac{3}{8})$ and the right $\vec{u}_{\mathrm{r}} = \vec{u}_2 = (\frac{1}{2}, \frac{1}{12})$. Fig.~\ref{fig:LeRoux_profiles_general} shows the molecular dynamics profiles, with the shock curve $S_{-1}$ indicated in red and the rarefaction wave $R_1$ in blue. One observes that the MC profile of $S_{-1}$ is less sharp than the shock in Fig.~\ref{fig:LeRoux_profiles}, presumably due to the higher shock speed.

\subsection{Entropy}

The thermodynamic entropy for the probability distribution of \eqref{eq:probLeRoux} is
\begin{equation}
\label{eq:LeRoux_entropy}
S(\rho, v) = - \sum_{\eta=-1}^1 \mathbbm{P}_{\rho,v}(\eta) \log \mathbbm{P}_{\rho,v}(\eta),
\end{equation}
using the standard physics convention for the sign. The definition in \cite{Bressan2013} uses a convex function which has the opposite sign.
\begin{figure}[!ht]
\centering
\subfloat[entropy $S(q,v)$]{\includegraphics[width=0.4\textwidth]{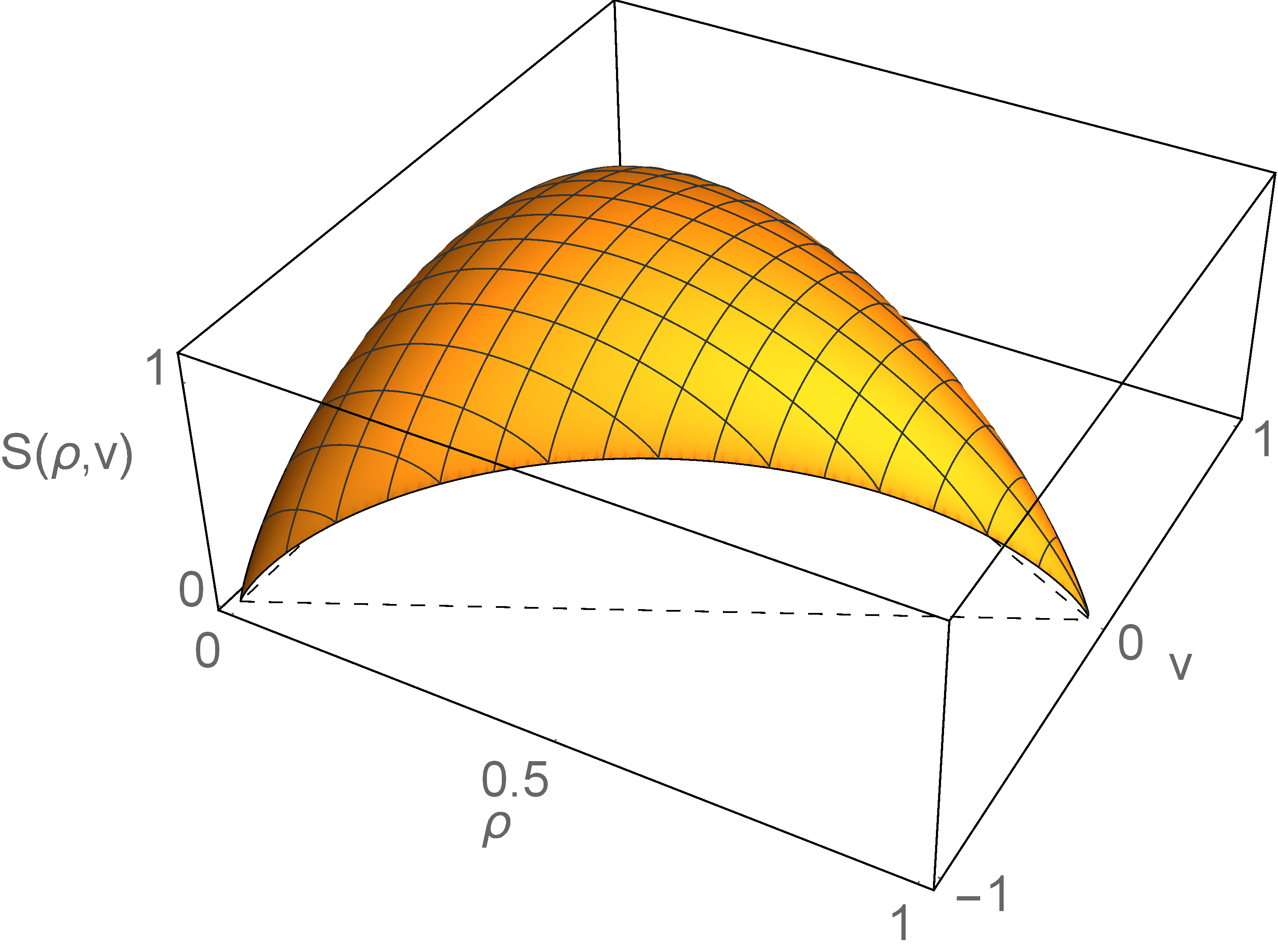}}
\hspace{0.1\textwidth}
\subfloat[entropy flux $q(\rho, v)$]{\includegraphics[width=0.4\textwidth]{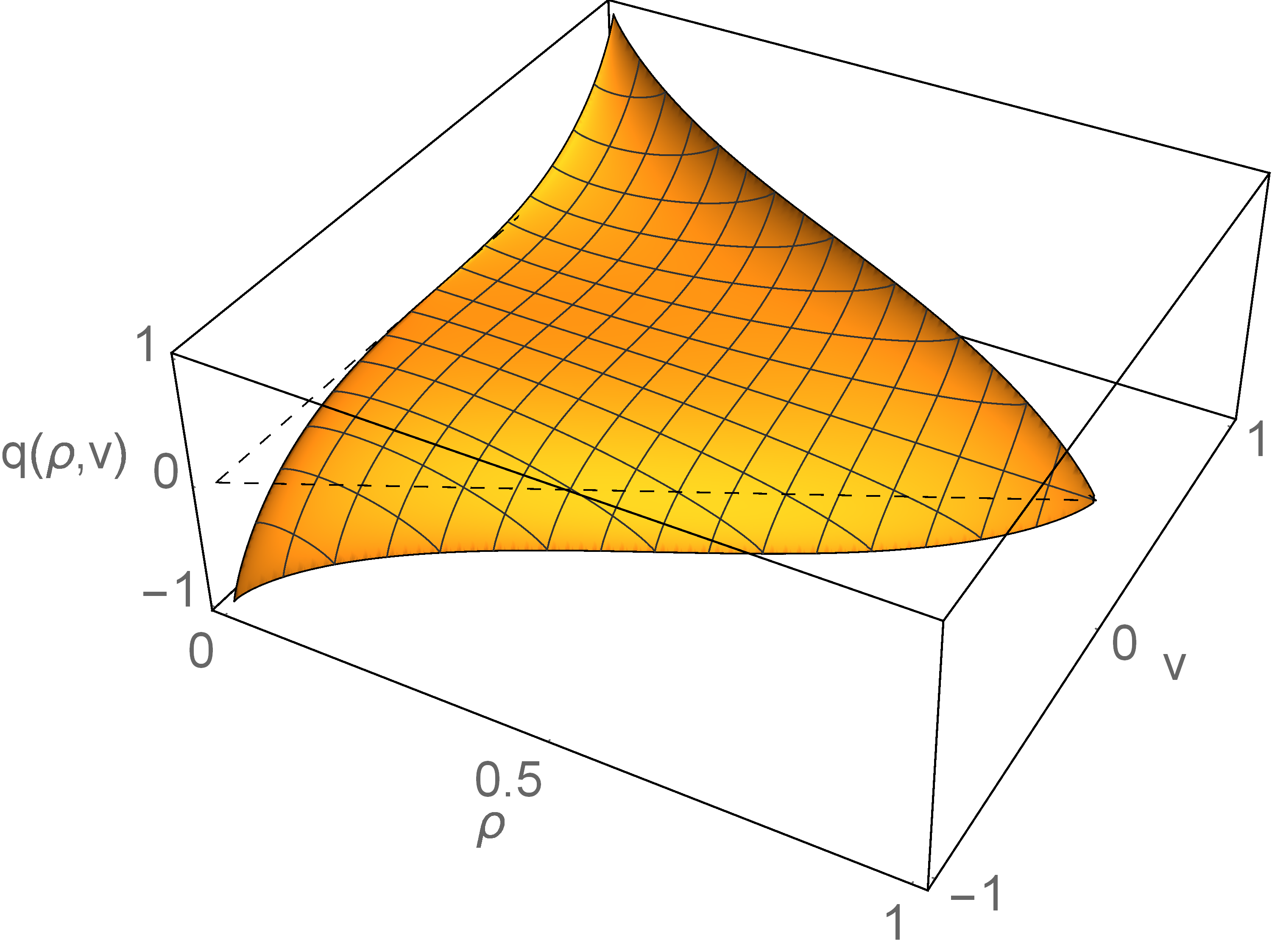}}
\caption{Entropy according to Eq.~\eqref{eq:LeRoux_entropy} and entropy flux according to Eq.~\eqref{eq:LeRoux_entropy_flux} of the LeRoux model.}
\label{fig:LeRoux_entropy}
\end{figure}
The corresponding entropy flux $q(\vec{u})$ has to satisfy
\begin{equation}
\label{eq:entropy_flux_relation_LeRoux}
D S(\vec{u}) \, A = D q(\vec{u})
\end{equation}
with $\vec{u} = (\rho, v)$ and $D = (\partial_{\rho}, \partial_{v})$. Up to a constant the solution for $q(\vec{u})$ is
\begin{equation}
\label{eq:LeRoux_entropy_flux}
q(\rho, v) = v + \sum_{\eta=-1}^1 (v - \eta) \mathbbm{P}_{\rho,v}(\eta) \log \mathbbm{P}_{\rho,v}(\eta).
\end{equation}
Note that $q(\rho, v)$ is an odd function in $v$. Fig.~\ref{fig:LeRoux_entropy} visualizes both entropy and entropy flux.

The entropy inequality admissibility condition states that
\begin{equation}
\label{eq:entropy_inequality_LeRoux}
\partial_t S(\vec{u}) + \partial_x q(\vec{u}) = \Delta S(\vec{u}) \ge 0
\end{equation}
in the sense of distributions. For continuously differentiable solutions due to \eqref{eq:entropy_flux_relation_LeRoux} one has $\Delta S(\vec{u}) = 0$ and 
\begin{equation}
\partial_t S(\vec{u}) + \partial_x q(\vec{u}) = 0.
\end{equation}
In particular, no entropy is produced at a rarefaction wave. On the other side, at shocks one can follow the same steps as for the Rankine-Hugoniot jump condition to derive from \eqref{eq:entropy_inequality_LeRoux} that
\begin{equation}
\label{eq:entropy_inequality_LeRoux_shock}
-\int \ud t \big(\lambda (S - S_0) - (q - q_0) \big) \phi(\lambda t, t) = \int \ud t\, \Delta S\, \phi(\lambda t, t) 
\end{equation}
for any continuously differentiable test function $\phi(x,t)$ with compact support. The integration in \eqref{eq:entropy_inequality_LeRoux_shock} proceeds along the shock curve, and $S_0$, $q_0$ are the values on the left side of the shock. Inserting the shock speed \eqref{eq:LeRoux_shock_speed} and the shock solution into \eqref{eq:LeRoux_entropy} and \eqref{eq:LeRoux_entropy_flux}, one obtains (with $b = b_{\sigma}$)
\begin{multline}
\Delta S_{\sigma} = \sigma \bigg( \frac{1}{b} \Big( \rho_0 - \rho + \rho\,\rho_0 \log\!\Big[\frac{\rho}{\rho_0} \Big]\Big) + \frac{1}{b} \left(1 - b^2 - (1-\rho)(1-\rho_0)\right) \frac{1}{2} \log\!\left[\frac{1 - (\rho/b)^2}{1 - (\rho_0/b)^2}\right] \\
- \left(1 - \rho - \rho_0 + \frac{\rho\,\rho_0}{b^2}\right) \left(\mathrm{arctanh}\Big(\frac{\rho_0}{b}\Big) - \mathrm{arctanh}\Big(\frac{\rho}{b}\Big)\right) \bigg).
\end{multline}
Note that this expression is invariant under the interchange $\rho \leftrightarrow \rho_0$ and simultaneously $\sigma \leftrightarrow -\sigma$, as expected from symmetry of the shock curves moving to the right and left. As required, the solution of the Riemann problem satisfies $\Delta S_{\sigma} \ge 0$.

\section{Riemann problem for anharmonic chains}
\label{sec:RiemannGeneral}

With the LeRoux lattice gas as guiding example, we turn to our central theme which is the time evolution for domain-wall initial conditions of a system of anharmonically coupled mechanical point particles. The $j$-th particle has mass $m$, position $q_j$, momentum $p_j$, and is coupled to its neighbors $j-1$ and $j+1$ through the potential $V$.
Then Newton's equations of motion are given by
\begin{equation}\label{3.1}
m \tfrac{\ud^2}{\ud t^2} q_j = V'(q_{j+1} - q_j) - V'(q_{j} - q_{j-1}) .
\end{equation}
We read this equation as a discretized wave equation with $q_j \in \mathbb{R}$ the displacement of the wave field at lattice site $j$. The hamiltonian of the chain is
\begin{equation}\label{3.2}
H = \sum_{j \in \mathbb{Z}}\big( \tfrac{1}{2m} p_j^2 + V(q_{j+1} - q_j) \big).
\end{equation}
For the harmonic potential, $V(x) = x^2$, Eq.~\eqref{3.1} reduces to a discrete linear wave equation. In the theoretical analysis we use $j \in \mathbb{Z}$. Numerically, $j \in [-\tfrac{L}{2},\dots,\tfrac{L}{2}-1]$ with periodic boundary conditions.

In contrast to the LeRoux lattice gas, the dynamics is deterministic. The initial conditions are however random, specifically to have a domain-wall state. As before, we expect the Euler equations to provide an accurate description on a macroscopic scale, provided the times are not too long. The validity of the Euler equations is based on maintaining local stationarity away from shocks. For a stochastic system the local approach to stationarity is in a certain sense build into the dynamics. For mechanical systems one relies on sufficiently strong dynamical chaos. Thereby integrable systems, as the harmonic and Toda chain are ruled out. In the LeRoux lattice gas the jumps are asymmetric. The steady states are non-equilibrium and the dynamics does not satisfy the principle of detailed balance. In contrast, for the anharmonic chain the domain-wall state is manufactured by joining two thermal equilibrium states. Because of momentum conservation, the thermal average defining the Euler currents does not vanish. If one broke this conservation law by adding in \eqref{3.2} an on-site potential, then the Euler currents would vanish identically, no Riemann problem ensues, and the first macroscopic time-scale is diffusive.

It is convenient to introduce the stretch $r_j = q_{j+1} - q_j$. Then the equations of motion turn into
\begin{equation}
\tfrac{\ud}{\ud t} r_j = \tfrac{1}{m}(p_{j+1} - p_j), \qquad
\tfrac{\ud}{\ud t} p_j = V'(r_j) - V'(r_{j-1}),
\end{equation}
from which one concludes that stretch and momentum are conserved. The respective currents are $-\tfrac{1}{m} p_j$ and $-V'(r_{j-1})$. In addition, we define the local energy
\begin{equation}
\label{3.3}
e_j = \tfrac{1}{2m} p_j^2 + V(r_j),
\end{equation}
which changes in time as
\begin{equation}
\label{eq:ej_ode}
\tfrac{\ud}{\ud t}e_j = \tfrac{1}{m} p_{j+1} V'(r_j) - \tfrac{1}{m} p_j V'(r_{j-1}).
\end{equation}
As anticipated, the energy is locally conserved. Its current equals $-\tfrac{1}{m} p_j V'(r_{j-1})$. More precisely than before, in our context non-integrable means that there are no further locally conserved fields. Unfortunately, this property is difficult to check. Besides the harmonic chain, the only known integrable system 
is the Toda chain with $V(x) = \mathrm{e}^{-x}$. [The Calogero-Moser chain has a two-sided decaying potential, which is not allowed in our context.]

The thermodynamic fields conjugate to stretch, momentum, and energy are the pressure $P$, the mean momentum, $m v$, and the inverse temperature, $\beta > 0$, respectively. From \eqref{3.2} we conclude that in thermal equilibrium the $p_j$'s and $r_j$'s are independent. The probability density function for $p_j$ is the shifted Maxwellian 
\begin{equation}
\label{eq:ensemble_pj}
\tfrac{1}{\sqrt{2 \pi m/\beta}} \,\mathrm{e}^{-\beta \frac{1}{2 m} (p_j - mv)^2}
\end{equation}
and the one for $r_j$ is given by
\begin{equation}
\label{eq:ensemble_rj}
Z^{-1}\, \mathrm{e}^{-\beta \left( V(r_j) + P r_j \right)}, \qquad Z(P,\beta) = \int_\mathbb{R} \ud x\, \mathrm{e}^{-\beta \left( V(x) + P x \right)}.
\end{equation}
To obtain a finite spatial partition function $Z$, we require that $V(x)$ is bounded from below and has at least a one-sided linearly growing lower bound as $\lvert x \rvert \to \infty$. Then $Z < \infty$ for $P$ in a suitably chosen interval. Equilibrium averages are denoted by $\langle \cdot \rangle_{P,v,\beta}$, the subscripts being omitted if obvious from the context. To assemble a domain-wall initial state, we set $(P,v,\beta) = (P_{\ell},v_{\ell},\beta_{\ell})$ for $j < 0$ and $(P,v,\beta) = (P_{\mathrm{r}}, v_{\mathrm{r}},\beta_{\mathrm{r}})$ for $j \geq 0$ in \eqref{eq:ensemble_pj} and \eqref{eq:ensemble_rj}. By construction these initial data are in thermal equilibrium except for the jump at the origin.

A little bit more of thermodynamics will be needed. The average stretch is given by 
\begin{equation}
\label{eq:avr_stretch}
r(P,\beta) = \langle r_j \rangle_{P,v,\beta} = Z^{-1} \int_\mathbb{R} \ud x\, x\, \mathrm{e}^{-\beta \left( V(x) + P x \right)},
\end{equation}
the average momentum is $\langle p_j \rangle = mv$, and the average \emph{internal} energy is
\begin{equation}
\label{eq:avr_internal_energy}
e(P,\beta) = \left\langle \tfrac{1}{2 m} p_j ^2 + V(r_j) \right\rangle_{P,v=0,\beta} = \tfrac{1}{2} \beta^{-1} + Z^{-1} \int_\mathbb{R} \ud x\, V(x)\, \mathrm{e}^{-\beta \left( V(x) + P x \right)}.
\end{equation}
The average total energy is then $\mathfrak{e} = \langle e_j \rangle_{P,v,\beta} = e + \tfrac{1}{2} mv^2$. Later on we will need $P(r,e)$ and $\beta(r,e)$ as the inverse to Eqs.~\eqref{eq:avr_stretch} and \eqref{eq:avr_internal_energy}. By convexity of the respective thermodynamic potential this inverse is uniquely defined.

The Euler equations are obtained by assuming that the conserved fields are slowly varying on the scale of the lattice. They thus become functions of space-time $(x,t)$. We now drop the prefix ``average'' and call the space-time fields simply stretch, momentum, and energy. Since all particles have the same mass, we will follow the standard convention in using velocity instead of momentum as hydrodynamic field. If local equilibrium is propagated, then the macroscopic Euler currents are the thermal average of the microscopic currents, explicitly,
\begin{equation}
\label{eq:avr_current}
\vec{\mathsf{j}} = \big( {- \tfrac{1}{m}}\left\langle p_j\right\rangle, - \tfrac{1}{m}\left\langle V'(r_{j-1})\right\rangle, - \tfrac{1}{m}\left\langle p_j V'(r_{j-1})\right\rangle\big) = \left(-v, \tfrac{1}{m} P, v P \right).
\end{equation}
Through Eq.~\eqref{eq:avr_current} the currents become functions of the local fields, where the pressure is evaluated as
\begin{equation}
\label{eq:pressure_eval}
P = P(r, \mathfrak{e} - \tfrac{1}{2} m v^2).
\end{equation}
Then the Euler equations read
\begin{equation}
\label{eq:anharmonicEuler1}
\partial_t r -\partial_x v = 0, \quad m\partial_t v +\partial_x P(r, \mathfrak{e} - \tfrac{1}{2} mv^2)=0,\quad \partial_t \mathfrak{e}+\partial_x \big(v P(r, \mathfrak{e} - \tfrac{1}{2} mv^2)\big) = 0.
\end{equation}
We combine the conserved fields as 3-vector $\vec{u}(x,t) = \big(r(x,t),v(x,t),\mathfrak{e}(x,t)\big)$. Then the Euler equations take the canonical form
\begin{equation}
\label{eq:anharmonicEuler}
\partial_t \vec{u} + \partial_x \vec{\mathsf{j}}(\vec{u}) = 0.
\end{equation}
Since the \emph{total} energy $\mathfrak{e}$ (instead of internal energy $e$) is locally conserved, $\mathfrak{e}$ has to be used when applying the theory of hyperbolic conservation laws. But for the Riemann problem, it will turn out to be more concise to use the internal $e$ as parameter.

In fact, Eq.~\eqref{eq:anharmonicEuler1} is identical to the Euler equations of a one-dimensional fluid in Lagrangian coordinates. Its Riemann problem has been studied in great detail starting with the pioneering work of Bethe \cite{Bethe1942}. The interest in one-dimensional fluids also served as a strong motivation to develop a mathematical theory of hyperbolic conservation laws with several components \cite{Lax1957, Glimm1965}. For the physics perspective we refer to the excellent review by Menikoff and Plohr \cite{MenikoffPlohr1989}. The tutorial by Bressan \cite{Bressan2013} provides the necessary mathematical background. So it appears that we only have to point at the relevant literature. From the physics side the main goal is to understand the qualitative link between the equation of state and the solution to the Riemann problem, in particular in case the equation of state allows for a phase transition. On the other hand, we plan to quantitatively compare the microscopic dynamics with the solution of the Riemann problem and thus need shock and rarefaction profiles in a fairly explicit form. Compared to one-dimensional fluids, anharmonic chains have the advantage that the grand-canonical potential is given in terms of a single one-dimensional integral. Static correlations of the conserved fields vanish except for coinciding points. No phase transition is possible. 

Before turning to our concrete examples, we have to recall a few general properties.
The linearization matrix is given by
\begin{equation}
\label{eq:A_general}
A = \frac{\partial \vec{\mathsf{j}}(\vec{u})}{\partial \vec{u}} =
\begin{pmatrix}
0 & -1 & 0 \\
\tfrac{1}{m} \partial_r P & -v \partial_e P & \tfrac{1}{m} \partial_e P \\
v \partial_r P & P - m v^2\,\partial_e P & v \partial_e P
\end{pmatrix}.
\end{equation}
The eigenvalues of $A$ are $(-c, 0, c)$, with the adiabatic sound speed $c$, $c >0$, defined by
\begin{equation}
\label{eq:sq_sound_speed}
c^2 = \tfrac{1}{m} (-\partial_r P + P\,\partial_e P ).
\end{equation}
The right eigenvectors of $A$ corresponding to the eigenvalues $0$ and $\sigma c$, $\sigma = \pm 1$, read
\begin{equation}
\label{eq:psi}
\psi_0 = Z_0^{-1} \begin{pmatrix} \partial_e P \\ 0 \\ - \partial_r P \end{pmatrix}, \qquad
\psi_{\sigma} = Z_{\sigma}^{-1} \begin{pmatrix} -\sigma \\ c \\ \sigma P + m v c \end{pmatrix},
\end{equation}
and the left eigenvectors of $A$ are
\begin{equation}
\label{eq:psi_tilde}
\tilde{\psi}_0 = \tilde{Z}_0^{-1} \begin{pmatrix} P \\ -mv \\ 1 \end{pmatrix}, \qquad
\tilde{\psi}_{\sigma} = \tilde{Z}_{\sigma}^{-1} \begin{pmatrix} \sigma \partial_r P \\ m \big(c - \sigma v \partial_e P\big) \\ \sigma \partial_e P \end{pmatrix}.
\end{equation}
By construction $\langle \tilde{\psi}_{\alpha} \vert \psi_{\beta} \rangle = 0$ for $\alpha \neq \beta$. For the solution of the Riemann problem, the positive normalization constants are not needed. As explained in appendix~\ref{appendix:general_anharm}, they are fixed by requiring that normal modes are orthonormal with respect to the equilibrium measure.

\subsection{Entropy}

Since in equilibrium the $p_j$'s and $r_j$'s are independent, we can use the generalisation of \eqref{eq:LeRoux_entropy} to probability densities in order to obtain the physical entropy, $S$, per volume. 
Inserting from \eqref{eq:ensemble_pj}, \eqref{eq:ensemble_rj} yields
\begin{equation}
\label{eq:entropy}
S(r, e) = \beta (r P + e) + \tfrac{1}{2} \log(2 \pi m) - \tfrac{1}{2} \log\beta + \log Z(P, \beta)
\end{equation}
with $P= P(r,e)$, $\beta = \beta(r,e)$. Denoting by $D =(\partial_r,\partial_{v},\partial_\mathfrak{e})$ the gradient in state space, one obtains
\begin{equation}\label{3.20}
D S\big(r, \mathfrak{e} - \tfrac{1}{2 } mv^2\big) = \beta(r,e) \big( P(r,e), -mv, 1\big)
\end{equation}
and for a smooth solution
\begin{equation}
\partial_t S = D S\cdot \partial_t \vec{u} = - D S\cdot\partial_x \vec{\mathsf{j}} = - \beta \big(- P\,\partial_x v - v\partial_x P + \partial_x( v P)\big) = 0.
\end{equation}
Anharmonic chains are special in the sense that the entropy current vanishes. Entropy may be produced at shock discontinuities, but is not propagated. This behavior should be contrasted with the LeRoux lattice gas, which has a non-zero entropy current, compare with \eqref{eq:LeRoux_entropy_flux}.

\subsection{Rarefaction curves}

The rarefaction curves are obtained by solving the following Cauchy problem in state space 
\begin{equation}
\label{eq:Cauchy}
\partial_{\tau} \vec{u} = \psi_{\alpha}(\vec{u}),
\end{equation}
$\alpha = 0, \pm 1$, where $\psi_{\alpha}$ are the right eigenvectors of $A$, see \cite{Bressan2013} for details,

\paragraph{Eigenvalue $0$:} This is a contact discontinuity, which can be thought of as a rarefaction wave in the limit of zero extension with a non-zero jump. Eq.~\eqref{eq:Cauchy} for $\alpha = 0$ reads
\begin{equation}
\label{eq:rarefaction0_ode}
\partial_{\tau} \begin{pmatrix} r \\ v \\ \mathfrak{e} \end{pmatrix} = \begin{pmatrix} \partial_e P \\ 0 \\ -\partial_r P \end{pmatrix}
\end{equation}
with $P$ evaluated at $(r, \mathfrak{e} - \tfrac{1}{2} m v^2)$. The normalization constant has been absorbed into $\tau$. Since for the velocity $v(\tau) = v_0$ and $\partial_{\tau} e = \partial_{\tau} \mathfrak{e} - m v\,\partial_{\tau} v = \partial_{\tau} \mathfrak{e}$, one obtains the closed system 
\begin{equation}
\label{eq:contact_r_e_ode}
\partial_{\tau} r = \partial_e P(r, e), \qquad
\partial_{\tau} e = - \partial_r P(r, e),
\end{equation}
which is of hamiltonian form with $P(r,e)$ as hamiltonian function. Across a contact discontinuity both pressure and velocity are conserved. By \eqref{3.20} and \eqref{eq:contact_r_e_ode} the entropy changes as
\begin{equation}
\partial_{\tau} S = \beta \left(P \partial_e P - \partial_r P\right) = \beta m c^2. 
\end{equation}
Thus, the change of entropy across the contact discontinuity equals
\begin{equation}
S - S_0 = \int_0^{\tau_{\max}} \ud \tau\, \beta(\tau) m c(\tau)^2 > 0,
\end{equation}
where $\beta(\tau) = \beta(r(\tau), e(\tau))$ and $c(\tau) = c(r(\tau), e(\tau))$.

\paragraph{Eigenvalue $\sigma c$:} Eq.~\eqref{eq:Cauchy} for $\sigma = \pm 1$ reads
\begin{equation}
\label{eq:rarefaction1_ode}
\partial_{\tau} \begin{pmatrix} r \\ v \\ \mathfrak{e} \end{pmatrix} = \begin{pmatrix} -\sigma \\ c \\ \sigma P + m v c \end{pmatrix}.
\end{equation}
The first equation of \eqref{eq:rarefaction1_ode} is solved by $r(\tau) = r_0 - \sigma \tau$. For the internal energy it follows that
\begin{equation}
\label{eq:rarefaction1_ode_eint}
\partial_{\tau} e = \partial_{\tau} \big(\mathfrak{e} - \tfrac{1}{2} m v^2 \big) = \sigma P + m v c - m v c = \sigma P.
\end{equation}
However, instead of the energy equation it is more convenient to use the conservation of entropy
\begin{equation}
S( r_0 - \sigma \tau, e(\tau)) = S_0.
\end{equation}
Inserting $e(\tau)$ into \eqref{eq:rarefaction1_ode}, the velocity is then determined by 
\begin{equation}
\label{eq:rarefaction1_ode2}
\partial_\tau v = c(r_0 - \sigma \tau, e(\tau)).
\end{equation}
The rarefaction curves can be obtained without actually solving differential equations.

The gradient of $\sigma c$ along trajectories of the vector field $\psi_{\sigma}$, for $\sigma = \pm 1$, is
\begin{equation}
\label{eq:gradient_c_rarefaction}
\sigma \psi_{\sigma} \cdot D c = Z_{\sigma}^{-1} \big(P\,\partial_e c - \partial_r c \big).
\end{equation}
A common simplifying assumption for hyperbolic conservation laws is genuine nonlinearity, i.e., $ \sigma \psi_{\sigma} \cdot D c > 0$ for any $r$, $e$. On general grounds, there is no reason why genuine nonlinearity should hold. A point in case are hard-point particles with alternating masses, discussed in Sect.~\ref{sec:hard_point_particles} below. But genuine nonlinearity does not hold for a square-well interaction potential, see Sect.~\ref{sec:square_well}.

\subsection{Shock curves}

According to the Rankine-Hugoniot jump condition, we search for nontrivial solutions of
\begin{equation}
\label{eq:rankine_hugoniot_general}
\lambda (\vec{u} - \vec{u}_0) = \vec{\mathsf{j}}(\vec{u}) - \vec{\mathsf{j}}(\vec{u}_0).
\end{equation}
Using the shorthand notation $P_0 = P(r_0, \mathfrak{e}_0 - \tfrac{1}{2}mv_0^2)$, more explicitly Eq.~\eqref{eq:rankine_hugoniot_general} reads
\begin{align}
\label{eq:rankine_hugoniot_stretch}
\lambda (r - r_0) &= -(v - v_0), \\
\label{eq:rankine_hugoniot_momentum}
\lambda (v - v_0) &= \tfrac{1}{m}\big(P - P_0\big), \\
\label{eq:rankine_hugoniot_energy}
\lambda (\mathfrak{e} - \mathfrak{e}_0) &= v P - v_0 P_0.
\end{align}
According to Eq.~\eqref{eq:rankine_hugoniot_stretch}, the shock speed is
\begin{equation}
\label{eq:shock_speed}
\lambda = - \frac{v - v_0}{r - r_0}.
\end{equation}
Eqs.~\eqref{eq:rankine_hugoniot_momentum}, \eqref{eq:rankine_hugoniot_energy} and the relation $\mathfrak{e} = e + \frac{1}{2} m v^2$ lead to
\begin{equation}
\lambda (e - e_0) = \tfrac{1}{2}(v - v_0) (P + P_0),
\end{equation}
and inserting Eq.~\eqref{eq:rankine_hugoniot_stretch}, one arrives at
\begin{equation}
\label{eq:shock_energy_condition}
e - e_0 = -\tfrac{1}{2} (r - r_0) (P + P_0).
\end{equation}
Multiplying Eqs.~\eqref{eq:rankine_hugoniot_stretch} and \eqref{eq:rankine_hugoniot_momentum} leads to the condition
\begin{equation}
\label{eq:shock_velocity_condition}
-m(v - v_0)^2 = (r - r_0) (P - P_0) .
\end{equation}

There is no general procedure to solve the Rankine-Hugoniot equations. Also the issue of stability can be discussed only once the solution is of a more explicit form.

\section{Hard-point particles with alternating masses}
\label{sec:hard_point_particles}

A widely studied anharmonic chain is the Fermi-Pasta-Ulam lattice with potential $V(x) = \tfrac{1}{2} x^2 + \tfrac{1}{3} \alpha x^3 + \tfrac{1}{4} \beta x^4$ in the historical notation \cite{FPU1955}. As follows from \eqref{eq:avr_stretch} and \eqref{eq:avr_internal_energy}, $r(P,\beta)$ and $e(P,\beta)$ are given by simple integrals. But one still has to invert the pair of functions. To our knowledge, the corresponding Riemann problem has never been studied. To simplify one looks for the factorized ideal gas ansatz $P(r,e) = 2 e h(r)$, which holds if the potential takes only the values $0$, $\infty$. This leaves the choice $V(x) = 0$ for $b \leq x \leq a$ and $V(x) = \infty$ for $x < b$ and $ a < x$. The parameter $b$ describes a hard core at which particles are specularly reflected from each other. The limiting hard-point, $b = 0$, is also allowed. There is an inward collision at separation $a$. Physically one can imagine that neighboring particles are connected by a massless string of maximal length $a$. From the simulation perspective such a potential has the advantage that no differential evolution equation has to be solved. One can simply proceed from collision to collision. At a collision, the momenta are exchanged. Thus $\sum_j g(p_j)$ with general $g$ is conserved. The dynamics is integrable. The standard resolution is to prescribe alternating masses, $m_0$ for even labels and $m_1$ for odd labels. Thereby the collisions become nontrivial and seem to generate sufficient chaos. Now the unit cell contains two particles. But, as discussed in \cite{Spohn2014}, the Euler equations still retain their form \eqref{eq:anharmonicEuler1} upon substituting for $m$ the average mass $\frac{1}{2}(m_0 + m_1)$, again denoted by $m$.

We specialize the results from Sect.~\ref{sec:RiemannGeneral} to the case of a hard-point potential, $V_\mathrm{hp}(x) = \infty$ for $x<0$ and $V_\mathrm{hp}(x) = 0$ for $x \geq 0$. Then the inverse temperature is $\beta = 1/(2e)$, the pressure is given by
\begin{equation}
\label{eq:hp_pressure}
P_{\text{hp}}(r, e) = \frac{2 e}{r},
\end{equation}
and the sound speed is obtained as
\begin{equation}
c_{\text{hp}} = \frac{1}{r} \sqrt{6 e/m} .
\end{equation}
The right eigenvectors of $A$ are 
\begin{equation}
\psi_{0,\text{hp}} = \sqrt{2/3} \begin{pmatrix} r \\ 0 \\ e \end{pmatrix}, \qquad \psi_{\sigma,\text{hp}} = \begin{pmatrix} -\tfrac{1}{\sqrt{6}}\,\sigma r \\ \sqrt{e/ m} \\ v \sqrt{em} + \sigma \sqrt{\tfrac{2}{3}}\,e \end{pmatrix}
\end{equation}
and the left eigenvectors
\begin{equation}
\label{eq:tilde_psi_hp}
\tilde{\psi}_{0,\text{hp}} = \frac{1}{\sqrt{6}} \begin{pmatrix} 2 / r \\ -m v / e \\ 1/e \end{pmatrix}, \qquad
\tilde{\psi}_{\sigma,\text{hp}} = \frac{1}{\sqrt{6}} \begin{pmatrix} -\sigma/r \\ m \big(\tfrac{1}{2} r\, c_{\text{hp}} - \sigma v\big) / e \\ \sigma/e \end{pmatrix} .
\end{equation}

\subsection{Rarefaction curves}

\paragraph{Eigenvalue $0$:}
For hard-point particles Eq.~\eqref{eq:rarefaction0_ode} reads 
\begin{equation}
\partial_{\tau} \begin{pmatrix} r \\ v \\ \mathfrak{e} \end{pmatrix} = \begin{pmatrix} 2/r \\ 0 \\ 2 e/r^2 \end{pmatrix}
\end{equation}
with initial state $(r_0, v_0, \mathfrak{e}_0)$. Note that $\partial_{\tau} \mathfrak{e} = \partial_{\tau} e$, since $\partial_{\tau} v = 0$, and the solution obeys
\begin{equation}
\label{eq:rarefaction0_hp_energy}
e(\tau) = \frac{e_0}{r_0} r(\tau).
\end{equation}
In particular the pressure is conserved, as required.

\paragraph{Eigenvalue $\sigma c$:}
For hard-point particles, Eqs.~\eqref{eq:rarefaction1_ode} and \eqref{eq:rarefaction1_ode_eint} lead to
\begin{equation}
\label{eq:rarefaction1_hp_energy}
e(\tau) = e_0 \left(\frac{r(\tau)}{r_0}\right)^{-2}
\end{equation}
and
\begin{equation}
\label{eq:rarefaction1_hp_momentum}
v(\tau) = v_0 + \sigma \sqrt{6 e_0/m} \left(\frac{1}{r(\tau)/r_0} - 1\right).
\end{equation}
In particular, 
\begin{equation}
c_{\text{hp}}(\tau) = c_{\text{hp},0} \left(\frac{r(\tau)}{r_0}\right)^{-2}.
\end{equation}
The gradient of $\sigma c$ along trajectories of the vector field $\psi_{\sigma}$, Eq.~\eqref{eq:gradient_c_rarefaction}, becomes then
\begin{equation}
\label{eq:gradient_c_rarefaction_hp}
\sigma\psi_{\sigma,\text{hp}}\cdot D c_{\text{hp}} = \sqrt{2/3}\, c_{\text{hp}} > 0,
\end{equation}
i.e., genuine nonlinearity holds.

\subsection{Shock curves}

For hard-point particles the condition \eqref{eq:shock_energy_condition} reads
\begin{equation}
e - e_0 = -\tfrac{1}{2} (r - r_0) \left(\frac{2 e}{r} + \frac{2 e_0}{r_0}\right),
\end{equation}
implying that
\begin{equation}
\label{eq:shock_hp_energy}
\frac{e}{e_0} = \frac{\hat{r} (2 - \hat{r})}{2 \hat{r} - 1}, \qquad \hat{r} = r/r_0
\end{equation}
for $\tfrac{1}{2} < \hat{r} \le 2$. Inserted into Eq.~\eqref{eq:shock_velocity_condition} leads to
\begin{equation}
\label{eq:shock_hp_velocity}
v = v_0 - \sigma \sqrt{6 e_0/m} \frac{\hat{r} - 1}{\sqrt{2 \hat{r} - 1}}, \qquad \hat{r} = r/r_0
\end{equation}
with $\sigma = \pm 1$. The shock speed is then
\begin{equation}
\label{eq:shock_speed_hp}
\lambda_{\text{hp}} = - \frac{v - v_0}{r - r_0} = \frac{\sigma c_{\text{hp},0}}{\sqrt{2 \hat{r} - 1}}
\end{equation}
with $c_{\text{hp},0} = \frac{1}{r_0} \sqrt{6 e_0/m}$ the sound speed of the initial state.

The \emph{Lax admissibility condition} states that characteristics must run ``towards'' the shock,
\begin{equation}
\sigma c_{\text{hp},0} \ge \lambda_{\text{hp}} \ge \sigma c_{\text{hp}} .
\end{equation}
Inserting the relation \eqref{eq:shock_hp_energy} gives
\begin{equation}
c_{\text{hp}} = \frac{1}{r} \sqrt{6 e/m} = \frac{1}{\sqrt{2 \hat{r} - 1}} \sqrt{(2 - \hat{r})/\hat{r}}\, c_{\text{hp},0} ,
\end{equation}
such that the Lax admissibility condition becomes
\begin{equation}
\sigma \ge \frac{\sigma}{\sqrt{2 \hat{r} - 1}} \quad \text{and} \quad \sigma \ge \sigma \sqrt{(2 - \hat{r})/\hat{r}}.
\end{equation}
For $\sigma = 1$, this is equivalent to $\hat{r} \ge 1$, and for $\sigma = -1$ equivalent to $\hat{r} \le 1$.

\begin{figure}[!ht]
\centering
\subfloat[energy in dependence of stretch]{\label{fig:hp_energy_stretch}%
\includegraphics[width=0.45\textwidth]{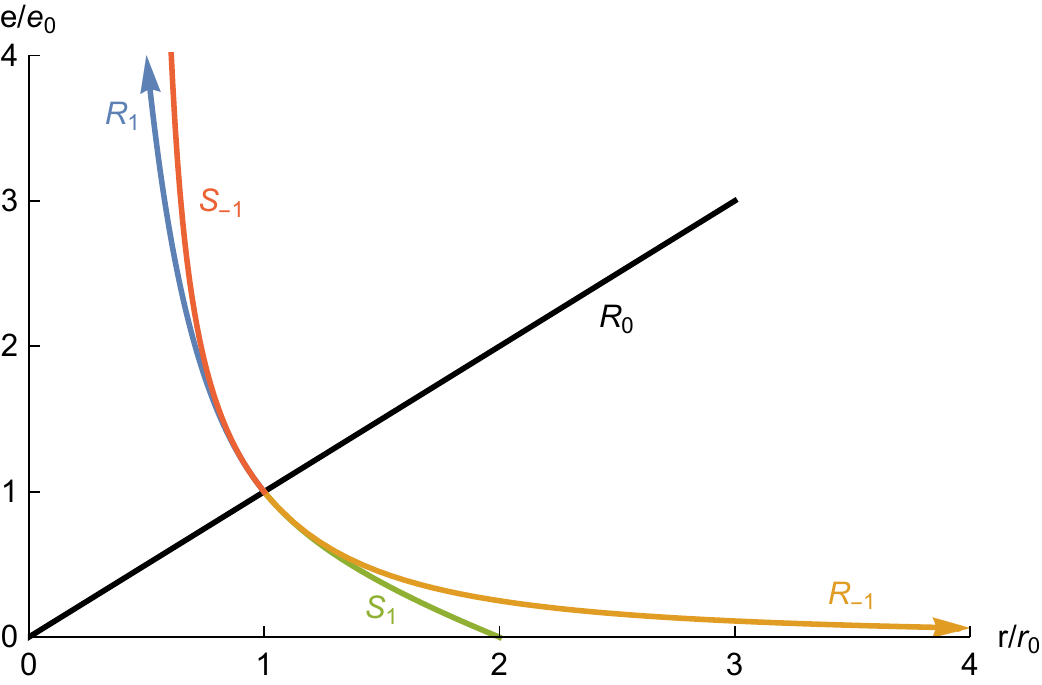}}
\hspace{0.05\textwidth}
\subfloat[velocity in dependence of stretch]{\label{fig:hp_velocity_stretch}%
\includegraphics[width=0.45\textwidth]{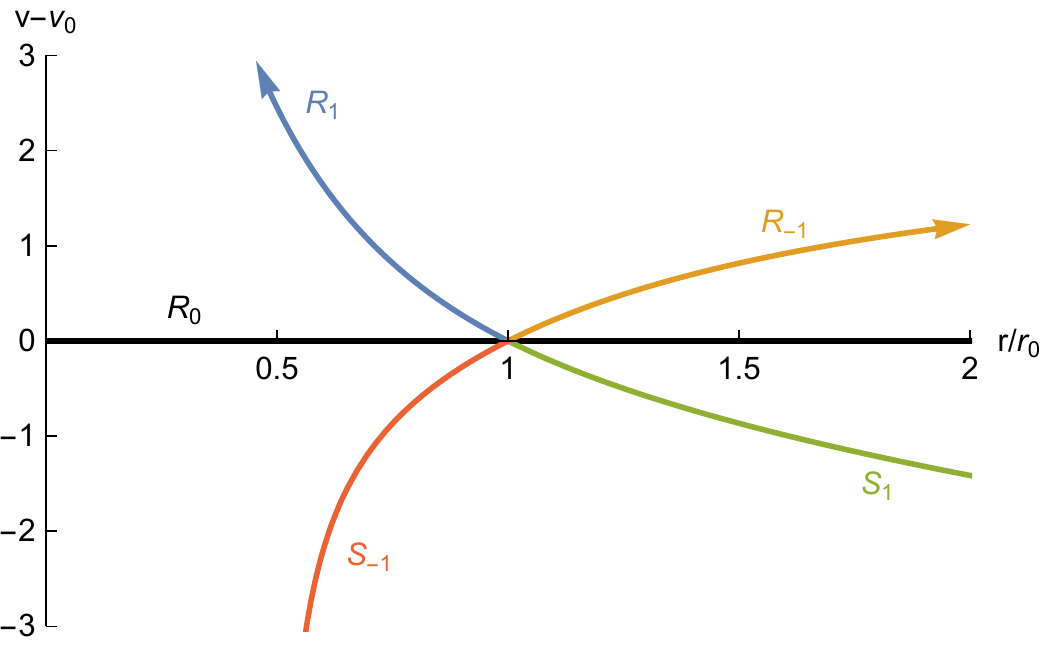}}
\caption{Integral curves for hard-point particles. (a) Internal energy in dependence of stretch for rarefaction and shock curves, see Eqs.~\eqref{eq:rarefaction0_hp_energy}, \eqref{eq:rarefaction1_hp_energy} and \eqref{eq:shock_hp_energy}, respectively. The black line is the identity function, and the red curve $S_{-1}$ diverges at $r/r_0 = 1/2$. (b) Velocity in dependence of stretch, see Eqs.~\eqref{eq:rarefaction1_hp_momentum} and \eqref{eq:shock_hp_velocity}, for $m = 1$ and $e_0 = 1$.}
\end{figure}
Fig.~\ref{fig:hp_energy_stretch} displays the internal energy in dependence of the stretch, both for the rarefaction and shock curves, denoting by $R_i$ the $i$-th rarefaction curve and by $S_i$ the $i$-th shock curve. Analogously, Fig.~\ref{fig:hp_velocity_stretch} displays the change of velocity in dependence of the stretch. These figures should be compared with Fig.~\ref{fig:LeRoux_rarefaction_shock}. Considering only rarefactions and shocks, the solution to both Riemann problems are qualitatively the same. The contact discontinuity is merely an independent additional feature.

\subsection{Entropy}

\begin{figure}[!ht]
\centering
\subfloat[shock $x$-$t$ profile]{\includegraphics[width=0.3\textwidth]{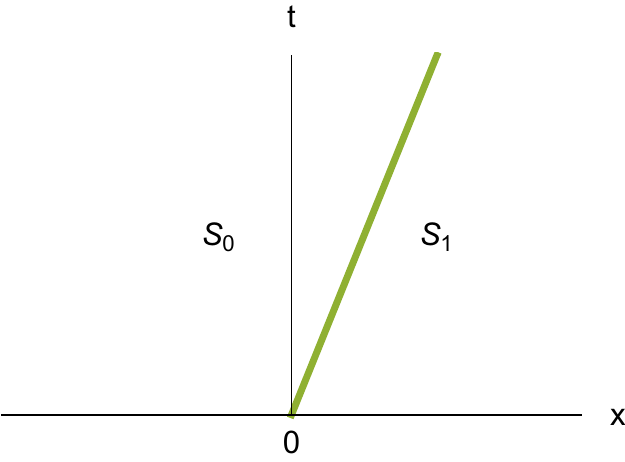}}
\hspace{0.04\textwidth}
\subfloat[entropy at $t=0$]{\includegraphics[width=0.3\textwidth]{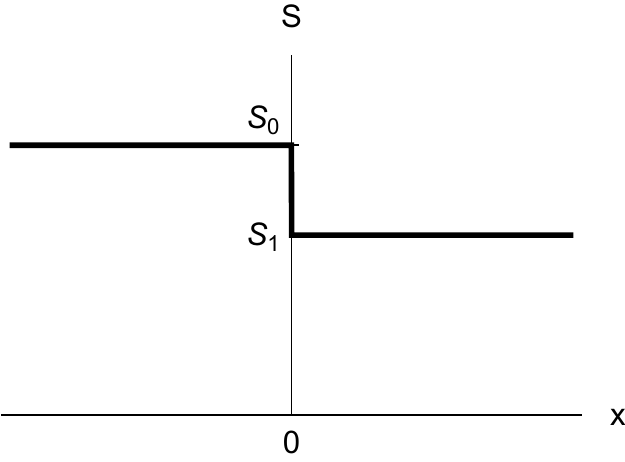}}
\hspace{0.04\textwidth}
\subfloat[entropy at $t>0$]{\includegraphics[width=0.3\textwidth]{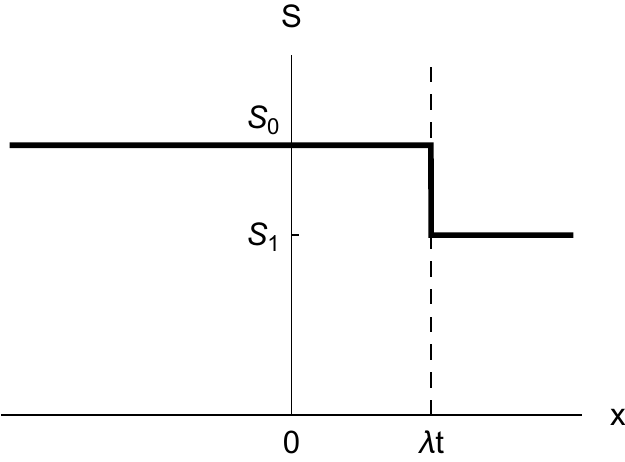}}
\caption{Schematic illustration of a shock curve with speed $\lambda > 0$ and corresponding time evolution of the entropy.}
\label{fig:entropy_shock}
\end{figure}

The entropy \eqref{eq:entropy} for the hard-point particles is
\begin{equation}
\label{eq:entropy_hp}
S_{\text{hp}}(r,e) = \log(r) + \tfrac{1}{2} \log(e) ,
\end{equation}
up to a constant shift by $\tfrac{3}{2} + \tfrac{1}{2} \log(4 \pi m)$. For a jump along a shock curve with speed $\lambda$, the entropy admissibility condition $\partial_t S(\vec{u}) \ge 0$ becomes
\begin{equation}
\lambda \big(S(r_1, e_1) - S(r_0, e_0)\big) \le 0,
\end{equation}
compare to Eq.~(47) in \cite{Bressan2013} with opposite sign. We follow the physics convention of a concave entropy function, while \cite{Bressan2013} prefers a convex function. In other words, at a shock a region with higher entropy invades a region with lower entropy. Fig.~\ref{fig:entropy_shock} schematically visualizes the time evolution of the entropy across a shock with speed $\lambda > 0$.

\subsection{Molecular dynamics}

\begin{figure}[!ht]
\centering
\includegraphics[width=0.5\textwidth]{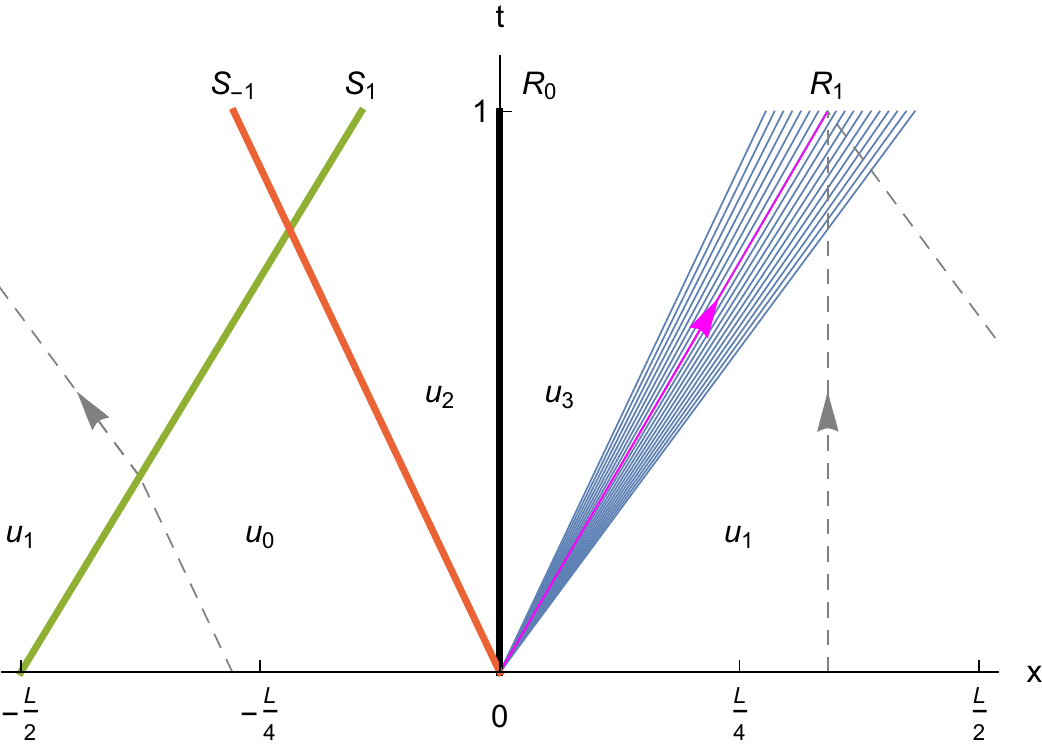}
\caption{Visualization of the theoretically predicted rarefaction and shock curves for a system of size $L$ with periodic boundary conditions and domain-wall initial conditions. The largest time corresponds to $L/4$.}
\label{fig:hardpoint_finite_vol_summary}
\end{figure}

For a molecular dynamics simulation we use a box of size $L$ with periodic boundary conditions, for which we adopt the domain $[-\frac{L}{2}, \frac{L}{2}-1]$. Imposing domain-wall initial conditions as $\vec{u}_0|\vec{u}_1$ at the origin implies that somewhere else one has the reversed initial condition $\vec{u}_1|\vec{u}_0$. We call this the periodic Riemann problem. The Riemann problem centered at $0$ is of key interest. But numerically we automatically realize two distinct Riemann problems. The longest time of simulation is limited by collisions between the two solution branches. A conventional choice for the masses is $m = 2$, corresponding to alternating masses with $m_0 = 1$ and $m_1 = 3$. As in \cite{TWMendlSpohn2016}, we prescribed $\vec{u}_1 = (r_1, v_1, \mathfrak{e}_1) = (1, 0, 1)$ and determine the entries of $\vec{u}_0$ such that the rarefaction waves and shocks as shown in Fig.~\ref{fig:hardpoint_finite_vol_summary} arise. In \cite{TWMendlSpohn2016} the goal was to have a wide rarefaction wave. Here we explain in detail how the solution of the periodic Riemann problem is constructed.

For conciseness, we denote the quotients of the stretches by $\mu = r_1 / r_3$, $\phi = r_3 / r_2$, $\omega = r_2 / r_0$ and $\xi = r_0 / r_1$. Considering the right rarefaction $R_1$, condition \eqref{eq:rarefaction1_hp_energy} for the internal energy reads $e_1 = \mu^{-2} e_3$, and \eqref{eq:rarefaction1_hp_momentum} for the velocity
\begin{equation}
v_1 = v_3 + \sigma \sqrt{6 e_3 / m} \left(\frac{1}{\mu} - 1\right), \quad \sigma = 1.
\end{equation}
We can still choose $\mu$, which we set to $\mu = \frac{4}{5}$ in Fig.~\ref{fig:hardpoint_finite_vol_summary}. Accordingly, $r_3 = \frac{5}{4}$, $v_3 = - \frac{1}{5} \sqrt{3}$ and $e_3 = \frac{16}{25}$. It remains to show that $\vec{u}_0$ and $\vec{u}_2$ can be chosen in accordance with Fig.~\ref{fig:hardpoint_finite_vol_summary}. The contact discontinuity $R_0$ at the origin implies that the velocity is conserved, i.e., $v_2 = v_3$. Furthermore $e_3/e_2 = r_3/r_2 = \phi$. Considering the shock curve $S_{-1}$, according to \eqref{eq:shock_hp_energy},
\begin{equation}
e_2 = \frac{\omega (2 - \omega)}{2\omega - 1} e_0,
\end{equation}
since $r_2 = \omega r_0$ as defined above, and according to \eqref{eq:shock_hp_velocity},
\begin{equation}
v_2 = v_0 + \sqrt{6 e_0/m} \frac{\omega -1}{\sqrt{2 \omega -1}}.
\end{equation}
Finally, \eqref{eq:shock_hp_energy} for the shock curve $S_1$ implies that
\begin{equation}
e_0 = e_1 \frac{\xi (2 - \xi)}{2\xi - 1}
\end{equation}
and \eqref{eq:shock_hp_velocity} that
\begin{equation}
v_0 = v_1 - \sqrt{6 e_1/m} \frac{\xi -1}{\sqrt{2 \xi -1}}.
\end{equation}
In summary, it holds that $r_1 = \mu\,r_3 = \mu\,\phi\,r_2 = \mu\,\phi\,\omega\,r_0 = \mu\,\phi\,\omega\,\xi\,r_1$, i.e.,
\begin{equation}
\label{eq:periodic_hp_stretch_condition}
\mu\,\phi\,\omega\,\xi = 1.
\end{equation}
Similarly,
\begin{equation}
\frac{e_1}{r_1} = \frac{1}{\mu^3}\, \frac{e_3}{r_3} = \frac{1}{\mu^3}\, \frac{e_2}{r_2} = \frac{1}{\mu^3}\, \frac{2 - \omega}{2 \omega -1}\, \frac{e_0}{r_0} = \frac{1}{\mu^3}\, \frac{2 - \omega}{2 \omega -1}\, \frac{2 - \xi}{2 \xi -1}\, \frac{e_1}{r_1},
\end{equation}
which enforces that
\begin{equation}
\label{eq:periodic_hp_energy_condition}
\frac{1}{\mu^3}\, \frac{2 - \omega}{2 \omega -1}\, \frac{2 - \xi}{2 \xi -1} = 1.
\end{equation}
The analogous procedure for the velocity leads to
\begin{equation}
\label{eq:periodic_hp_momentum_condition}
- \frac{\xi - 1}{\sqrt{2 \xi -1}} - \frac{1}{\sqrt{\phi}} \mu \frac{\omega - 1}{\sqrt{\omega (2 - \omega)}} + (1-\mu ) = 0.
\end{equation}
Solving \eqref{eq:periodic_hp_stretch_condition}, \eqref{eq:periodic_hp_energy_condition}, and \eqref{eq:periodic_hp_momentum_condition} for $\phi$, $\omega$ and $\xi$ with the help of a computer algebra program leads to rather lengthy expressions in terms of roots of certain polynomials, which we do not write down explicitly. Instead, we report the numerical values for $\mu = \frac{4}{5}$, namely $\phi = 1.00728$, $\omega = 0.99875$ and $\xi = 1.2425$. Note that $\phi$ and $\omega$ are close to $1$. The numerical values for $\vec{u}_2$ are $(r_2, v_2, e_2) = (1.241, -0.3464, 0.63537)$, and for $\vec{u}_0$ the values read $(r_0, v_0, e_0) = (1.2425,-0.34469,0.63379)$. The shock $S_{-1}$ is hardly visible, while $R_{1}$ and $S_{1}$ are well-developed. The rounding of the edges is presumably a finite size effect. A similar feature is noted in Fig.~\ref{fig:LeRoux_profiles_general} for the LeRoux lattice gas. An alternative explanation would be rounding because of diffusion. But then this should show also in Fig.~\ref{fig:LeRoux_profiles}. To decide larger size systems would have to be simulated.

\begin{figure}[!t]
\centering
\subfloat[stretch, $t = 0$]{\includegraphics[width=0.225\textwidth]{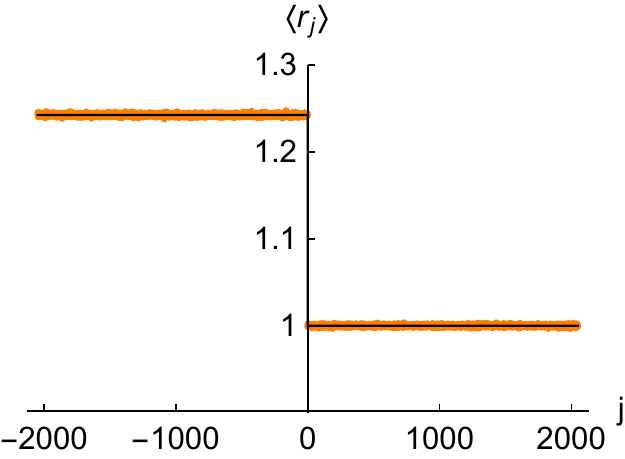}}
\hspace{0.02\textwidth}
\subfloat[stretch, $t = 256$]{\includegraphics[width=0.225\textwidth]{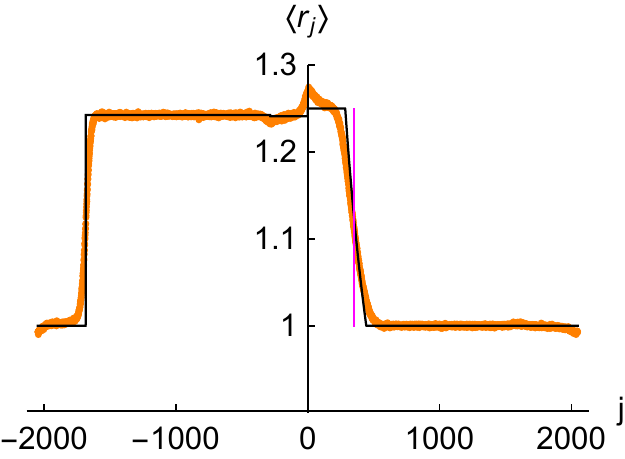}}
\hspace{0.02\textwidth}
\subfloat[stretch, $t = 512$]{\includegraphics[width=0.225\textwidth]{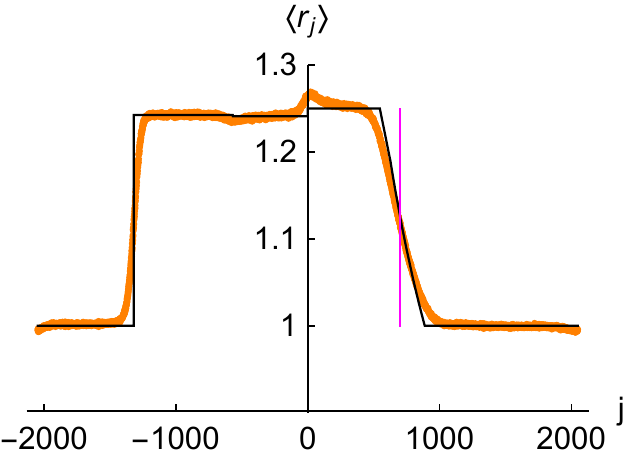}}
\hspace{0.02\textwidth}
\subfloat[stretch, $t = 1024$]{\includegraphics[width=0.225\textwidth]{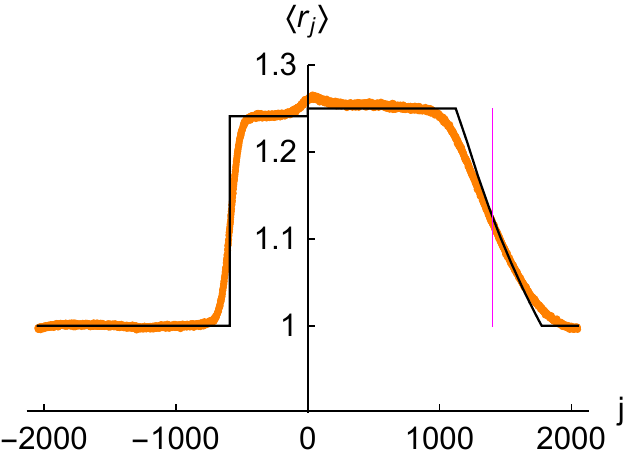}}\\
\subfloat[velocity, $t = 0$]{\includegraphics[width=0.225\textwidth]{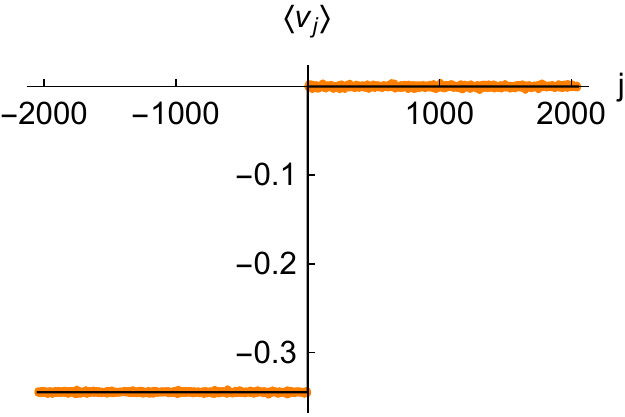}}
\hspace{0.02\textwidth}
\subfloat[velocity, $t = 256$]{\includegraphics[width=0.225\textwidth]{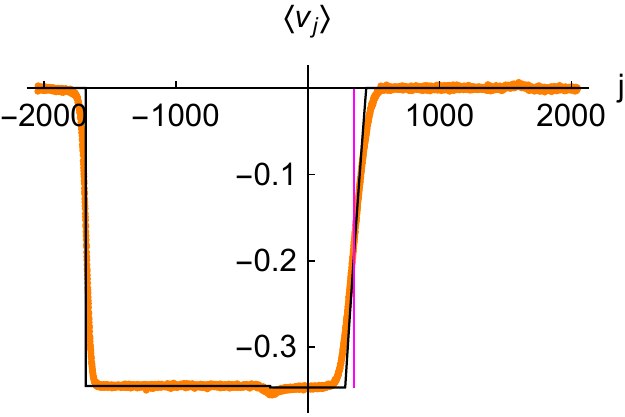}}
\hspace{0.02\textwidth}
\subfloat[velocity, $t = 512$]{\includegraphics[width=0.225\textwidth]{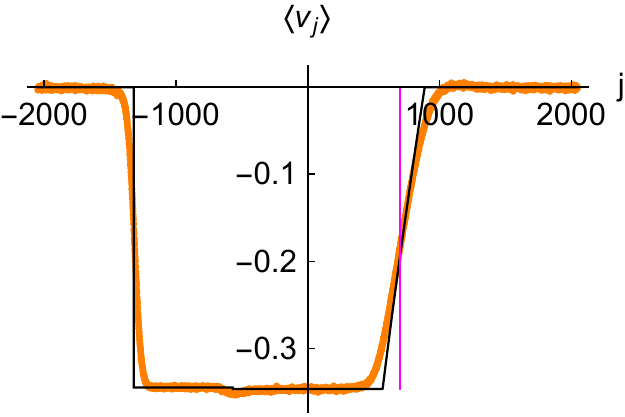}}
\hspace{0.02\textwidth}
\subfloat[velocity, $t = 1024$]{\includegraphics[width=0.225\textwidth]{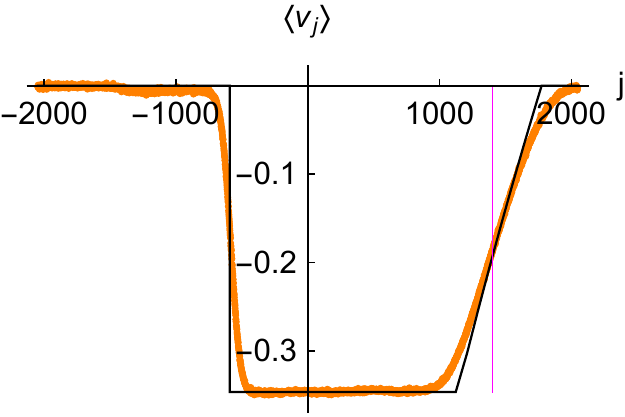}}\\
\subfloat[energy, $t = 0$]{\includegraphics[width=0.225\textwidth]{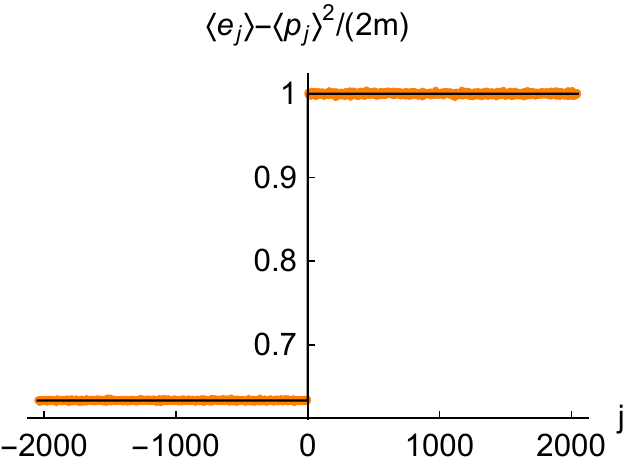}}
\hspace{0.02\textwidth}
\subfloat[energy, $t = 256$]{\includegraphics[width=0.225\textwidth]{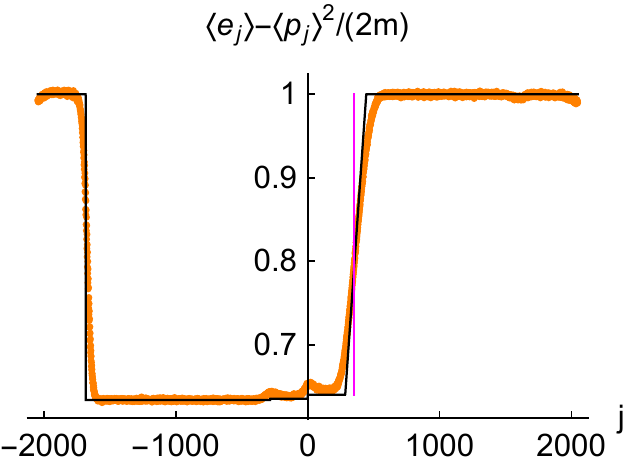}}
\hspace{0.02\textwidth}
\subfloat[energy, $t = 512$]{\includegraphics[width=0.225\textwidth]{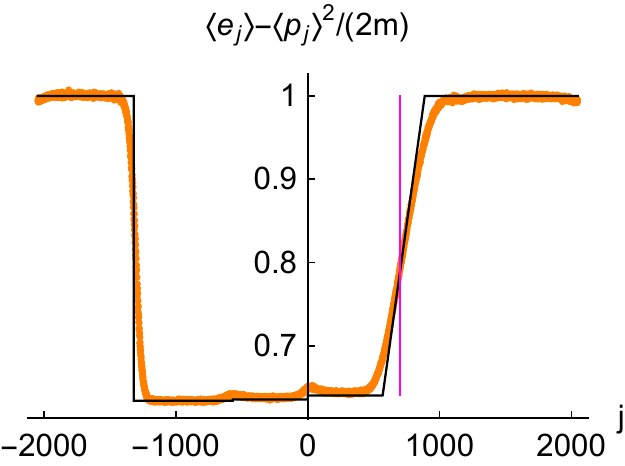}}
\hspace{0.02\textwidth}
\subfloat[energy, $t = 1024$]{\includegraphics[width=0.225\textwidth]{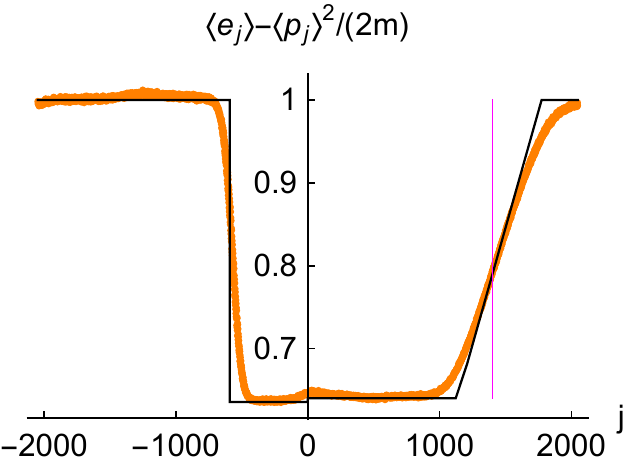}}
\caption{Hard-point particle stretch, velocity and internal energy profiles at various time points, corresponding to the periodic Riemann problem of Fig.~\ref{fig:hardpoint_finite_vol_summary} with system size $L = 4096$. The orange dots are molecular dynamics results and the black thin lines show the theoretically predicted profiles.}
\label{fig:hp_profiles}
\end{figure}

Fig.~\ref{fig:hp_profiles} shows stretch, velocity, and internal energy profiles obtained by molecular dynamics simulations, after averaging over $10^6$ simulation runs with initial states chosen according to $\vec{u}_0$ and $\vec{u}_1$ in Fig.~\ref{fig:hardpoint_finite_vol_summary}. A shock generated through rapid compression is studied in \cite{Hurtado2006}.

\section{Square-well potential}
\label{sec:square_well}

\begin{figure}[!ht]
\centering
\includegraphics[width=0.5\textwidth]{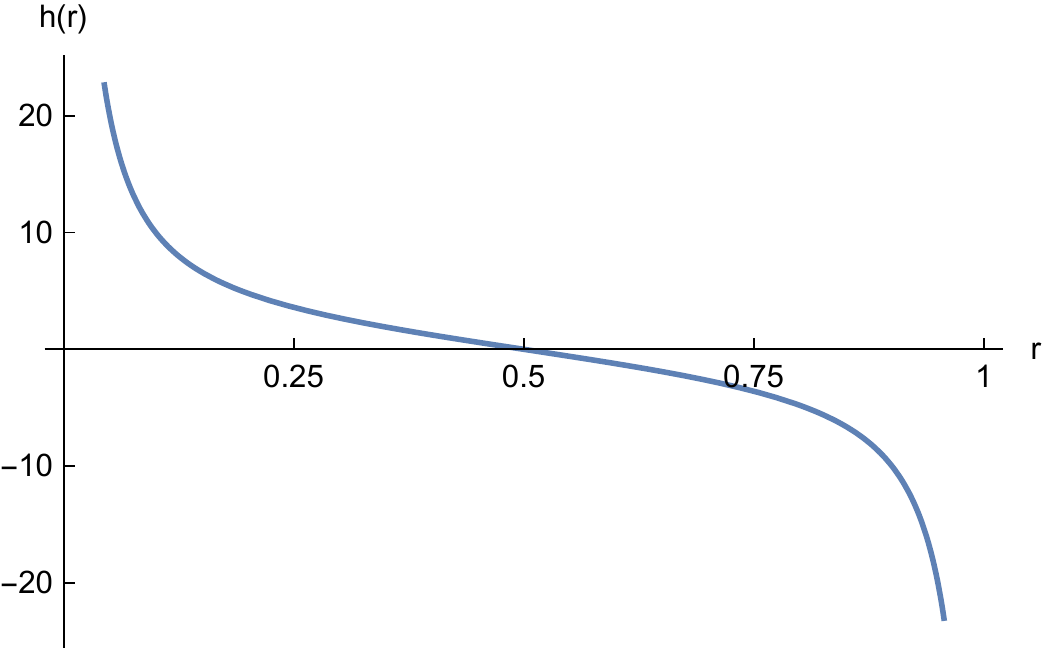}
\caption{Inverse function to $y \mapsto y^{-1} - (\mathrm{e}^y -1)^{-1}$, defining the pressure $2eh(r)$.}
\label{fig:squarewell_h}
\end{figure}

Hard-point particles are an example of a genuinely nonlinear hyperbolic conservation law. As discussed in \cite{MenikoffPlohr1989}, in general, one-dimensional fluids do not have such a property and the structure of solutions to the Riemann problem is considerably richer than for hard-points. A similar observation is well known for stochastic dynamics. For a single component genuine nonlinearity corresponds to a convex flux function, which imposes an additional constraint. A two-component stochastic system is studied in \cite{Balazs2016}. For anharmonic chains we still want to comply with the factorized ideal gas law $P(r,e) = 2 e h(r)$, we consider the hard-core potential with core diameter $b = 0$ and inward reflection at $a$. This defines the \emph{square-well interaction potential} (see also \cite{MendlSpohn2014})
\begin{equation}
\label{eq:sqw_V}
V_\mathrm{sw}(x) = 0 \quad \mathrm{for} \quad 0 \le x \le a\,,\quad
V_\mathrm{sw}(x) = \infty \hspace{4pt}\mathrm{otherwise}.
\end{equation}
Since the potential is zero within the well,
\begin{equation}
e = \frac{1}{2\beta}
\end{equation}
and the pressure factorizes as
\begin{equation}
\label{eq:squarewell_pressure_factorization}
a\beta P = h(r/a)\,,
\end{equation}
where $h$ is the inverse function of $y \mapsto y^{-1} - (\mathrm{e}^y -1)^{-1}$. The unit length can be chosen such that $a = 1$, which we adopt in the following. Then \eqref{eq:squarewell_pressure_factorization} is rewritten as
\begin{equation}
\label{eq:pressure_h}
P(r,e) = 2 e h(r).
\end{equation}
Hard-point particles are obtained in the limiting case $a \to \infty$, which corresponds to setting $h(r) = 1/r$.
$h(r)$ is visualized in Fig.~\ref{fig:squarewell_h}.
Note that derivatives and indefinite integrals can be expressed through $h$ itself, for example
\begin{equation}
h'(r) = -\left(h(r)^2 - \tfrac{1}{4} \sinh\!\big(\tfrac{1}{2}h(r)\big)^{-2}\right)^{-1}
\end{equation}
and
\begin{equation}
\int \ud r\, h(r) = h(r) + \frac{h(r)}{\mathrm{e}^{h(r)} - 1} + \log\!\Big(\frac{h(r)}{\mathrm{e}^{h(r)} - 1}\Big) - 1.
\end{equation}

It is more transparent to keep for a while a general $h$, specializing to the square-well potential at the end. The only constraint is $h'(r) < 0$, ensuring thermodynamic stability. Wendroff \cite{Wendroff1972} discusses a two component model, which in essence corresponds to such a choice upon dropping the contact discontinuity. Inserting \eqref{eq:pressure_h} into \eqref{eq:A_general} results in the linearized currents
\begin{equation}
\label{5.7}
A_{h} =
\begin{pmatrix}
0 & - 1 & 0 \\
 \tfrac{2}{m}e h' & -2 v h & \tfrac{2}{m}h \\
 2 ve h' & 2 \big(e - mv^2 \big) h & 2 v h
\end{pmatrix}
\end{equation}
and the square of the sound speed
\begin{equation}
\label{eq:c_h_sq}
c_h^2 = \tfrac{1}{m} (-\partial_r P + P\,\partial_e P ) = \tfrac{1}{m} 2 e (2 h^2 - h').
\end{equation}
The right eigenvectors of $A$ are
\begin{equation}
\label{5.9}
\psi_{0,h} = Z_{0,h}^{-1} \begin{pmatrix} 2 h \\ 0 \\ -2 e h'\end{pmatrix}, \qquad
\psi_{\sigma,h} = Z_{\sigma,h}^{-1} \begin{pmatrix} -\sigma \\ c_h \\ \sigma 2 e h + m v c_h \end{pmatrix}.
\end{equation}

The gradient of $\sigma c_h$ along trajectories of the vector field $\psi_{\sigma,h}$ for $\sigma = \pm 1$, i.e., \eqref{eq:gradient_c_rarefaction} for the special case \eqref{eq:pressure_h}, is
\begin{equation}
\label{eq:gradient_c_rarefaction_h}
\sigma \psi_{\sigma,h} \cdot D c_h = \tfrac{1}{2} \sqrt{e/m} \, \frac{4 h^3 - 6 h h' + h''}{2 h^2-h'}.
\end{equation}
For the hard-point case, $h(r) = 1/r$, this simplifies to \eqref{eq:gradient_c_rarefaction_hp}. Note that the right hand side of \eqref{eq:gradient_c_rarefaction_h} is independent of $\sigma$, which is achieved by an appropriate choice of the sign of $\psi_{\sigma,h}$. The square-well potential turns out to violate genuine nonlinearity, since the expression \eqref{eq:gradient_c_rarefaction_h} changes sign at $r = \frac{1}{2}$. By a suitable choice of $h$, presumably one can generate shocks and rarefactions of the same richness as
 in \cite{Wendroff1972}. Compared to the genuinely nonlinear case the main novel
feature is to have a shock at the borderline of a rarefaction wave. For the square-well potential only the case of a left bordering shock is realized. In principle, the shock could also switch to the opposite side, but it cannot lie in the interior of the rarefaction wave.

\subsection{Rarefaction curves}

\paragraph{Eigenvalue $0$:}

According to Eq.~\eqref{eq:rarefaction0_ode}
\begin{equation}
\partial_{\tau} \begin{pmatrix} r \\ v \\ \mathfrak{e} \end{pmatrix} = \begin{pmatrix} \partial_e P \\ 0 \\ - \partial_r P \end{pmatrix}
\end{equation}
with the pressure $P(r,e)$ conserved. Thus from \eqref{eq:pressure_h}, it follows that
\begin{equation}
\label{eq:rarefaction0_stretch_energy_h}
e(\tau) = e_0 \frac{h(r_0)}{h(r(\tau))}.
\end{equation}

\paragraph{Eigenvalue $\sigma c$:} According to Eqs.~\eqref{eq:rarefaction1_ode}, \eqref{eq:rarefaction1_ode_eint}, the stretch obeys $r(\tau) = r_0 - \sigma \tau$ and the internal energy $\partial_{\tau} e = \sigma P$. Together with \eqref{eq:pressure_h}, one obtains
\begin{equation}
\label{eq:rarefaction1_energy_h}
e(\tau) = e_0 \exp\!\Big[-2 \int_{r_0}^{r(\tau)} \ud\rho\, h(\rho) \Big],
\end{equation}
which only depends on $\tau$ via $r(\tau)$. Inserting this relation into the differential equation $\partial_{\tau} v = c_h$, the sound speed depending on $r(\tau)$ and $e(\tau)$ via \eqref{eq:c_h_sq}, leads to
\begin{equation}
\partial_{\tau} v = -\sigma\,\sqrt{2 e_0/m}\,\exp\!\Big[-\int_{r_0}^{r(\tau )} \ud\rho\, h(\rho) \Big] \sqrt{2 h(r(\tau))^2 - h'(r(\tau ))} \, r'(\tau)
\end{equation}
and integrates to
\begin{equation}
\label{eq:rarefaction1_velocity_h}
v(\tau) = v_0 - \sigma\,\sqrt{2 e_0/m} \int_{r_0}^{r(\tau)} \ud s\, \exp\!\Big[-\int_{r_0}^s \ud\rho\, h(\rho) \Big] \sqrt{2 h(s)^2 - h'(s)}.
\end{equation}

\subsection{Shock curves}

\begin{figure}[!ht]
\centering
\subfloat[energy in dependence of stretch]{\label{fig:sqw_energy_stretch}%
\includegraphics[width=0.45\textwidth]{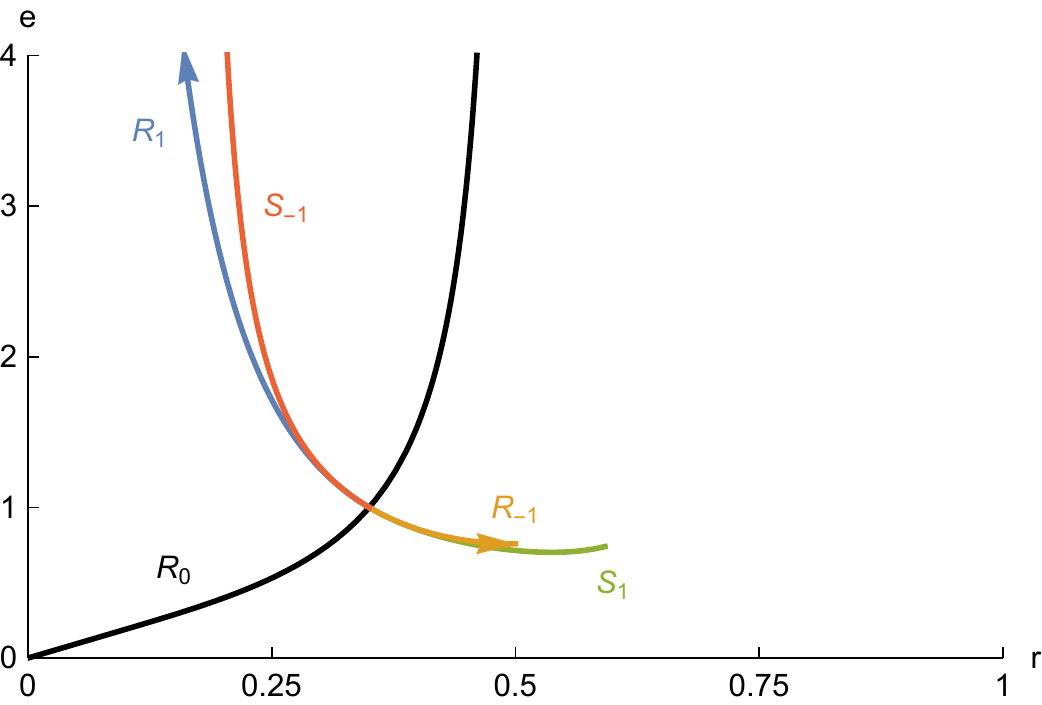}}
\hspace{0.05\textwidth}
\subfloat[velocity in dependence of stretch]{\label{fig:sqw_velocity_stretch}%
\includegraphics[width=0.45\textwidth]{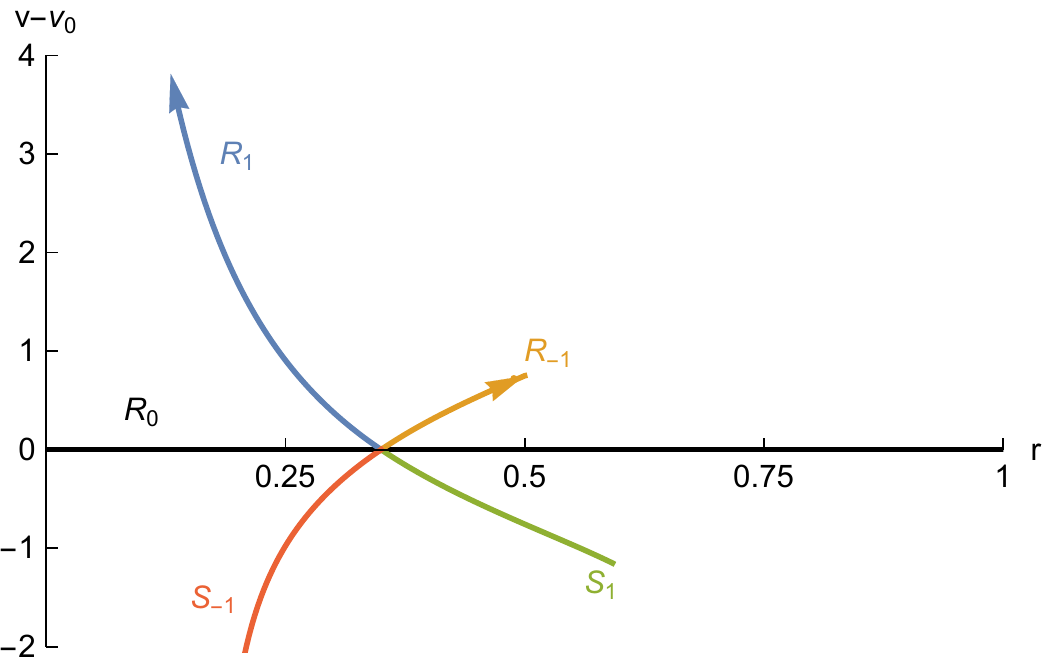}}
\caption{Integral curves for the square-well potential \eqref{eq:sqw_V} with $m = 1$, $r_0 = 0.35$, $e_0 = 1$. (a) Internal energy in dependence of stretch for the rarefaction and shock curves, according to Eqs.~\eqref{eq:rarefaction0_stretch_energy_h}, \eqref{eq:rarefaction1_energy_h} and \eqref{eq:shock_energy_h}, respectively. (b) Velocity in dependence of stretch, see Eqs.~\eqref{eq:rarefaction1_velocity_h} and \eqref{eq:shock_momentum_h}.}
\end{figure}

For $P$ as in \eqref{eq:pressure_h}, Eq.~\eqref{eq:shock_energy_condition} leads to
\begin{equation}
\label{eq:shock_energy_h}
\frac{e}{e_0} = \frac{1 - h(r_0) (r - r_0)}{1 + h(r) (r - r_0)}
\end{equation}
for $r \le r_0 + 1/h(r_0)$ if $h(r_0) > 0$ and $r \ge r_0 + 1/h(r_0)$ if $h(r_0) < 0$. The condition \eqref{eq:shock_velocity_condition} leads to
\begin{equation}
\label{eq:shock_momentum_h}
v = v_0 - \sigma\,\mathrm{sign}(r - r_0) \sqrt{2 e_0/m}\,\sqrt{-(r - r_0) \big(h(r) e/e_0 - h(r_0)\big)}.
\end{equation}
For the hard-point particles with $h(r) = 1/r$, Eq.~\eqref{eq:shock_momentum_h} simplifies to the expression \eqref{eq:shock_hp_velocity}. Inserting \eqref{eq:shock_momentum_h} into \eqref{eq:shock_speed} results in the shock speed
\begin{equation}
\label{eq:shock_speed_h}
\lambda_h = \sigma \sqrt{2 e_0/m} \sqrt{-\frac{h(r) e/e_0 - h(r_0)}{r - r_0}}.
\end{equation}
For the square-well potential Fig.~\ref{fig:sqw_energy_stretch} visualizes an integral curve of the internal energy in dependence of the stretch, both for the rarefaction and shock curves. Analogously, Fig.~\ref{fig:sqw_velocity_stretch} visualizes the velocity in dependence of the stretch.

\begin{figure}[!ht]
\centering
\includegraphics[width=0.45\textwidth]{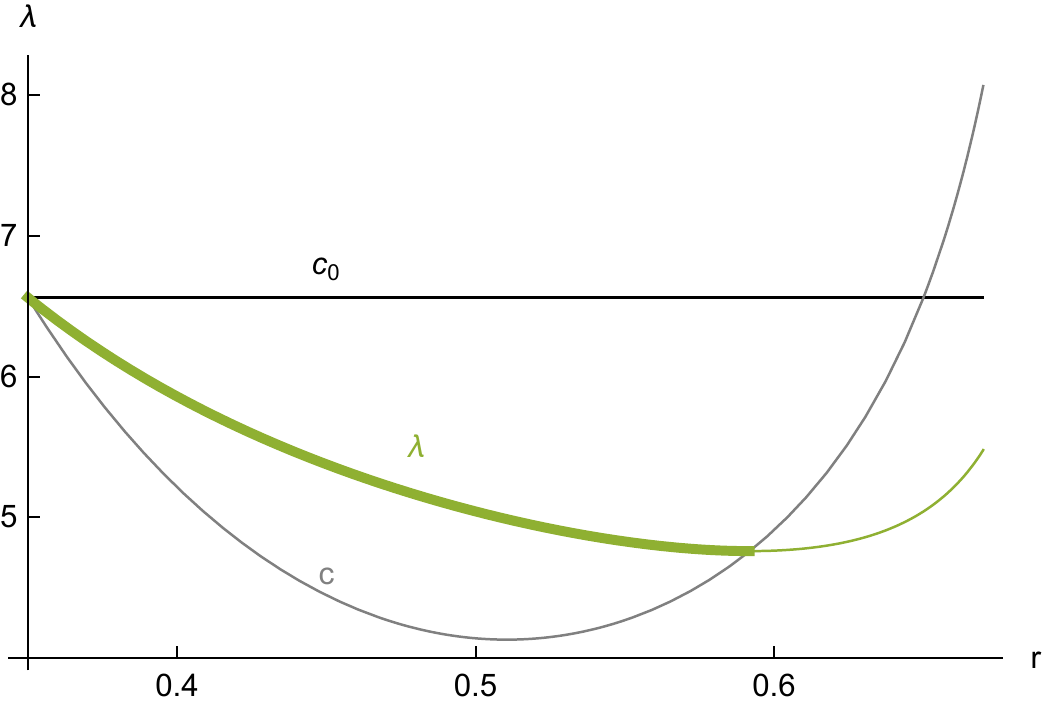}
\caption{Illustration of the Lax admissibility condition $c_{h,0} \ge \lambda_h \ge c_h$ for the square-well potential with $m = 1$, $r_0 = 0.35$, $e_0 = 1$. The green curve shows the shock speed \eqref{eq:shock_speed_h} in dependence of the stretch; the condition only holds within the thick curve segment.}
\label{fig:Lax_condition}
\end{figure}
The Lax admissibility condition
\begin{equation}
\sigma c_{h,0} \ge \lambda_h \ge \sigma c_h
\end{equation}
can be tested numerically for the square-well $h(r)$. Fig.~\ref{fig:Lax_condition} shows an example with $r_0 = 0.35$ where the condition is satisfied within an interval $r_0 \le r \le 0.59$. For larger $r$, the sound speed (evaluated along the shock solution \eqref{eq:shock_energy_h}) becomes larger than $\lambda_h$.
\begin{figure}[!ht]
\centering
\includegraphics[width=0.45\textwidth]{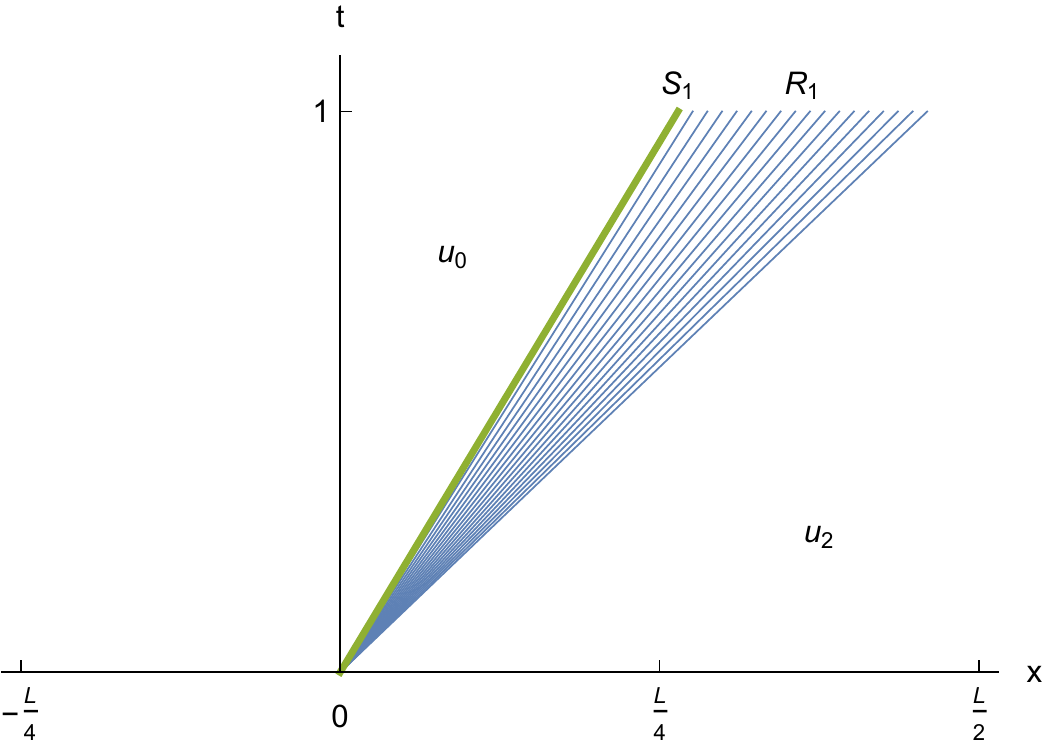}
\caption{A solution of the Euler equation with square-well interaction potential, where a shock wave is followed immediately by a rarefaction wave.}
\label{fig:squarewell_shock_rarefaction}
\end{figure}
Thus a shock wave is followed immediately by a rarefaction wave, as shown in Fig.~\ref{fig:squarewell_shock_rarefaction} with $L = 18t$. The state to the left of the shock curve is $\vec{u}_0 = (r_0, v_0, e_0) = (0.35, 0, 1)$, and the transition from shock to rarefaction wave appears at $r_1 = 0.59$ with $\vec{u}_1 = (0.59, -1.15, 0.74)$. The state $\vec{u}_1$ is connected by a rarefaction wave to $\vec{u}_2 = (0.7, -1.82, 1.108)$.

\subsection{Entropy}

Specifically for the square-well interaction potential, the entropy equals
\begin{equation}
S_{h}(r,e) = r h(r) - 1 + \tfrac{1}{2} \log(e) - \log\!\big(1 + (1 - r) h(r)\big),
\end{equation}
up to a constant shift by $\tfrac{3}{2} + \tfrac{1}{2} \log(4 \pi m)$. As expected, for the hard-point case $h(r) = 1/r$, this expression simplifies to \eqref{eq:entropy_hp}.

\subsection{Molecular dynamics}

\begin{figure}[!t]
\centering
\subfloat[stretch, $t = 0$]{\includegraphics[width=0.225\textwidth]{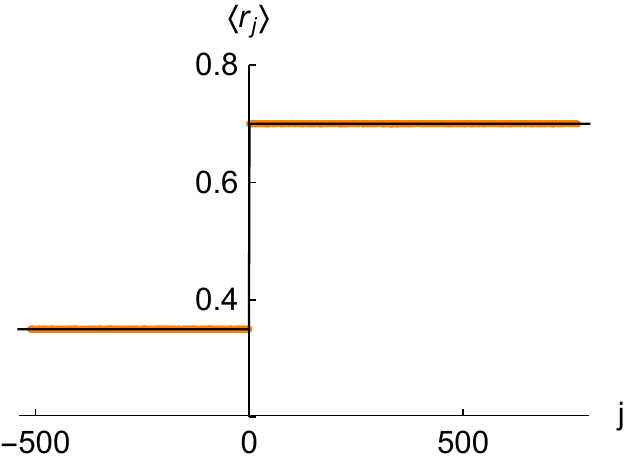}}
\hspace{0.02\textwidth}
\subfloat[stretch, $t = 16$]{\includegraphics[width=0.225\textwidth]{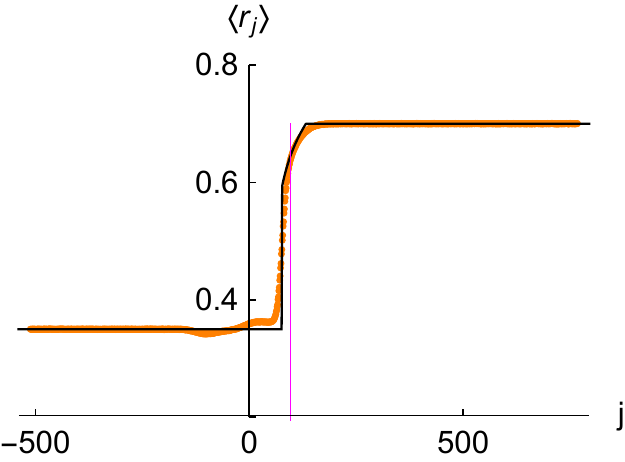}}
\hspace{0.02\textwidth}
\subfloat[stretch, $t = 32$]{\includegraphics[width=0.225\textwidth]{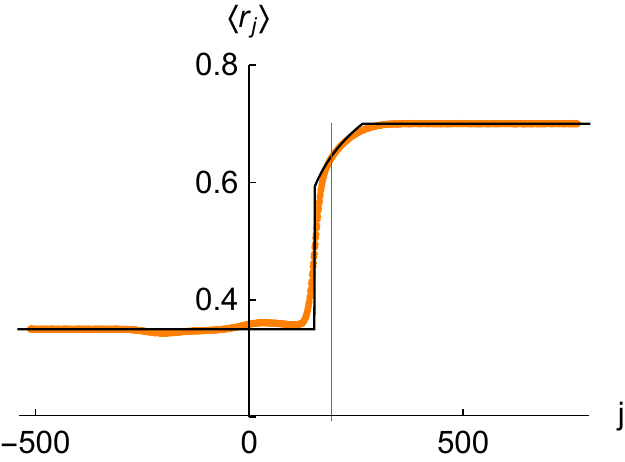}}
\hspace{0.02\textwidth}
\subfloat[stretch, $t = 64$]{\includegraphics[width=0.225\textwidth]{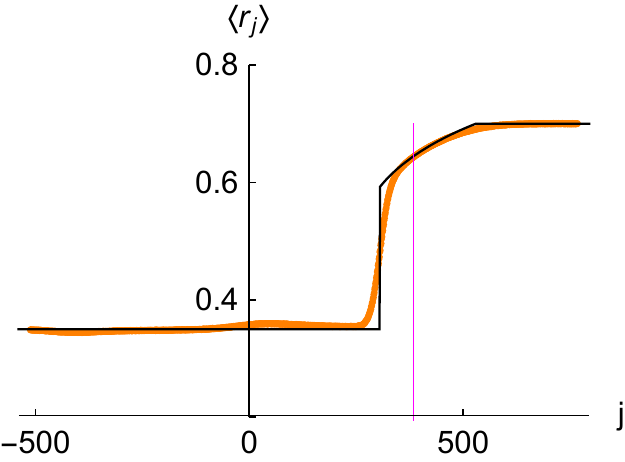}}\\
\subfloat[velocity, $t = 0$]{\includegraphics[width=0.225\textwidth]{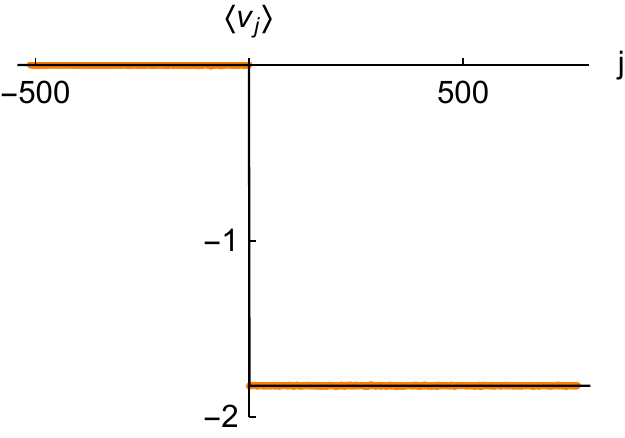}}
\hspace{0.02\textwidth}
\subfloat[velocity, $t = 16$]{\includegraphics[width=0.225\textwidth]{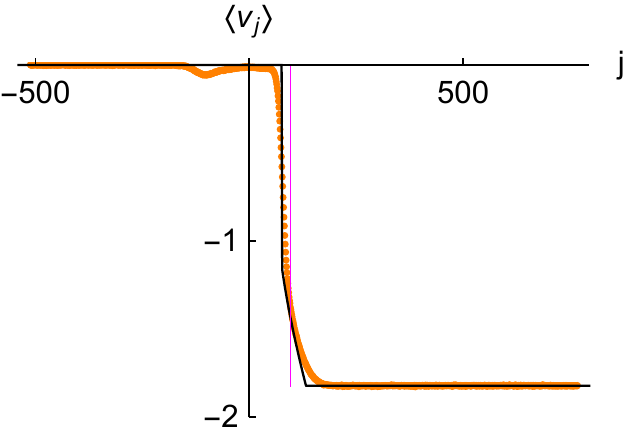}}
\hspace{0.02\textwidth}
\subfloat[velocity, $t = 32$]{\includegraphics[width=0.225\textwidth]{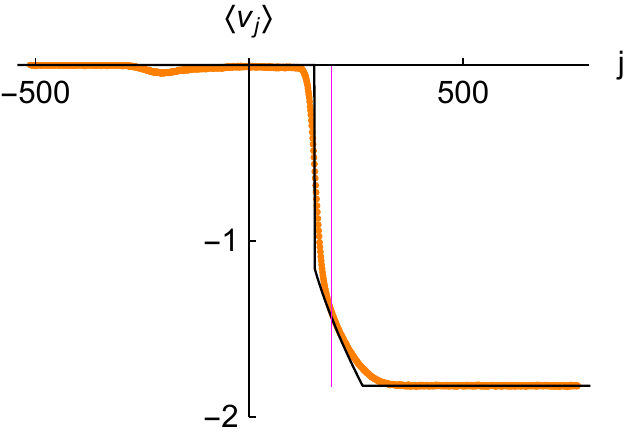}}
\hspace{0.02\textwidth}
\subfloat[velocity, $t = 64$]{\includegraphics[width=0.225\textwidth]{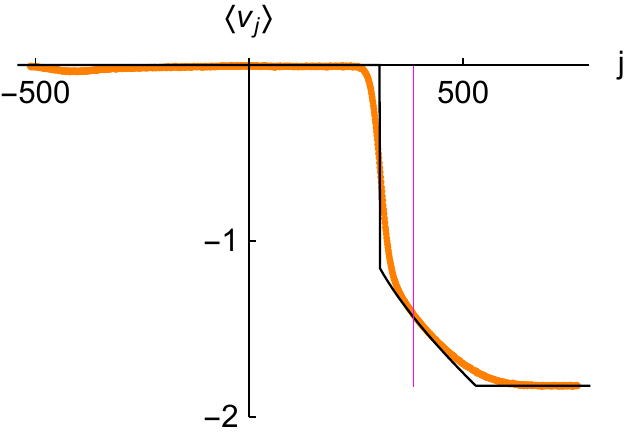}}\\
\subfloat[int.\ energy, $t = 0$]{\includegraphics[width=0.225\textwidth]{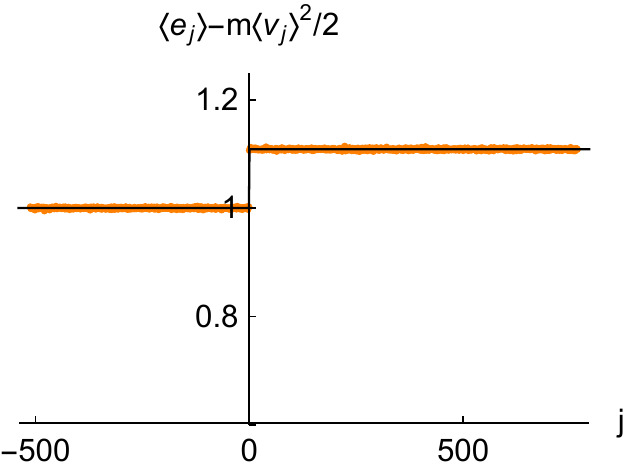}}
\hspace{0.02\textwidth}
\subfloat[int.\ energy, $t = 16$]{\includegraphics[width=0.225\textwidth]{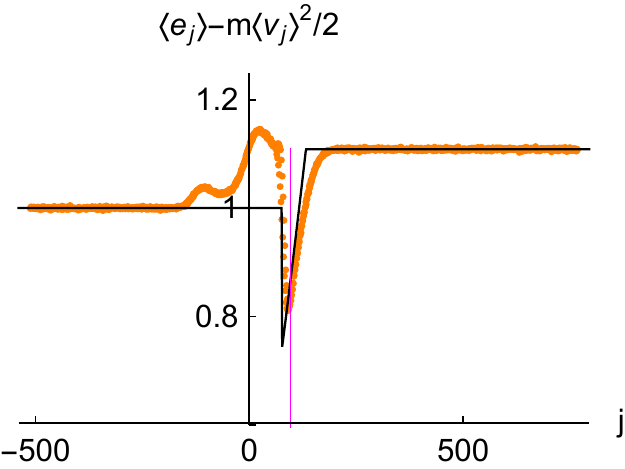}}
\hspace{0.02\textwidth}
\subfloat[int.\ energy, $t = 32$]{\includegraphics[width=0.225\textwidth]{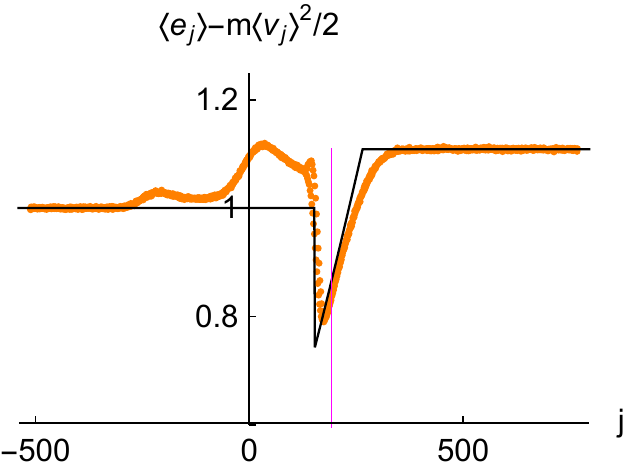}}
\hspace{0.02\textwidth}
\subfloat[int.\ energy, $t = 64$]{\includegraphics[width=0.225\textwidth]{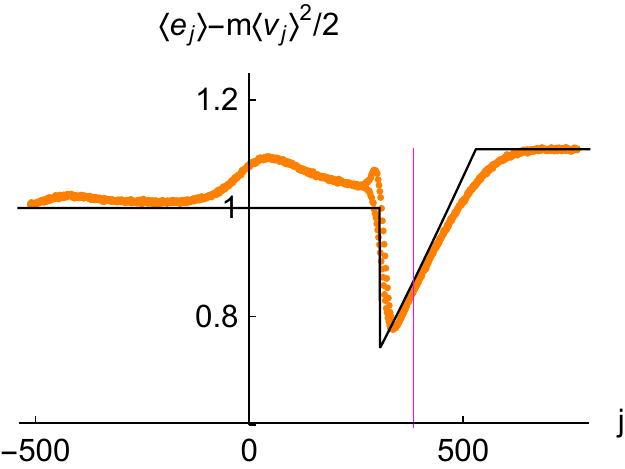}}\\
\subfloat[total energy, $t = 0$]{\includegraphics[width=0.225\textwidth]{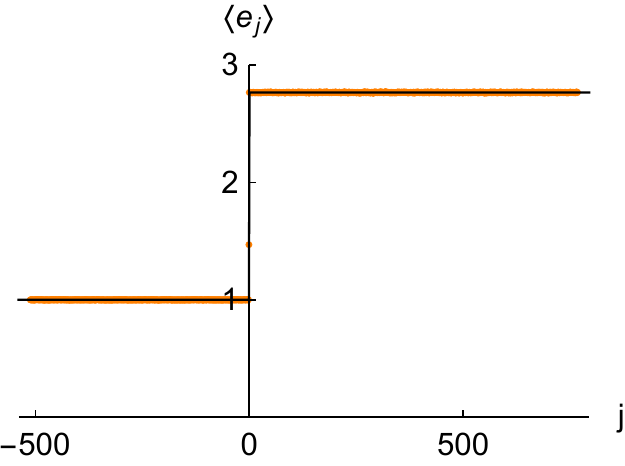}}
\hspace{0.02\textwidth}
\subfloat[total energy, $t = 16$]{\includegraphics[width=0.225\textwidth]{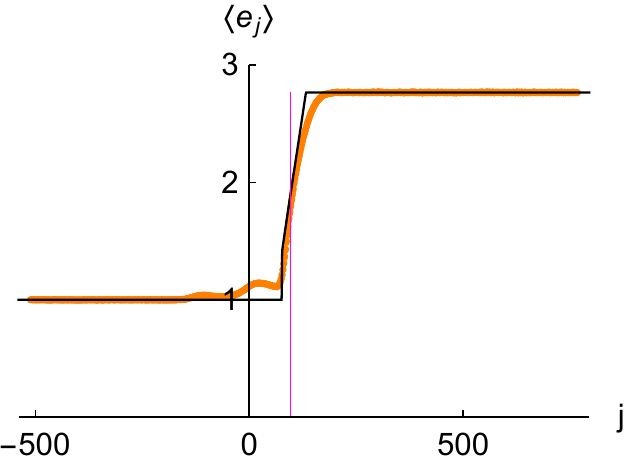}}
\hspace{0.02\textwidth}
\subfloat[total energy, $t = 32$]{\includegraphics[width=0.225\textwidth]{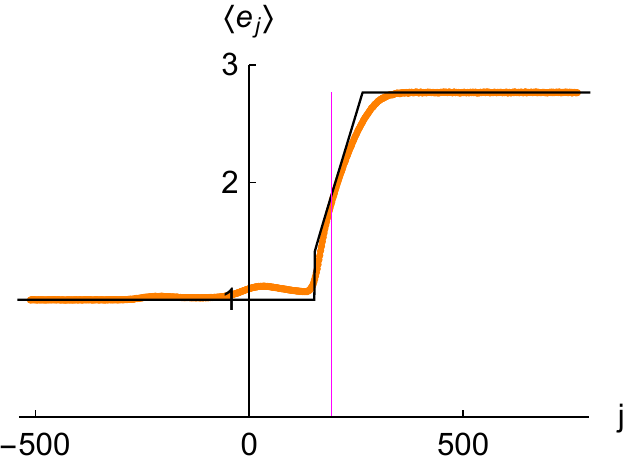}}
\hspace{0.02\textwidth}
\subfloat[total energy, $t = 64$]{\includegraphics[width=0.225\textwidth]{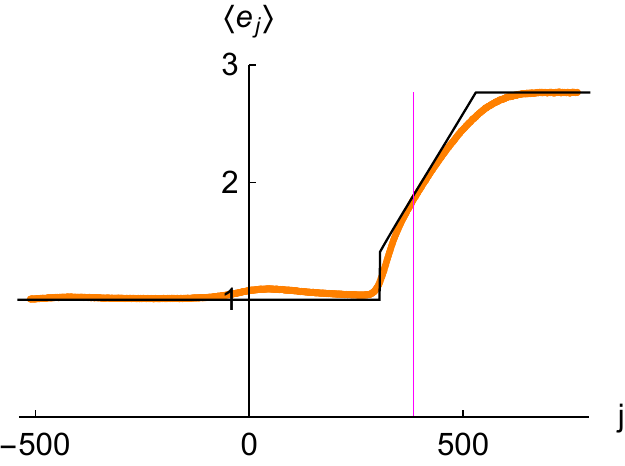}}
\caption{Stretch, velocity, internal energy and total energy profiles for the hard-point particle chain with square-well interaction potential, corresponding to the Riemann problem in Fig.~\ref{fig:squarewell_shock_rarefaction} with system size $L = 4096$. To increase visibility, only a lattice interval around the origin is shown.}
\label{fig:squarewell_profiles}
\end{figure}

Fig.~\ref{fig:squarewell_profiles} shows stretch, velocity, internal and total energy profiles obtained by molecular dynamics simulations with square-well interaction potential, after averaging over $10^6$ simulation runs and initial states chosen according to $\vec{u}_0$ and $\vec{u}_2$ in Fig.~\ref{fig:squarewell_shock_rarefaction}. The alternating masses are set as $m_0 = 1/2$ and $m_1 = 3/2$, such that the average mass $m = 1$. For stretch and velocity the shock followed by a rarefaction wave is well reproduced. For the total energy the shock is not so strong and hence hardly visible. The internal energy is not conserved. Our prediction is based on local equilibrium which apparently is not so accurate close to the shock. The entire system shows still further shocks and rarefaction waves due to periodic boundary conditions. The maximum time $t = 64$ in Fig.~\ref{fig:squarewell_profiles} is chosen prior to their collision with the structure shown in Fig.~\ref{fig:squarewell_shock_rarefaction}.

\section{Fluctuations of the time-integrated current}

In a famous contribution Johansson \cite{Johansson2000} considered the TASEP with $0|1$ step initial conditions and particles hopping only to the left. He proved that the time-integrated current along a given ray $\{x = \mathsf{v}t\}$ with $\lvert\mathsf{v}\rvert < 1$ behaves for large $t$ as
\begin{equation}\label{7.1}
\Phi(\mathsf{v}t,t) \simeq c_\mathsf{v}t + \kappa_{\mathsf{v}}(\Gamma_{\mathsf{v}} t)^{1/3}\xi_{\mathrm{GUE}}.
\end{equation}
Following standard conventions, the time scale is denoted by $\Gamma > 0$, in our particular case $\Gamma_{\mathsf{v}}$. $\kappa = \pm 1$ is the overall sign of the amplitude. The amplitude itself, $\xi_{\mathrm{GUE}}$, is a Tracy-Widom GUE distributed random variable, which was originally obtained as the distribution of the largest eigenvalue of a GUE random matrix \cite{TracyWidom1993, TracyWidom1994}. In formulas,
\begin{equation}\label{7.2}
\mathbb{P}(\xi_\mathrm{GUE} \leq s) = \det(1 - K_s)\vert_{L^2(\mathbb{R}_+)}
\end{equation}
with the Airy kernel
\begin{equation}\label{7.3}
K_s(x,x') = \int_0^\infty \ud \lambda\,\mathrm{Ai}(x+s+\lambda) \mathrm{Ai}(x'+s+\lambda),
\end{equation}
$\mathrm{Ai}$ denoting the standard Airy function. $c_\mathsf{v}$, $\kappa_\mathsf{v}$, and $\Gamma_\mathsf{v} $ are computed, model-dependent parameters, while the exponent $1/3$ and $\xi_\mathrm{GUE}$ are universal. Using distinct methods, later the result was extended to a general initial step \cite{PS2002, BenArousCorwin2011} and also to the ASEP \cite{TracyWidom2009}.

The Euler equation for the TASEP reads
\begin{equation}\label{7.4}
\partial_t u - \partial_x \big(u(1- u)\big) = 0,
\end{equation}
$u$ the particle density. To have Tracy-Widom fluctuations the solution to the $u_\ell | u_\mathrm{r}$ Riemann problem for \eqref{7.4} has to develop a rarefaction wave and the ray of integration must lie in the interior of the wave. In fact, the rarefaction profile happens to be linear, as for the LeRoux lattice gas. In general the profile will be nonlinear. Still, provided $\mathsf{v}$ is properly chosen, asymptotically the fluctuations of the time-integrated current are expected to have the same probability law as in \eqref{7.2}. We regard this observation as a strong indication that also the LeRoux lattice gas, even more ambitiously anharmonic chains, has Tracy-Widom statistics for the time-integrated current. There is one immediate difficulty with such a conjecture. The current is a vector. So which linear combination has a statistics governed by $\xi_{\mathrm{GUE}}$?

We will first study the fluctuations of time-integrated currents abstractly and then specialize to the LeRoux lattice gas and anharmonic chains with square-well type potential including the hard-point limit.

\subsection{Time-integrated currents}

To define the time-integrated current, in general, let us start from a conservation law of the form
\begin{equation}\label{7.5}
\partial_t u(x,t) + \partial_x \mathcal{J}(x,t) = 0.
\end{equation}
Thus, as a property special for one dimension, the vector field $(-u, \mathcal{J})$ is curl-free and hence admits a potential, $\Phi(x,t)$, up to a constant which we fix by $\Phi(0,0) = 0$. $\Phi(x,t)$ is then the current integrated from $(0,0)$ to $(x,t)$ along an arbitrary integration path. Numerically a convenient choice, to be used later on, is
\begin{equation}\label{7.6}
\Phi(x,t) = \int _0^t \ud t'\, \mathcal{J}(x,t') - \int_0^x \ud x'\, u(x',0),
\end{equation}
assuming $x> 0$, $t>0$. For a system with $n$ components, the same definition applies to each component separately and we set $\vec{\Phi} = (\Phi_1,\dots,\Phi_n)$. Clearly, the same argument works also for a spatial lattice with the $x'$-integration replaced by a lattice sum.

We now consider an $n$ component hyperbolic conservation law in the form
\begin{equation}\label{7.7}
\partial_t \vec{u} + \partial_x \vec{\mathsf{j}}(\vec{u}) = 0 \quad\mathrm{equivalently}\quad \partial_t \vec{u} + A(\vec{u})\partial_x \vec{u} = 0.
\end{equation}
The linearization matrix $A$ has eigenvalues $c_\sigma$, left eigenvectors, $\tilde{\psi}_\sigma$, and right eigenvectors, $\psi_\sigma$, $\sigma = 1,\dots,n$. The eigenvalues are assumed to be non-degenerate. We consider the $\vec{u}_\ell | \vec{u}_\mathrm{r}$ Riemann problem such that its solution contains a rarefaction wave across which $\vec{u}(x)$ increases (or decreases) smoothly for $x_\mathrm{min} < x < x_\mathrm{max}$. The current is integrated along the ray $\{x = \mathsf{v}t\}$, which has to lie inside the rarefaction wave, i.e. 
$x_\mathrm{min} < \mathsf{v} < x_\mathrm{max}$. The rarefaction wave is associated with a particular eigenvalue, whose label is denoted by $\sigma$ and regarded as fixed in the following. The quantity of interest is the distribution of integrated current $\vec{\Phi}(\mathsf{v}t,t)$. Along $\{x = \mathsf{v}t\}$ the fields take the value $\vec{u}_\mathsf{v}$ and $c_\sigma(\vec{u}_\mathsf{v}) = \mathsf{v}$. Hence, averaging 
\eqref{7.6} in local equilibrium to leading order in $t$,
\begin{equation}\label{7.8}
\vec{\Phi}(\mathsf{v}t,t) \simeq \big(\,\vec{\mathsf{j}}(\vec{u}_{\mathsf{v}}) - \mathsf{v} \vec{u}_{\mathsf{v}}\big) t.
\end{equation}

To access fluctuations we consider a point on $\{x = \mathsf{v}t\}$, field value $\vec{u}_\mathsf{v}$, and want to study small fluctuations with shape function $\vec{f}(x)$ which varies on a scale small compared to the variation of the rarefaction wave. Thus we have to linearize
\eqref{7.7} relative to a homogeneous background $\vec{u}_\mathsf{v}$. The resulting time evolution is given by 
\begin{equation}
(\mathrm{e}^{-A\partial_x t}\vec{f}\,)(x) = \sum_{\sigma'=1}^n|\psi_{\sigma'}\rangle\langle \tilde{\psi}_{\sigma'}|\vec{f}(x - c_{\sigma'}t)\rangle,
\end{equation}
where $\langle\cdot|\cdot\rangle$ denotes the scalar product for $n$-vectors and $A = A(\vec{u}_\mathsf{v})$. Since 
$c_\sigma(\vec{u}_\mathsf{v}) = \mathsf{v}$, only the term with $\sigma' = \sigma$ propagates along the ray $\mathsf{v}t$, while all other components separate from it linearly in time. Hence only $\langle \tilde{\psi}_{\sigma}|\vec{\Phi}(\mathsf{v}t,t) \rangle$ can build up anomalous fluctuations. In case of a single component, the quadratic term is responsible for the $t^{1/3}$ fluctuations. For several components, one first has to transform to normal modes. The strength of the self-coupling is denoted by $G^\sigma_{\sigma \sigma}$. The slope of the rarefaction profile does not vanish identically, hence classified to be in the KPZ universality class of curved profiles. For it the time scale of the anomalous fluctuations is set by $\Gamma_\sigma = \lvert G^\sigma_{\sigma \sigma}\rvert$. Thus we conjecture that for large $t$
\begin{equation}\label{7.9}
\big\langle \tilde{\psi}_{\sigma} \vert\vec{\Phi}(\mathsf{v}t,t) - t (\vec{\mathsf{j}}(\vec{u}_{\mathsf{v}}) - \mathsf{v} \vec{u}_{\mathsf{v}})\big\rangle
\simeq \kappa_\sigma(\Gamma_\sigma t)^{1/3} \xi_\mathrm{GUE}.
\end{equation}
Any linear combination of currents other than in Eq.~\eqref{7.9} encounters to some part almost independent contributions. Hence, if $\chi$ is not parallel to $\tilde{\psi}_{\sigma}$, the standard central limit theorem should apply in the form
\begin{equation}\label{7.10}
\big\langle \chi \vert\vec{\Phi}(\mathsf{v}t,t) - t (\vec{\mathsf{j}}(\vec{u}_{\mathsf{v}}) - \mathsf{v} \vec{u}_{\mathsf{v}})\big\rangle
\simeq (\Gamma_{\chi} t)^{1/2}\xi_\mathrm{G},
\end{equation}
where $\xi_\mathrm{G}$ is a standard Gaussian random variable. We have no theoretical prediction for the value of $\Gamma_{\chi}$. Of course it has to vanish as $\chi$ tends to $\tilde{\psi}_{\sigma}$.

As explained in Appendix~\ref{appendix:G}, the universal scale factor $\Gamma_\sigma = \lvert G^\sigma_{\sigma \sigma}\rvert$ can be computed from thermal averages. There we also establish that 
\begin{equation}
\label{eq:Gsigma_gradient_c}
G_{\sigma\sigma}^{\sigma} = \tfrac{1}{2} \sigma \psi_{\sigma} \cdot D c, \quad \sigma = \pm 1.
\end{equation}
Thus genuine nonlinearity is equivalent to $G_{\sigma\sigma}^{\sigma}$ having a definite sign for all admissible $r$, $e$.

\subsection{Monte-Carlo and molecular dynamics simulations}

To shorten notation, we define the projected current components in \eqref{7.9}, with their asymptotic value subtracted, as
\begin{equation}
\label{eq:Phi_sharp_def}
\Phi^{\sharp}_{\sigma}(t) = \big\langle \tilde{\psi}_{\sigma} \vert\vec{\Phi}(\mathsf{v}t,t) - t (\vec{\mathsf{j}}(\vec{u}_{\mathsf{v}}) - \mathsf{v} \vec{u}_{\mathsf{v}})\big\rangle.
\end{equation}
These are referred to as normal modes of the current.

\begin{figure}[!ht]
\centering
\subfloat[]{\includegraphics[width=0.3\textwidth]{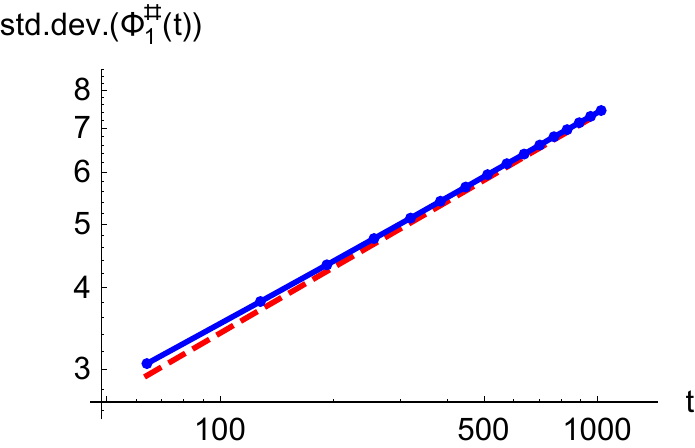}}
\hspace{0.15\textwidth}
\subfloat[]{\includegraphics[width=0.3\textwidth]{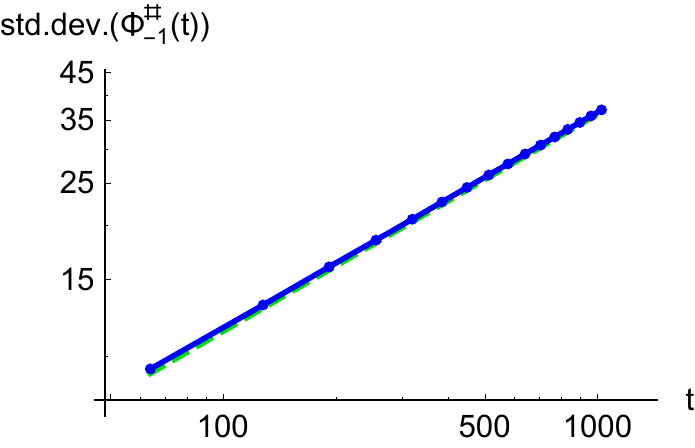}}\\
\subfloat[$(\Gamma_1 t)^{-1/3}\,\Phi^{\sharp}_1(t)$]{\includegraphics[width=0.3\textwidth]{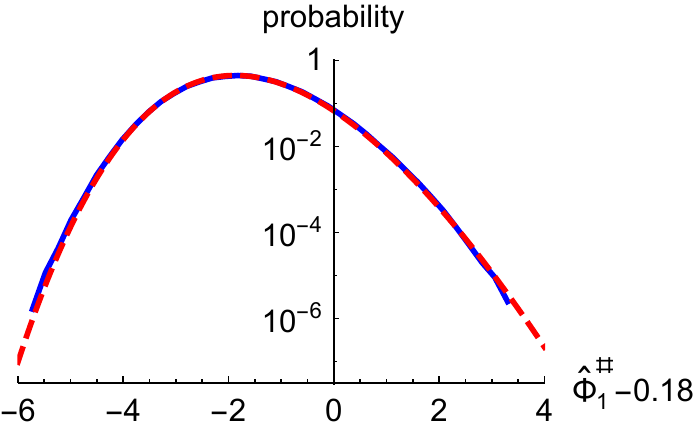}}
\hspace{0.15\textwidth}
\subfloat[$(\Gamma_{-1} t)^{-1/2}\,\Phi^{\sharp}_{-1}(t)$]{\includegraphics[width=0.3\textwidth]{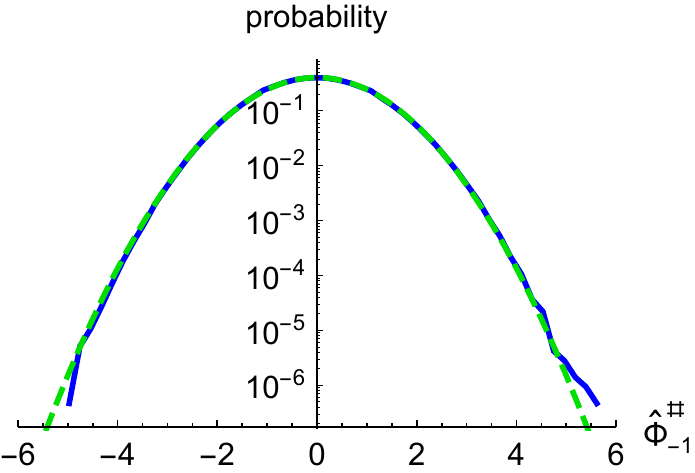}}
\caption{(a) Standard deviation of $\Phi^{\sharp}_1(t)$ for the LeRoux model as a function of time, compared with the theoretical prediction $\sigma_{\mathrm{GUE}} (\Gamma_1 t)^{1/3}$. (b) The standard deviation of $\Phi^{\sharp}_{-1}(t)$ shows central limit type fluctuations, scaling as $t^{1/2}$ (green dashed). (c) PDF of $(\Gamma_1 t)^{-1/3}\,\Phi^{\sharp}_1(t)$ at $t = 1024$ compared with the Tracy-Widom distribution (red dashed) and (d) PDF of $(1.34 t)^{-1/2}\,\Phi^{\sharp}_{-1}(t)$ compared with a normalized Gaussian (green dashed).}
\label{fig:LeRoux_scaling_distribution}
\end{figure}

\paragraph{LeRoux model.}
We record the integrated current $\vec{\Phi}(\mathsf{v}t,t)$ for the simulation parameters as in Fig.~\ref{fig:LeRoux_profiles} above, with $\mathsf{v} = \frac{2}{5}$ and $\mathsf{v}t$ marked as purple vertical line in Fig.~\ref{fig:LeRoux_profiles}. The integrated current is then transformed to normal modes via Eq.~\eqref{eq:Phi_sharp_def}, using the theoretical values for $\tilde{\psi}_{\sigma}$, $\vec{u}_{\mathsf{v}}$ and $\vec{\mathsf{j}}(\vec{u}_{\mathsf{v}})$. The resulting $\Phi^{\sharp}_{\sigma}(t)$ is shown in Fig.~\ref{fig:LeRoux_scaling_distribution} and compared with the theoretical predictions \eqref{7.9} and \eqref{7.10}. The top row shows the standard deviation of $\Phi^{\sharp}_{\sigma}(t)$ as a function of time, in comparison with $\sigma_{\mathrm{GUE}} (\Gamma_1 t)^{1/3}$ for $\sigma = 1$ and $\sim t^{1/2}$ for $\sigma = -1$, where $\sigma_{\mathrm{GUE}}$ denotes the standard deviation of the Tracy-Widom distribution. The corresponding probability density functions of $\Phi^{\sharp}_{\sigma}(t)$ in the bottom row of Fig.~\ref{fig:LeRoux_scaling_distribution} are reproduced from \cite{TWMendlSpohn2016}, and accurately match the predicted Tracy-Widom and Gaussian distributions, respectively. Note that the rescaling uses the theoretical value $\Gamma_1 = \lvert G^1_{1 1} \rvert = 0.539$, see also appendix~\ref{appendix:LeRoux}. However there is still a global shift by $0.18$. Such a shift is familiar from one-component models. The higher cumulants have all relaxed, while the mean is still drifting.

\begin{figure}[!ht]
\centering
\subfloat[std.~dev.~of $\Phi^{\sharp}_1(t)$]{\includegraphics[width=0.3\textwidth]{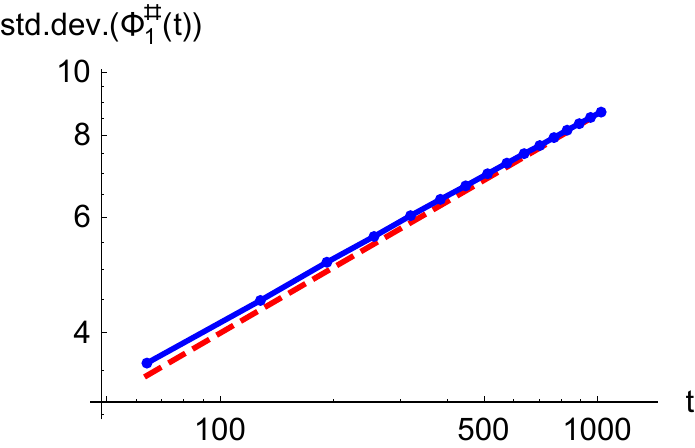}}
\hspace{0.04\textwidth}
\subfloat[std.~dev.~of $\Phi^{\sharp}_0(t)$]{\includegraphics[width=0.3\textwidth]{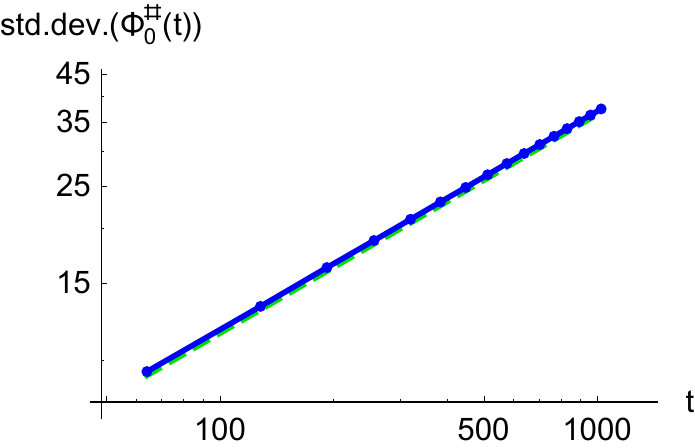}}
\hspace{0.04\textwidth}
\subfloat[std.~dev.~of $\Phi^{\sharp}_{-1}(t)$]{\includegraphics[width=0.3\textwidth]{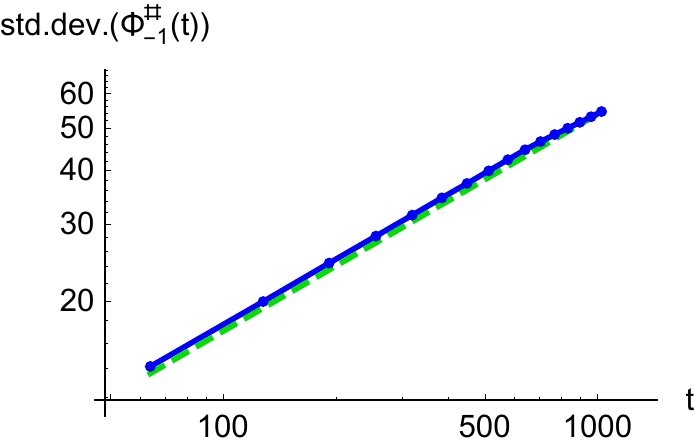}}\\
\subfloat[$(\Gamma_1 t)^{-1/3}\,\Phi^{\sharp}_1(t)$]{\includegraphics[width=0.3\textwidth]{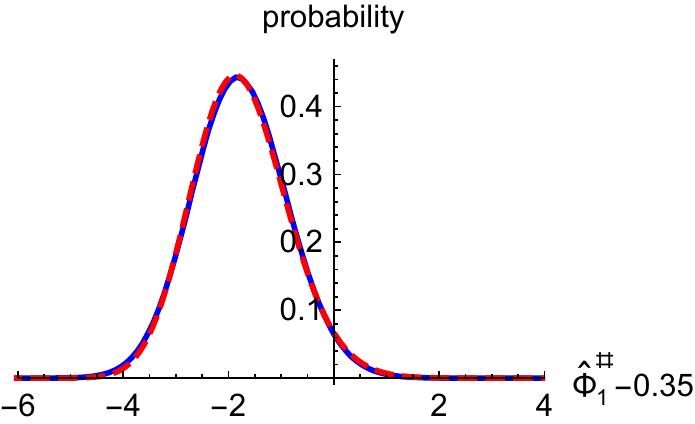}}
\hspace{0.04\textwidth}
\subfloat[$(\Gamma_0 t)^{-1/2}\,\Phi^{\sharp}_0(t)$]{\includegraphics[width=0.3\textwidth]{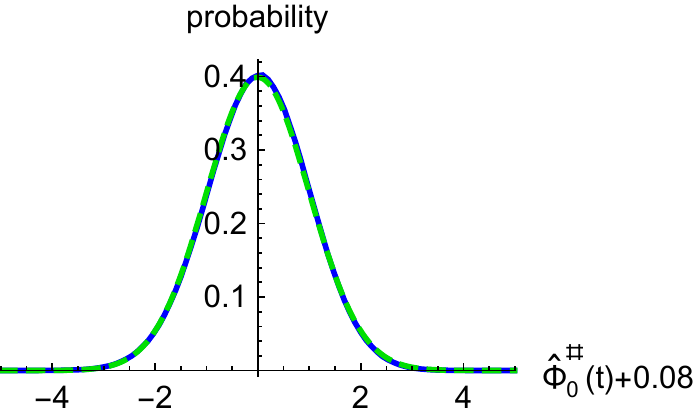}}
\hspace{0.04\textwidth}
\subfloat[$(\Gamma_{-1} t)^{-1/2}\,\Phi^{\sharp}_{-1}(t)$]{\includegraphics[width=0.3\textwidth]{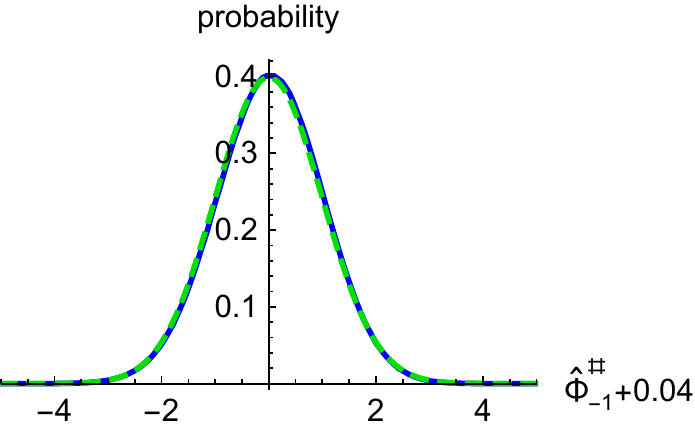}}\\
\subfloat[$(\Gamma_1 t)^{-1/3}\,\Phi^{\sharp}_1(t)$]{\includegraphics[width=0.3\textwidth]{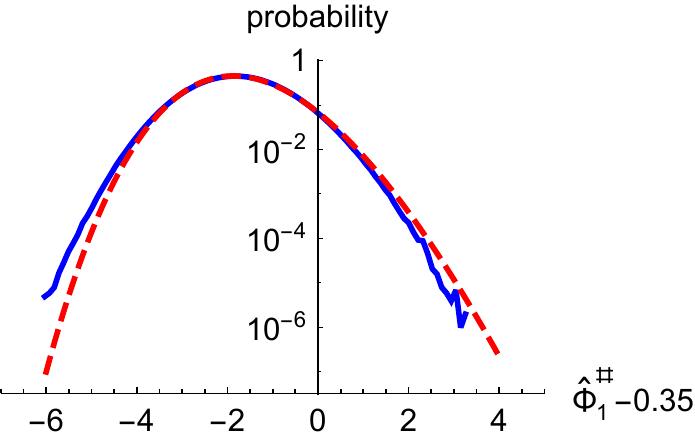}}
\hspace{0.04\textwidth}
\subfloat[$(\Gamma_0 t)^{-1/2}\,\Phi^{\sharp}_0(t)$]{\includegraphics[width=0.3\textwidth]{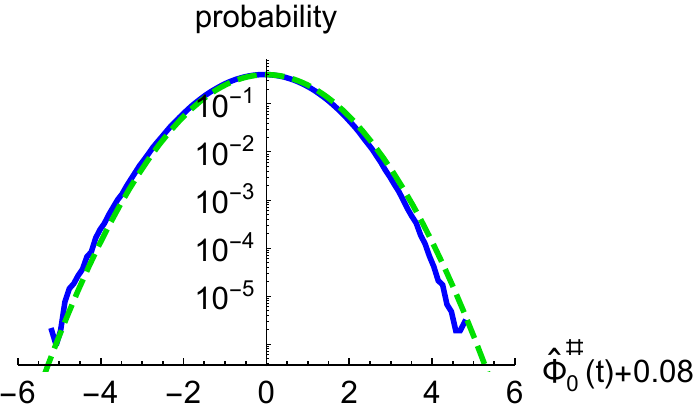}}
\hspace{0.04\textwidth}
\subfloat[$(\Gamma_{-1} t)^{-1/2}\,\Phi^{\sharp}_{-1}(t)$]{\includegraphics[width=0.3\textwidth]{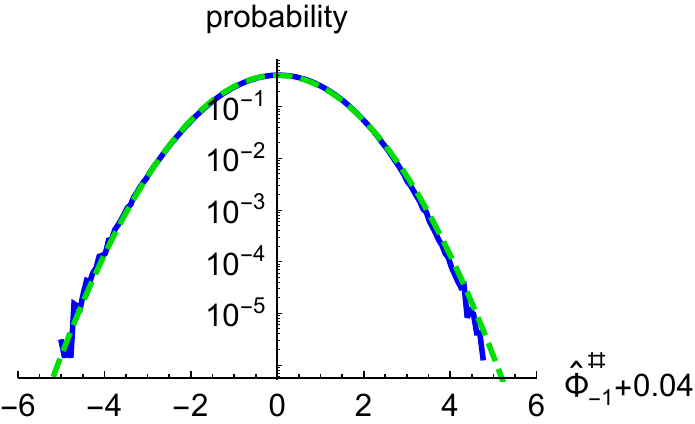}}
\caption{Top row: standard deviation of $\Phi^{\sharp}_{\sigma}(t)$ as a function of time for the hard-point particle chain with alternating masses. The $\sigma = 1$ component in (a) scales as $t^{1/3}$ (red dashed) and the standard deviations of the $\sigma = 0, -1$ components in (b) and (c) scale as $t^{1/2}$ (green dashed). Middle and bottom row: statistical distribution of the rescaled $\hat{\Phi}^{\sharp}_{\sigma}(t)$ at $t = 1024$. The red dashed curve in (d) and (g) is the predicted Tracy-Widom PDF. The projections for $\sigma = 0, -1$ follow a Gaussian distribution (green dashed).}
\label{fig:hardpoint_scaling_phi_distribution}
\end{figure}

\paragraph{Hard-point particles with alternating masses.}
Analogous to the LeRoux model, for the hard-point particle chain with alternating masses we integrate stretch, velocity, and energy currents in MD simulations along the purple ray in Figs.~\ref{fig:hardpoint_finite_vol_summary} and \ref{fig:hp_profiles}, to obtain $\vec{\Phi}(\mathsf{v}t,t)$. For anharmonic chains there are three normal modes $\Phi^{\sharp}_{\sigma}(t)$, $\sigma = -1, 0, 1$, which we compute from $\vec{\Phi}(\mathsf{v}t,t)$ via \eqref{eq:Phi_sharp_def}, again using the theoretical values for $\tilde{\psi}_{\sigma}$, $\vec{u}_{\mathsf{v}}$ and $\vec{\mathsf{j}}(\vec{u}_{\mathsf{v}})$. Only the $\sigma = 1$ mode is expected to follow a Tracy-Widom distribution. Fig.~\ref{fig:hardpoint_scaling_phi_distribution} shows the simulation results for $\Phi^{\sharp}_{\sigma}(t)$ in comparison with $\xi_\mathrm{GUE}$ for $\sigma = 1$ and normal distributions for $\sigma = 0, -1$. As before, a small correction to the mean values is indicated at the $1$-axis labels. The numerical fit uses $\Gamma_1=0.86$, while the theoretical prediction is $\lvert G^1_{1 1} \rvert = 0.559$.

\begin{figure}[!ht]
\centering
\subfloat[]{\includegraphics[width=0.3\textwidth]{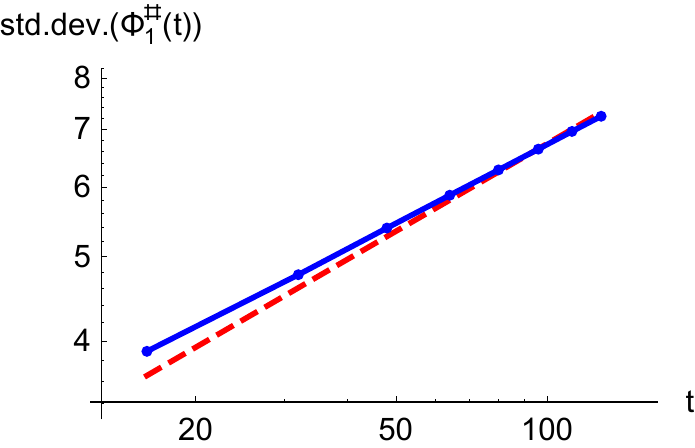}}
\hspace{0.04\textwidth}
\subfloat[]{\includegraphics[width=0.3\textwidth]{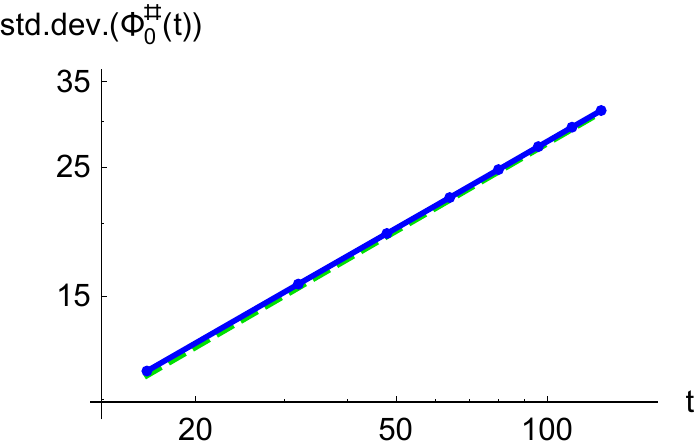}}
\hspace{0.04\textwidth}
\subfloat[]{\includegraphics[width=0.3\textwidth]{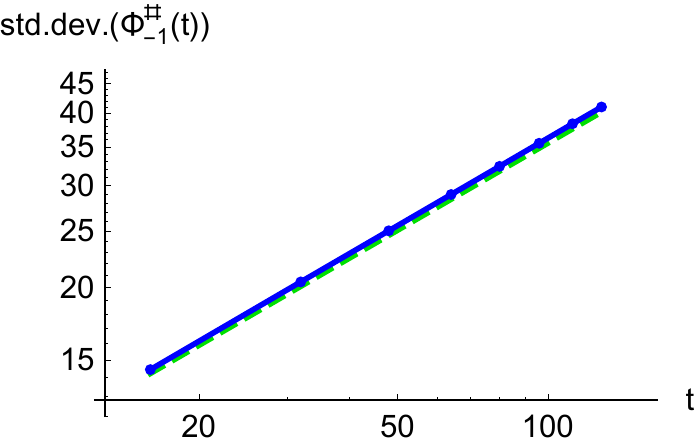}}\\
\subfloat[$(\Gamma_1 t)^{-1/3}\,\Phi^{\sharp}_1(t)$]{\includegraphics[width=0.3\textwidth]{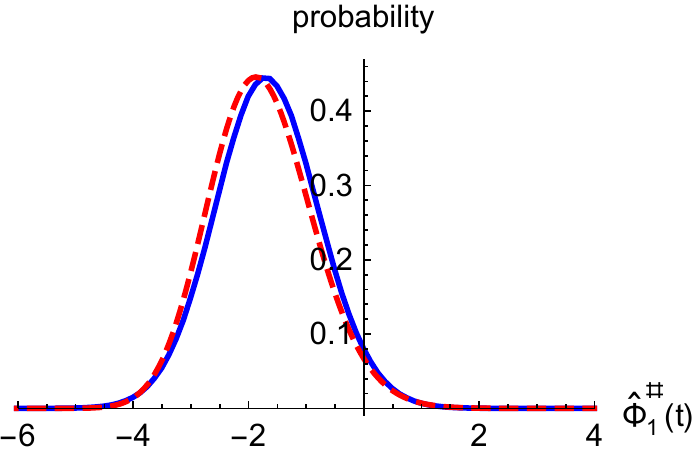}}
\hspace{0.04\textwidth}
\subfloat[$(\Gamma_0 t)^{-1/2}\,\Phi^{\sharp}_0(t)$]{\includegraphics[width=0.3\textwidth]{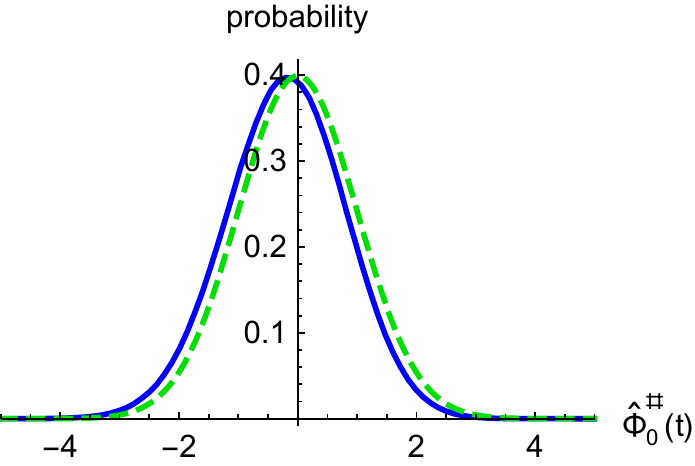}}
\hspace{0.04\textwidth}
\subfloat[$(\Gamma_{-1} t)^{-1/2}\,\Phi^{\sharp}_{-1}(t)$]{\includegraphics[width=0.3\textwidth]{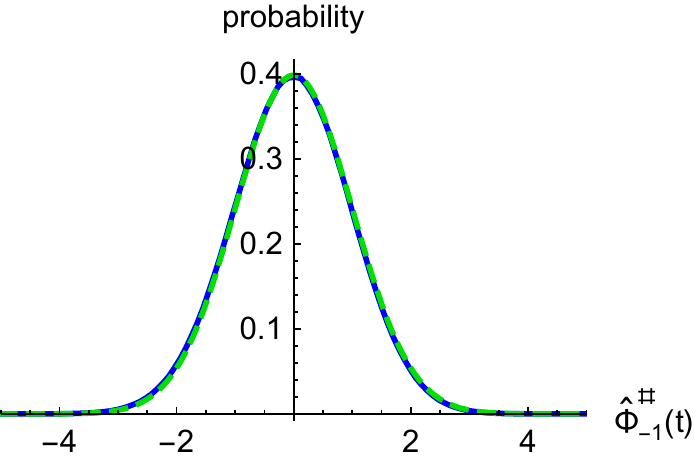}}\\
\subfloat[$(\Gamma_1 t)^{-1/3}\,\Phi^{\sharp}_1(t)$]{\includegraphics[width=0.3\textwidth]{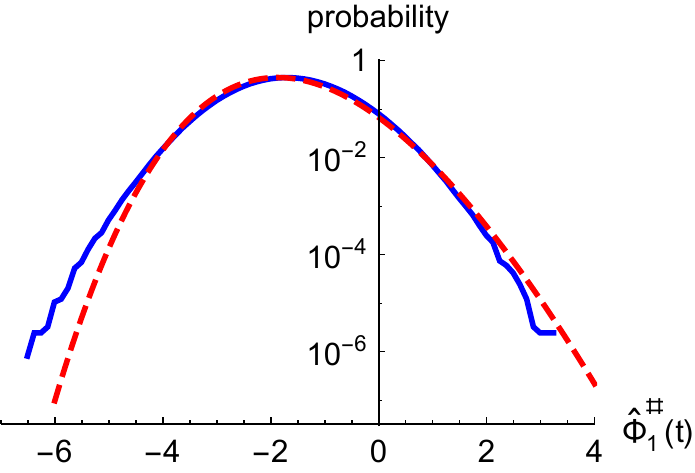}}
\hspace{0.04\textwidth}
\subfloat[$(\Gamma_0 t)^{-1/2}\,\Phi^{\sharp}_0(t)$]{\includegraphics[width=0.3\textwidth]{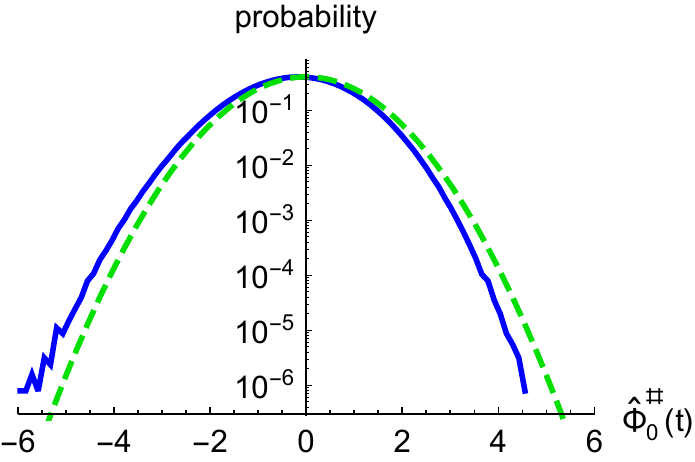}}
\hspace{0.04\textwidth}
\subfloat[$(\Gamma_{-1} t)^{-1/2}\,\Phi^{\sharp}_{-1}(t)$]{\includegraphics[width=0.3\textwidth]{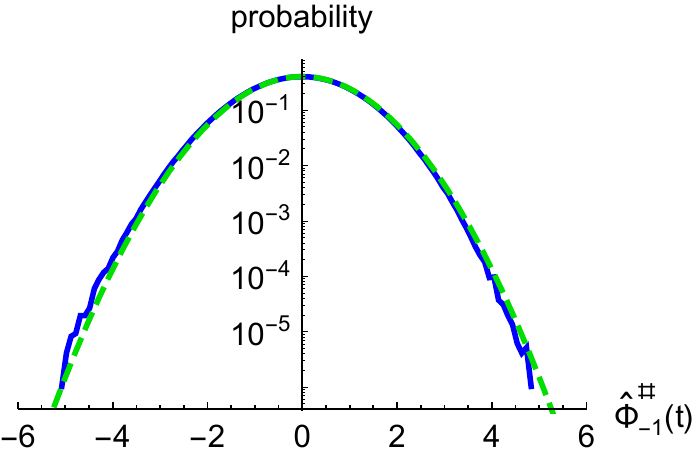}}
\caption{Top row: standard deviation of $\Phi^{\sharp}_{\sigma}(t)$ as a function of time for the square-well interaction potential. The $\sigma = 1$ component in (a) scales as $t^{1/3}$ (red dashed), while the standard deviations of the $\sigma = 0, -1$ components in (b) and (c) scale almost perfectly as $t^{1/2}$ (green dashed). Middle and bottom row: statistical distribution of the rescaled $\hat{\Phi}^{\sharp}_{\sigma}(t)$ at $t = 128$. The red dashed curve in (d) and (g) is the predicted Tracy-Widom PDF. The projections for $\sigma = 0, -1$ approximately follow a Gaussian distribution (green dashed).}
\label{fig:squarewell_scaling_phi_distribution}
\end{figure}

\paragraph{Square-well potential.}
To also have an example where genuine nonlinearity is violated, we repeat the analogous analysis for MD simulations with square-well interaction potential, using the parameters as in Figs.~\ref{fig:squarewell_shock_rarefaction} and \ref{fig:squarewell_profiles}. The current is integrated along the ray $\{x = \mathsf{v}t\}$ with $\mathsf{v} = 6$. To avoid collisions due to the periodic boundary conditions, the largest simulation time is relatively short, $t = 128$. The results for $\Phi^{\sharp}_{\sigma}(t)$ are shown in Fig.~\ref{fig:squarewell_scaling_phi_distribution} in comparison with the theoretical predictions. Different from the previous two examples, we show the simulation results without mean value correction. The agreement is slightly less precise as for the alternating masses. One reason could be due to the mentioned short simulation time. Nevertheless, one clearly observes that the mean value of $\Phi^{\sharp}_1(t)$ is close to that of $\xi_{\mathrm{GUE}}$, whereas $\Phi^{\sharp}_{\sigma}(t)$ for $\sigma = 0, -1$ are approximately centered around zero, in accordance with the theoretical prediction. The numerically fitted coefficient $\Gamma_1 = 4$, whereas the theoretical prediction is $\lvert G^1_{1 1} \rvert = 2.376$. The other two numerical coefficients are $\Gamma_{0} = 7.6$ and $\Gamma_{-1} = 13$. In Fig.~\ref{fig:squarewell_scaling_phi_distribution}, the sign of $\tilde{\psi}_1$ for the transformation to normal modes is flipped, in accordance with the changing sign in Eq.~\eqref{eq:gradient_c_rarefaction_h}.

\pagebreak

\section{Summary and conclusions}

We studied anharmonic chains with nonequilibrium initial conditions, specifically domain-wall. Two rather distinct theoretical predictions have been tested against molecular dynamics simulations.\\
(i) The validity of the macroscopic Euler equations, which predict flat profiles interrupted by shocks and rarefaction waves. Our system size is moderate, roughly 4000 particles, which we try to compensate by averaging over order $10^7$ configurations sampled according to the initial domain-wall state. For times up to the crossing time for a sound wave, the agreement between the solution of the Euler equations and the simulation is fairly accurate. Exceptions are shock fronts, where additional oscillatory structures may appear. Also the borders of rarefaction waves are somewhat rounded.\\
(ii) We measured currents integrated along a ray in space-time. If this ray lies in a domain with flat profile, one expects $\sqrt{t}$ size Gaussian fluctuations. We confirmed such a behavior in a related set-up \cite{MendlSpohn2014, MendlSpohnCurrent2015} and did not repeat it here. However, for a ray inside a rarefaction wave we do obtain $t^{1/3}$ size fluctuations with Tracy-Widom GUE distributed random amplitude, provided the current is projected onto the respective left eigenvector. Such a behavior is observed even in case the rarefaction wave does not perfectly approximate the Euler solution.

Turning to stochastic particle dynamics with several components, as to be expected, the precision of the approximation is improved. Our working example is the LeRoux lattice gas. In this model, even strong shocks are realized by the dynamics with a width of a few lattice sites. The Tracy-Widom distribution is confirmed at a precision known for the one-component ASEP.

\paragraph{Acknowledgements.} The work of HS has been supported as a Simons Distinguished Visiting Scholar, when visiting the KITP early 2016. CM acknowledges support from the Alexander von Humboldt Foundation and computing resources of the Leibniz-Rechenzentrum.

\appendix

\section{Coupling matrices}
\label{appendix:G}

We compute the $G$ coupling matrices for the LeRoux model and anharmonic chains, following the derivation in \cite{Spohn2014}.

\subsection{LeRoux model}
\label{appendix:LeRoux}

The linearization matrix $A$ and its left and right eigenvectors are stated in Eqs.~\eqref{eq:LeRoux_A} and \eqref{eq:psiLeRoux}. The transformation to normal modes is accomplished through the matrix $R$ defined by
\begin{equation}
R = \begin{pmatrix} \langle \tilde{\psi}_{-1} \rvert \\ \langle \tilde{\psi}_{1} \rvert \end{pmatrix}, \quad
R^{-1} = \Big( \lvert \psi_{-1} \rangle \ \lvert \psi_{1} \rangle \Big).
\end{equation}
By construction one has
\begin{equation}
R A R^{-1} = \mathrm{diag}(c_{-1}, c_1).
\end{equation}
As usual, the static susceptibility matrix, $C$, is given by
\begin{equation}
\label{eq:covarianceLeRoux}
C = \begin{pmatrix}
\langle 1 - \lvert\eta_j\rvert ; 1 - \lvert\eta_j \rvert \rangle & \langle 1 - \lvert\eta_j \rvert ; \eta_j\rangle \\
\langle 1 - \lvert\eta_j \rvert; \eta_j \rangle & \langle \eta_j ; \eta_j\rangle
\end{pmatrix}
=
\begin{pmatrix}
\rho (1 - \rho) & -\rho v \\
-\rho v & 1 - \rho - v^2 \\
\end{pmatrix}.
\end{equation}
In addition we then require
\begin{equation}
R C R^{\mathrm{T}} = \mathbbm{1},
\end{equation}
thereby fixing the normalizations $\tilde{Z}_{\sigma}$, $Z_{\sigma}$ in \eqref{eq:psiLeRoux} to
\begin{equation}
\tilde{Z}_{\sigma} = \sqrt{2} \left( 4 \rho (1 - \rho ) + v^2 \big(1 - 5 \rho -v^2\big) + \sigma\,v \big(3 \rho + v^2 - 1\big)\sqrt{4 \rho + v^2}\right)^{1/2},
\end{equation}
\begin{equation}
Z_{\sigma} = 2\,\tilde{Z}_{\sigma}^{-1} \left(4 \rho - v \big(\sigma \sqrt{4 \rho + v^2} - v\big)\right).
\end{equation}

To obtain the nonlinear couplings $G$, in particular $G^1_{11}$, we first compute the Hessians of the current
as second derivatives of $\vec{\mathsf{j}}(\vec{u})$,
\begin{equation}
H^{\rho} = - \begin{pmatrix} 0 & 1 \\ 1 & 0\end{pmatrix}, \qquad H^{v} = - \begin{pmatrix} 0 & 0 \\ 0 & 2\end{pmatrix}.
\end{equation}
In normal coordinates 
\begin{multline}
\langle \psi_{\sigma} \vert H^{\rho} \vert \psi_{\tau} \rangle
= \rho\,v \begin{pmatrix} 1 & 0 \\ 0 & 1\end{pmatrix} + \frac{2 \rho}{\sqrt{4 \rho + v^2}} \big(1 - \rho - \tfrac{1}{2} v^2\big) \begin{pmatrix} -1 & 0 \\ 0 & 1\end{pmatrix} \\
+ \frac{v}{\sqrt{4 \rho + v^2}} \sqrt{\rho \big((1 - \rho)^2 - v^2\big)} \begin{pmatrix} 0 & 1 \\ 1 & 0\end{pmatrix}
\end{multline}
and
\begin{multline}
\langle \psi_{\sigma} \vert H^{v} \vert \psi_{\tau} \rangle
= -\big(1 - \rho - v^2\big) \begin{pmatrix} 1 & 0 \\ 0 & 1\end{pmatrix} + \frac{v}{\sqrt{4 \rho + v^2}} \big(1 - 3 \rho - v^2\big) \begin{pmatrix} -1 & 0 \\ 0 & 1\end{pmatrix} \\
- \frac{2}{\sqrt{4 \rho + v^2}} \sqrt{\rho \big((1 - \rho )^2 - v^2\big)} \begin{pmatrix} 0 & 1 \\ 1 & 0\end{pmatrix}.
\end{multline}
The coupling matrices are thus obtained as
\begin{equation}
\begin{split}
G^{\sigma}
&= \tfrac{1}{2} \sum_{i=\{\rho,v\}} R_{\sigma i} \, R^{-\mathrm{T}} H^{i} R^{-1} \\
&= Z_{\sigma}^{-1} \big(\sqrt{4 \rho +v^2} - \sigma v\big) \begin{pmatrix} \tfrac{1}{2}(1 - \sigma) & 0 \\ 0 & \tfrac{1}{2}(1 + \sigma)\end{pmatrix} + \tilde{Z}_{\sigma}^{-1} \sqrt{\rho \big((1 - \rho)^2 - v^2\big)} \begin{pmatrix} 0 & 1 \\ 1 & 0\end{pmatrix}.
\end{split}
\end{equation}
In particular, comparison with \eqref{eq:Leroux_gradient_c_rarefaction} shows that
\begin{equation} 
G^{\sigma}_{\sigma\sigma} = \tfrac{1}{2} \psi_\sigma \cdot D c_{\sigma}.
\end{equation}

\subsection{General anharmonic chain}
\label{appendix:general_anharm}

To set the scale for the Tracy-Widom distribution, one has to compute $G_{11}^1$. For a general anharmonic chain, in the special case $v = 0$, the coupling matrices are derived in \cite{Spohn2014}, a result which should be extended to $v \neq 0$. In fact, it turns out that the coupling matrices do not depend on $v$.

Following the notation of \cite{Spohn2014}, the static susceptibility matrix is given by
\begin{equation}
C = \begin{pmatrix}
\langle r_j; r_j \rangle & \langle r_j; v_j \rangle & \langle r_j; e_j \rangle \\
\langle r_j; v_j \rangle & \langle v_j; v_j \rangle & \langle v_j; e_j \rangle \\
\langle r_j; e_j \rangle & \langle v_j; e_j \rangle & \langle e_j; e_j \rangle
\end{pmatrix}
=
\begin{pmatrix}
\langle y; y \rangle & 0 & \langle y; V \rangle \\
0 & 1/(m \beta) & v/\beta \\
\langle y; V \rangle & v/\beta & \frac{1}{2}\beta^{-2} + m v^2 \beta^{-1} + \langle V; V\rangle
\end{pmatrix}.
\end{equation}
The linearization matrix $A$ in \eqref{eq:A_general} and its right and left eigenvectors in \eqref{eq:psi} and \eqref{eq:psi_tilde}, respectively, define the transformation to normal modes via
\begin{equation} 
R = \begin{pmatrix} \langle \tilde{\psi}_{-1} \rvert \\ \langle \tilde{\psi}_{0} \rvert \\ \langle \tilde{\psi}_{1} \rvert \end{pmatrix}, \quad
R^{-1} = \Big( \lvert \psi_{-1} \rangle \ \lvert \psi_{0} \rangle \ \lvert \psi_{1} \rangle \Big)
\end{equation}
such that
\begin{equation}
R A R^{-1} = \mathrm{diag}(-c, 0, c), \quad R C R^{\mathrm{T}} = \mathbbm{1}.
\end{equation}
To have $R R^{-1} = \mathbbm{1}$, the normalization constants of the eigenvectors must satisfy
\begin{equation}
Z_{0} \tilde{Z}_{0} = m c^2, \qquad Z_{\sigma} \tilde{Z}_{\sigma} = 2 m c^2 \ \ \text{for}\ \ \sigma = \pm 1.
\end{equation}
An explicit computation of the diagonal entries of $R C R^{\mathrm{T}}$ shows that the velocity terms cancel. Hence the relations
\begin{equation}
\label{eq:Z_explicit}
\tilde{Z}_0 = \sqrt{m \Upsilon}\,c, \qquad \tilde{Z}_{\sigma} = \sqrt{2 m/\beta}\,c \ \ \text{for}\ \ \sigma = \pm 1
\end{equation}
from \cite{Spohn2014} remain valid in general, where $\Upsilon = \beta \left( \langle y; y \rangle\langle V; V \rangle - \langle y; V \rangle^2\right) + \frac{1}{2} \beta^{-1} \langle y; y \rangle$.

As in \cite{Spohn2014}, we denote the Hessian matrices of the average current by
\begin{equation}
H^{i}_{\alpha \beta} = \partial_{u_\alpha} \partial_{u_\beta}\,\mathsf{j}_{i}
\end{equation}
with the conserved fields $\vec{u} = (r, v, \mathfrak{e})$ and the current vector defined in \eqref{eq:avr_current}. The coupling matrices are then given by
\begin{equation}
G^{\sigma} = \tfrac{1}{2} \sum_{i=1}^3 R_{\sigma i} \, R^{-\mathrm{T}} H^{i} R^{-1}
\end{equation}
for $\sigma = -1, 0, 1$. While the Hessian matrices $H^{i}$ depend on $v$, the coupling matrices are actually independent of $v$. Thus using the formulas in \cite{Spohn2014} one arrives at
\begin{equation}
\label{eq:G0_general}
G^0 = \frac{1}{2 \beta \sqrt{m \Upsilon}}
\begin{pmatrix}
-1 &  0 &  0 \\
 0 &  0 &  0 \\
 0 &  0 &  1
\end{pmatrix}
\end{equation}
and for $\sigma = \pm 1$
\begin{equation}
\begin{split}
\label{eq:Gsigma_general}
G^{\sigma} &=
\frac{P\,\partial_e c - \partial_r c }{2 \sqrt{2 m \beta}\,c}
\begin{pmatrix}
 1 &  0 & -1 \\
 0 &  0 &  0 \\
-1 &  0 &  1
\end{pmatrix}
- \frac{\partial_e P}{\sqrt{2 m \beta}}
\begin{pmatrix}
 \frac{1}{2}(1+\sigma) &  0 &  0 \\
 0                     &  0 &  0 \\
 0                     &  0 &  \frac{1}{2}(1-\sigma)
\end{pmatrix} \\
&\quad + \frac{\Upsilon}{2 \sqrt{2m/\beta}\,m c^2} \left[ (\partial_r P)^2 (\partial_e^2 P) - 2 (\partial_r P) (\partial_r \partial_e P) (\partial_e P) + (\partial_r^2 P) (\partial_e P)^2 \right]
\begin{pmatrix}
 0 & 0 & 0 \\
 0 & 1 & 0 \\
 0 & 0 & 0
\end{pmatrix} \\
&\quad + \frac{\sqrt{\Upsilon}}{2 \sqrt{m}\,c} \left[(\partial_r P) (\partial_e c) - (\partial_e P) (\partial_r c) \right]
\begin{pmatrix}
 0 &  1 &  0 \\
 1 &  0 & -1 \\
 0 & -1 &  0
\end{pmatrix}.
\end{split}
\end{equation}
The relation \eqref{eq:Gsigma_gradient_c} follows by using on both sides the expressions provided above. 

Note that the signs of some entries in $G^{\sigma}$ are flipped compared to \cite{Spohn2014}, which is due to different sign conventions for the eigenvectors of $A$.

\subsection{Hard-point and square-well potential}

The coupling constants for these models have been discussed already in Appendix A of \cite{MendlSpohn2014}. For completeness, here we adapt to the current sign convention for the eigenvectors and using the velocity (instead of momentum) as field variable. The linearization matrix $A$ and its right eigenvectors are stated in Eqs.~\eqref{5.7} and \eqref{5.9}. The corresponding left eigenvectors of $A$ are
\begin{equation}
\label{eq:tilde_psi_h}
\tilde{\psi}_{0,h} = \tilde{Z}_{0,h}^{-1} \begin{pmatrix} 2 e h \\ -m v \\ 1 \end{pmatrix}, \qquad
\tilde{\psi}_{\sigma,h} = \tilde{Z}_{\sigma,h}^{-1} \begin{pmatrix} 2 e \sigma h' \\ m(c_h - 2 \sigma v h) \\ 2 \sigma h \end{pmatrix}.
\end{equation}
Since the interaction potential is either zero or infinite, $\Upsilon_h = \frac{1}{2} \beta^{-1} \langle y; y \rangle = -e/h'$, and the normalization constants in \eqref{eq:Z_explicit} become
\begin{equation}
\tilde{Z}_{0,h} = \sqrt{m e}\,c_h/\sqrt{-h'}, \qquad \tilde{Z}_{\sigma,h} = 2 \sqrt{m e}\,c_h \,.
\end{equation}

Specializing \eqref{eq:G0_general} and \eqref{eq:Gsigma_general} to the square-well interaction potential leads to the coupling matrices
\begin{equation}
G_h^0 = \sqrt{- h'\,e/m} \begin{pmatrix}
-1 &  0 &  0 \\
 0 &  0 &  0 \\
 0 &  0 &  1
\end{pmatrix}
\end{equation}
and for $\sigma = \pm 1$
\begin{equation}
G_h^{\sigma} = \frac{1}{2} \sqrt{e/m} \left[\frac{1}{2(2 h^2 - h')}
\begin{pmatrix}
 a_3 &  a_1 & -a_3 \\
 a_1 &  a_2 & -a_1 \\
-a_3 & -a_1 &  a_3
\end{pmatrix}
- 4 h
\begin{pmatrix}
 \frac{1}{2}(1+\sigma) &  0 &  0 \\
 0                     &  0 &  0 \\
 0                     &  0 &  \frac{1}{2}(1-\sigma)
\end{pmatrix}
\right]
\end{equation}
with
\begin{equation}
\begin{split}
a_1 &= 2 (-h')^{-1/2} \big( h h'' - h'^2 - 2 h^2 h' \big), \\
a_2 &= 4 h (-h')^{-1}\big( h h'' - 2 h'^2 \big),\\
a_3 &= 4 h^3 - 6 h h' + h'' .
\end{split}
\end{equation}
As above, the signs of some entries in $G_h^{\sigma}$ are flipped compared to \cite{MendlSpohn2014}, due to different sign conventions for the eigenvectors of $A$.

{\small

}

\end{document}